\documentclass[prx,longbibliography,aps,twocolumn,showpacs,superscriptaddress,10pt]{revtex4-1}
\usepackage{comment}
\usepackage{amsmath}
\usepackage{amsfonts}
\usepackage{amssymb}
\usepackage{graphicx}
\usepackage{color}
\usepackage{bm}
\usepackage{soul}
\usepackage{xcolor}
\usepackage{braket}
\usepackage{footmisc}
\usepackage{todonotes}
\usepackage{enumitem}

\newcommand{\dd}{\mathrm{d}}

\newcommand{\kk}{\mathbf{k}}

\newcommand{\pp}{\mathbf{p}}
\newcommand{\vecphi}{\mathbf{\Phi}}

\newcommand{\beq}{\begin{equation}}
\newcommand{\eeq}{\end{equation}}

\usepackage{braket}
\usepackage{soul}
\usepackage{color}
\makeatother
\usepackage{placeins}

\newcommand{\bsub}{\begin{subequations}}
\newcommand{\esub}{\end{subequations}}

\DeclareMathOperator{\Tr}{Tr}
\DeclareMathOperator{\sgn}{sgn}

\DeclareMathOperator{\dn}{dn}
\newcommand{\ord}[1]{\bm{\mathit{O}}\left(#1\right)}

\newcommand{\vex}[1]{\bm{\mathrm{#1}}}

\newcommand{\e}{\varepsilon}
\newcommand{\Rh}{\mathsf{R}}
\newcommand{\re}{\mathrm{Re}}
\newcommand{\im}{\mathrm{Im}}

\begin{document}

\title{Dynamical phase transitions in the collisionless pre-thermal states\\ of isolated quantum systems: theory and experiments }

\author{Jamir Marino}
\affiliation{Institut f\"{u}r Physik, Johannes Gutenberg-Universit\"{a}t Mainz, D-55099 Mainz, Germany}

\author{Martin Eckstein}
\affiliation{Department of Physics, University of Erlangen-N\"{u}rnberg, 91058 Erlangen, Germany}

\author{Matthew S. Foster}
\affiliation{Department of Physics and Astronomy, Rice University, Houston, Texas 77005, USA}
\affiliation{Rice Center for Quantum Materials, Rice University, Houston, Texas 77005, USA}

\author{Ana Maria Rey}
\affiliation{
JILA, National Institute of Standards and Technology,
and Department of Physics, University of Colorado, Boulder, CO 80309}
\affiliation{ Center for Theory of Quantum Matter, University of Colorado, Boulder, CO 80309}

\date{\today}

\begin{abstract}
 
We overview the concept of 
  dynamical phase transitions    in isolated quantum systems quenched out of equilibrium. 
  We focus on non-equilibrium transitions characterized by an order parameter, which features qualitatively distinct    temporal behaviour on the two sides of a  certain dynamical critical point. 
  Dynamical phase transitions are currently mostly understood as long-lived prethermal phenomena in a regime  where inelastic collisions are incapable to thermalize the system. The latter enables the dynamics to substain  phases that explicitly break detailed balance and therefore cannot be encompassed by traditional thermodynamics.
  Our presentation covers both cold atoms as well as condensed matter  systems. 
  We revisit a broad plethora of platforms exhibiting pre-thermal DPTs, which become theoretically tractable in a certain limit, such as for a large number of particles, large number of order parameter components, or large spatial dimension. 
The  systems we explore  include, among others,  quantum magnets with collective interactions, $\phi^4$ quantum field theories, and Fermi-Hubbard models. 
A section dedicated to experimental explorations  of DPTs in condensed matter and AMO systems connects this large variety of theoretical models.

\end{abstract}

%\tableofcontents

\maketitle

% Dynamics  of correlations play an ubiquitous and central role in many-particle physics. They    control the buildup of   quantum fluctuations~\cite{cheneau2012light,calabrese2006time} and   repconventional use of un
%%%%%%%%%%%%%%%%%%%%%%%%%%%%%%%%%%%%%%%%%%%%%%%%%%%%%
\tableofcontents

\section{Introduction}

Dynamical  phase transitions  occur whenever the observables of an isolated quantum system   feature  distinct qualitative temporal behavior as a  function of a control parameter measuring the deviation of the system from equilibrium. %, which plays the analogue role of the temperature in equilibrium transitions.
In this review, we are interested in a particular class of {dynamical phase transitions} (DPTs)  which   are  characterized by an order parameter    that features   non-analytic behavior at   the non-equilibrium  critical point separating distinct dynamical phases \cite{Henkel2008}.  The DPTs we discuss here manifest during the so called pre-thermal stage of  dynamics of  a quantum many body system~\cite{Berges2004b,gring2012relaxation,Langen2016}. 
The notion of {pre-thermalization} is broad  encompassing high energy physics and condensed matter. A system is said to pre-thermalize when its observables approach a long-lived quasi-steady state at intermediate times, before inelastic collissions set in and the system reaches  thermodynamic equilibrium.
In this regard, the behaviour of the order parameter witnessing the DPT will manifest  after a transient where non-universal effects attributed to microscopic details of the model, such as lattice properties, or short-time inhomogeneous dephasing, have faded away.

Pre-thermalization is mostly studied in the context of small interaction quenches that can result in a long, perturbative, window of the dynamics, where the quasi-particles of some pre-quench integrable model are weakly deformed and still free to scatter elastically~\cite{Kollar2011,Robinson2014,Bertini2015,marcuzzi2016prethermalization,PhysRevB.94.245117,PhysRevLett.127.130601,PhysRevA.95.023621}. Interactions   renormalise the energy of these excitations,  and only at later times they induce inelastic processes responsible for the redistribution of energy and for the eventual  equilibration. Accordingly, we will refer to  pre-thermalization as a \emph{collisionless regime}, which is a shorthand to describe that inelastic collisions are still weak and ineffective. 

In order to access DPTs, one can perform  quenches in Hamiltonian parameters. In the last years this is becoming accessible   in
 AMO and to some extent also in condensed matter  experiments. Up to date the main focus has been on DPTs that take place  in  some  limit such that interactions can  still be treated non-perturbatively at the expense of solving dynamics which are mean-field like, gaussian or semi-classical. 
The study of DPTs in this regime is the central topic of this review. 
We will revisit a broad plethora of systems exhibiting     collisionless behaviour within   a long lived pre-thermal window, enabled by taking a large  intrinsic parameter in the system such as  the number of particles ($N$) -- in Sec.~\ref{sec:lmg} or \ref{sec:LaxMethod}, and   the number of  components of a field ($\mathcal{N}$) -- in Sec.~\ref{sec:ON}. Further, some aspects of DPTs beyond the collisionless regime can be studied in a controlled manner in limit of infinite dimensions, which is reviewed in  Sec.~\ref{sec:dmft}.

The focus on a non-equilibrium order parameter which characterizes the DPT,  distinguishes the subject of this review from the notion of dynamical transitions explored in the context of Loschmidt echoes~\cite{heyl2013dynamical,Jurcevic2017,Heyl2019}. We will also not review the broad topic of non-thermal fixed points in cold atoms~\cite{ber2008,super2011,pruf2018,erne2018,eigen2018} (see~\cite{gasen2019} for a review), which although appear akin to a pre-thermal phenomenon,  exhibits a different mechanism from the  large $N$, $\mathcal{N}$, or $d$  pre-thermal states discussed in this review.

Regarding its experimental observation, broadly speaking, non-equilibrium dynamical phases  are observed  in much wider contexts than their equilibrium counterparts. They are relevant for a  variety of disciplines beyond AMO and condensed matter physics,  including  chemistry, biology and even sociology \cite{Haken1975}.
For example, non-equilibrium critical phenomena have been  seen in, 
 polymers, colloidal gels, molecular glasses and  spin glasses  as they are   cooled  down below  a critical  temperature, ``glassy transition  point,'' at which the system’s evolution slows
down so much that it falls out of equilibrium \cite{Henkel2008}.
 Lasers also show dynamical transitions,   where the  light  emitted by an array of atoms changes abruptly  from incoherent to
extremely coherent when  the input power
(the control parameter in this case) exceeds a certain threshold \cite{Haken1975}.
Another example are liquid layers,  where the heat transport changes from conduction to convection when the
temperature-gradient between the lower and the upper surfaces of the liquid  is increased above a critical point \cite{Chandrasekhar1968}.
Finally, different dynamical phases occur in the   population dynamics or ecology when two interacting 'kinds' of species, the predator and the prey,    start to feature a  perpetual  cyclic pattern when the food supply for one species is  maintained at or beyond a certain limit \cite{Chakrabarti1995}. Similar behavior is seen in chemical reactions  where  a chemical instability triggers oscillatory patterns in which the colour of the reactants changes periodically from red to blue and vice versa \cite{Glansdorff1971}.
Although it is clear that in any of these situations  the concepts of temperature, and thermal equilibrium,   do not have a clear  meaning, all of them share the fact that the familiar notion of the order parameter, introduced by Landau for a description of second order phase transitions, can be used to   describe the distinct change of behavior of the system as an external control parameter is varied. This is profitable since it allows to   treat disparate systems   at a macroscopic level with  similar concepts and mathematical tools,  and allow us to make fascinating analogies between them invoking the concept of universality. 

Nevertheless,  a key point shared by all the systems described in the above paragraph is that they belong  to  the realm of non-equilibrium open systems, and therefore  their   dynamics are fundamentally irreversible \cite{Haken1975}. A more restricted scenario, which is the one discussed in this review, deals with evolution under fully unitary, coherent  and reversible dynamics. Under this context experimental observations of non-equilibrium phase transitions have  remained more elusive. Only in  the recent years new opportunities in many-body  physics have been opened up thanks to the increased degree of control  and   ability to synthesize, manipulate, and detect ultra cold atomic systems as well as new developments in THz pump-probe experiments.  The novel experimental capabilities are  stimulating  new  methods to benchmark and understand DPTs   in general, and they are leading to  exciting investigations of  non-equilibrium behaviors. 
 
 The review is structured as follows. In Sec.~\ref{sec:lmg} and Sec.~\ref{sec:LaxMethod} we present the dynamics of exactly solvable fully-connected spin models in the limit of large number of spins, $N\to\infty$. The associated DPTs include the Ising universality class and Richardson-Gaudin magnets. In Sec.~\ref{longrange}  we discuss different dynamical behaviours associated to interactions which are not all-to-all, but  decay spatially in a power law fashion.
 In Sec.~\ref{sec:ON}, we consider the limit of large number of components ($\mathcal{N}$) of a $\phi^4$ field theory. This serves as an instructive toy model to show how Gaussian fluctuations can modify the physical picture of Secs.~\ref{sec:lmg} and~\ref{sec:LaxMethod}.    
  In Sec.~\ref{sec:dmft}  we consider DPTs arising in strongly correlated systems in the large dimensionality limit $d\to\infty$, where an exact solution is possible within  
 dynamical mean field theory (DMFT); interestingly the model studied in this limit exhibits ergodic behaviour and eventually thermalizes, allowing to inspect aspects of DPTs beyond pre-thermalization.    Emphasis is placed on the similarities between the DPTs observed in this class of models and those discussed in Sec.~\ref{sec:lmg}.
  Finally, Sec.~\ref{sec:exp} discusses    the experimental platforms where DPTs have been observed from solid state to cold atoms. This section  serves as  further connector among  the other three theory sections.

\section{Dynamical phase transitions \\and the Landau-Ginzburg paradigm}
\label{intro}

The Landau-Ginzburg description of thermal equilibrium phase transitions is one of the cornerstones of criticality and universality~\cite{landau1969statistical} in a vast plethora of systems~\cite{hohenberg2015introduction} including   soft matter (e.g. liquid-gas critical points), condensed matter (e.g. magnets,   superfluidity, and superconductivity), as well as applications to cosmology~\cite{Berges2004a}. Together with a systematic inclusion of thermal or quantum fluctuations via a renormalization group analysis,   Landau-Ginzburg theory lays the foundations of critical phenomena in equilibrium statistical mechanics~\cite{,ZinnJustinbook,Amit/Martin-Mayor,Cardy1996}. 
Given its success in explaining equilibrium~\cite{Hohenberg1977} and  relaxational dynamics of classical~\cite{Tauberbook2014} and open quantum systems~\cite{Kamenevbook2011,Gagel2014,Sieberer2016review}, it was not  surprising to witness in the last decade   a surge of   works exploring  dynamical phase transitions after a quench, which can be   understood via   an effective   $\phi^4$-field theory description. %, as the archetype of dynamical phase transition in an isolated quantum many body system.  
Our review will therefore start considering those DPTs which possess an effective description in terms   of a double well picture of the energy landscape of the underlying physical model.

We will first consider a fully connected spin model (the Lipkin-Meshkov-Glick model~\cite{1965NucPh..62..211G,1965NucPh..62..188L,1965NucPh..62..199M,1999PhLB..451....1P, 2004RvMP...76..643D,1983PhRvB..28.3955B, 2005PhRvA..71f0304D, PhysRevLett.99.050402,lmg2008,defenu2021long,latorre2005entanglement,orus2008equivalence,2020JPhA...53a3001M}) which is exactly solvable through a mean-field analysis  in the thermodynamic limit (Sec.~\ref{sec:lmg}). 
Its dynamical phases can be understood using a double well picture, providing the simplest analytically tractable instance of a dynamical critical point.  

In Sec.~\ref{sec:ON}, we will then consider  the dynamics of a field theory with an $\mathcal{N}$-component scalar order parameter~\cite{ZinnJustinbook}, which in the $\mathcal{N}\to\infty$ limit admits an exact solution in terms of Gaussian fluctuations self-consistently coupled to the motion of the  order parameter.
The leading $1/\mathcal{N}$ correction controls non-integrable scattering among different  modes and therefore the disappearance of the pre-thermal plateau where the DPT of the field theory can occur. 
This is   analogous to the effect of $1/N$ corrections (where now $N$ is the system size) on the DPTs of LMG and Richardson models discussed in Secs.~\ref{sec:lmg} and~\ref{sec:RGviaLax}, and it merges with  the unifying theme of this review: DPTs occur in collisionless non-equilibrium   states of dynamics~ \cite{Berges2004a,Moeckel2008,Kollar2011,gring2012relaxation,marcuzzi2013prethermalization, Bertini2015,Langen2016,PhysRevLett.113.210402, PhysRevX.9.021027}. When collisions due to non-linearities  set it, these pre-thermal states are destabilized and the system is attracted towards a thermal fixed point where only equilibrium phase transitions can occur (if permitted by dimensionality and symmetries). A further example of this phenomenology will be provided in Sec.~\ref{sec:dmft}, where DPTs are obtained in strongly correlated systems through an exact dynamical mean field treatment  valid in the $d\to\infty$ limit (here $d$ are the physical dimensions).

\subsection{Fully-connected Ising model in transverse field}
 \label{sec:lmg}

%%%%%%%%%%%%%%%%%%%%
%   FIG 1  %%%%%%%%%%%%%%
%%%%%%%%%%%%%%%%%%%%%
\begin{figure*}[t!]
\includegraphics[width=17.5cm]{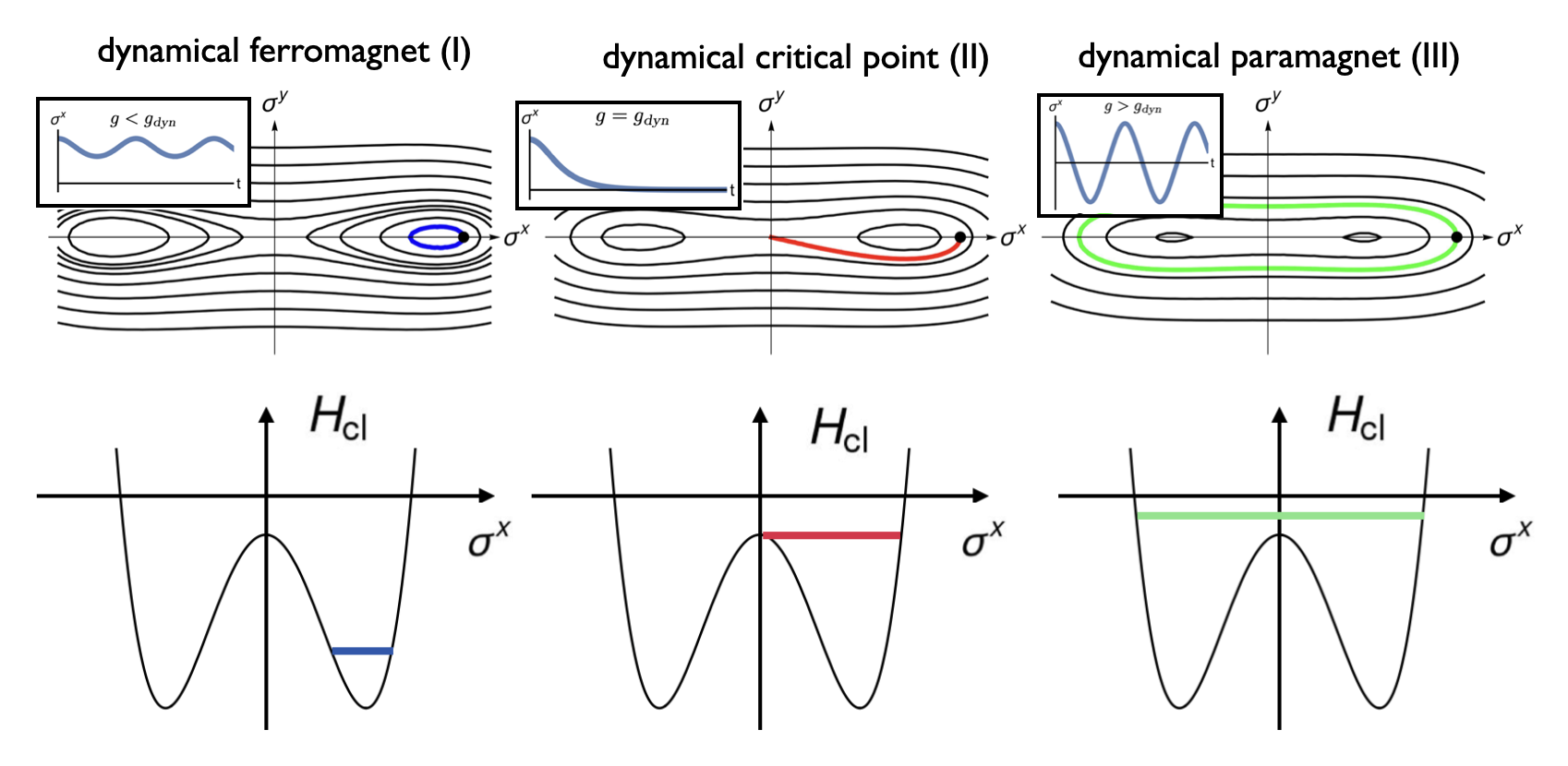}
\caption{  
Upper  row: Quantum quench of the transverse field in the   LMG model:   dynamical ferromagnetism (I) with oscillations around a non-vanishing value of $\sigma^x$; critical relaxation (II) at the dynamical critical point;   dynamical paramagnetism (III) with persistent oscillations around a zero mean value. 
Lower row:   energy $H_{cl}$  of the collective classical spin configuration, $\vec{\sigma}$,    as a function of the magnetization $\sigma^x$.
For the dynamical ferromagnet the energy of the initial state (cf.\ main text) is not sufficiently large to escape the minima of the potential (blue line); for the paramagnetic phase the energy is above the double-well barrier (green line), and therefore the orbit of the magnetization can encircle both minima and average out to zero over long time windows.
 The figure is adapted from Ref.~\cite{lerose2019impact}. }
\label{figLMG}
\end{figure*}
%%%%%%%%%%%%%%%%%%%%%
%%%%%%%%%%%%%%%%%%%%%
% 

In this section, we consider a  class of quantum Ising   systems with  spins $s$ on a $d$-dimensional lattice, interacting via a ferromagnetic coupling, $J_{\lvert\mathbf{r}-\mathbf{r'}\rvert} $, and subject to a transverse magnetic field pointing along the $z$ direction   
\beq\label{eq:modellmg}
H = - \sum_{\mathbf{r},\mathbf{r'}} J_{\lvert\mathbf{r}-\mathbf{r'}\rvert} \, \hat{\sigma}_{\mathbf{r}}^x \hat{\sigma}_{\mathbf{r'}}^x - g \sum_{\mathbf{r}} \hat{\sigma}_{\mathbf{r}}^z.
\eeq
Here the sums run over   $r=1, 2, ..., N$, with $N$ the total system size. %, labelled by $i,j=1,\dots,N$. 
 In Eq.~\eqref{eq:modellmg},  
 $\hat{\sigma}^\alpha_\mathbf{r}$ are the operators corresponding to the normalized spin components in the $\alpha=x,y,z$ direction.  For concreteness we consider in the following the $s=1/2$ case, although most of the results remain qualitatively unchanged for arbitrary spin length $s$. The parameter $g$ is the Larmor frequency induced by the transverse field.  This class of systems is of importance for     experimental realizations of DPTs in trapped ions~\cite{Zhang2017}, cavity QED systems~\cite{Muniz2020}, and neutral atoms arrays \cite{chu2020}, as we  discuss in Sec.~\ref{sec:exp}.

For all-to-all interactions $J_{\lvert\mathbf{r}-\mathbf{r'}\rvert} = \lambda/N$, for all $\mathbf{r}$, $\mathbf{r'}$,   the Hamiltonian in Eq.~\eqref{eq:modellmg} corresponds to the infinite-range or fully-connected Lipkin-Meshkov-Glick (LMG) model~\cite{1965NucPh..62..211G,1965NucPh..62..188L,1965NucPh..62..199M,1999PhLB..451....1P, 2004RvMP...76..643D,1983PhRvB..28.3955B, 2005PhRvA..71f0304D, PhysRevLett.99.050402,lmg2008,defenu2021long,latorre2005entanglement,orus2008equivalence,2020JPhA...53a3001M}  %of the hamiltonian~\eqref{eq:model}, 
\beq
\label{eq:MFH}
\hat{H} = - \frac{\lambda}{N} \sum_{i,j=1}^N \hat{\sigma}_i^x \hat{\sigma}_j^x - g \sum_{i=1}^N \hat{\sigma}_i^z,
\eeq
where each of the $N$ spins interacts with all the others  with the same ferromagnetic coupling strength $\propto\lambda$.
The $1/N$ scaling of %the infinite-range interaction coupling 
the  coupling $\lambda$ in Eq.~\eqref{eq:MFH} is adopted   to make the energy extensive in the  thermodynamic limit (notice that this scaling is not always satisfied in experimental implementations, e.g. cavity QED systems, see Sec.~\ref{sec:exp}).
As $N\to\infty$ the mean-field approximation  becomes exact for the Hamiltonian~\eqref{eq:MFH}, and therefore the model is   solvable in the thermodynamic limit: 
 the   properly normalized commutation relations of these collective spin operators, ${\hat{\sigma}}^\alpha \equiv \sum_{i=1}^N {\hat{\sigma}}_i^\alpha/N$, approach the classical limit for $N\to\infty$. In fact, %by rescaling the size of the collective spin by its magnitude $Ns$, one gets 
these  operators   have  a spectrum in $[-1,1]$ and they satisfy 
\beq\label{commut}
\big[\hat{\sigma}^{\alpha},\hat{\sigma}^{\beta}\big]= \frac{1}{Ns}i \epsilon^{\alpha\beta\gamma} \hat{\sigma}^{\gamma},
\eeq
which can be used to define an effective Planck's constant $\hbar_{\text{eff}} \equiv (N/2)^{-1}$. This property is instrumental to    investigate the corrections to the $N\to\infty $ limit via a semi-classical expansion in inverse powers of $N$. Such corrections include~\cite{lerose2019impact}  wavepacket spreading with deviations from the classical trajectory on Ehrenfest time scales $\propto\mathcal{O}(\sqrt{N})$, recurrences   on times $\sim \mathcal{O}(N)$ and tunnelling events between the two ferromagnetic minima on time scales  $\propto \mathcal{O}(e^{cN})$.

From Eq.~\eqref{commut}, it results that, for $N\to\infty$, the model can be exactly solved in terms of a classical, continuous spin $\vec{\sigma}$ of  normalized  length $\rho$, by neglecting   the above mentioned finite $N$ effects which require exact diagonalization or a semi-classical approximation to be observed~\cite{polkovnikov2010phase}. The three-dimensional vector $\vec{\sigma}=\rho(\sin\theta\cos\varphi,\sin\theta\sin\varphi,\cos\theta)$ is   a classical spin whose phase space is the surface of a Bloch sphere of radius $0<\rho\le1$,   with polar, $\theta$, and azimuth angles, $\varphi$. The LMG model then reduces to the classical Hamiltonian
\beq
\begin{split}
\label{eq:classicalH}
%\frac{H}{N} \quad \leadsto \quad 
&\mathcal{H}_{\text{cl}}(\vec{\sigma}) = - \lambda ( \sigma^x )^2  - g \sigma^z=\\
&=-\lambda \rho^2\sin^2\theta\cos^2\varphi-g\rho\cos\theta.
\end{split}
\eeq
For   ferromagnetic interactions   the system exhibits an equilibrium zero-temperature phase transition from a   paramagnetic ground state with $\braket{\hat{\sigma}^x}=0$ for $g>g_{\text{cr}}\equiv2\lambda$, to a pair of ferromagnetic ground states with $\braket{\hat{\sigma}^x}_{\pm}=\pm m\ne0$ for $g<g_{\text{cr}}$, characterized by the spontaneous  breaking of the $\mathbb{Z}_2$-symmetry. These two ferromagnetic minima can be derived from the Bloch sphere representation of Eq.~\eqref{eq:classicalH} and they are given by $(\theta^*,\varphi = 0)$ and $(\theta^*,\varphi = \pi)$ with $\cos\theta^*=g/g_{cr}$. Accordingly, the value of the order parameter is $\sigma^x=\pm\rho\sin\theta^*=\pm\rho\sqrt{1-(g/g_{cr})^2}$, where we have kept $\rho$ fixed since the total spin length of the system is conserved.

%Such phase transition can survive 
%
 % at   finite energy density (i.e. finite temperature or out-of-equilibrium), depending on the dimensionality and on the range of the interactions in $J_{\lvert\mathbf{r}-\mathbf{r'}\rvert}$ (see for instance~\cite{Sachdevbook,campa2009statistical,fisher1972critical}).  
  The associated classical energy landscape of the collective spin $\vec{\sigma}$ (along the plane $\sigma^y=0$) is portrayed in Fig.~\ref{figLMG} 
  as a function of the magnetization $\sigma^x$ for the ferromagnetic phase $g<g_{cr}=2\lambda\rho$. At equilibrium and in the thermodynamic limit, the degenerate ground state wave-functions of the collective
spin are localized at the two classical minima respectively,
and $\vec{\sigma}$ behaves like a classical particle at rest at the bottom
of one of the two wells.  \\
 
 DPTs in the LMG model have been studied extensively in the last decades~\cite{das06,kelly2019detecting,kelly2020thermalization,sciolla2011dynamical,PhysRevLett.121.240403,PhysRevB.78.104426,lerose2018chaotic,lerose2019prethermal}. 
We now  overview the dynamical phase diagram     summarized in Fig.~\ref{figLMG}, primarily following Ref.~\cite{lerose2019impact} for the sake of concreteness. 
 The non-equilibrium evolution of $\Braket{\vec{\sigma}_i(t)}$   is  given by the   classical   equations of motion associated to the Hamiltonian in Eq. ~\eqref{eq:classicalH}, $\dot{\sigma}^{\alpha} = \{ \sigma^{\alpha}, \mathcal{H}_{\text{cl}} \}$, where the evolution is given in terms of Poisson brackets.

We prepare the spin chain   in a ferromagnetic ground state of the Hamiltonian Eq.~\eqref{eq:classicalH} with   transverse field $g_0 < g_{\text{cr}} = 2\lambda\rho$.  This choice of  initial state is instrumental to illustrate the archetypal features of the dynamical phase diagram, and it    corresponds to a quantum quench of  the transverse field $g$. We prepare the spin chain in the ground state of the hamiltonian $H(g_0)$ at $t=0$, and then abruptly vary the transverse field up to the value $g$ so that the evolution  at times $t>0$ would occur  under the Hamiltonian $H(g)$.

For a   quench, $H(g_0)\to H(g)$, with  $g<g_{\text{dyn}}\equiv (g_0 + g_{\text{cr}})/2$, the   energy of the system remains below  the barrier that separates the two ferromagnetic sectors. Indeed, the value of $g_{\text{dyn}}$ can be determined~\cite{sciolla2011dynamical,Zunkovic} by equating the energy of the initial state with the height of energy 'hill' separating the two minima in Fig.~\ref{figLMG}. Correspondingly, the spin will precess within the starting ferromagnetic sector (trajectory I in Fig.~\ref{figLMG}). As the strength of the quench increases, the precession period   increases, until for $g \to g_{\text{dyn}} $ it takes an infinite time to complete one cycle, and the unstable point at the top of the energy barrier is approached exponentially (trajectory II in Fig.~\ref{figLMG}). For deep quenches above this threshold, $g>g_{\text{dyn}}$, the   energy of the system is larger than the energy  barrier separating the two minima of the double well.   The orbit of the collective spin  encircles both minima, such that the symmetry is  restored after taking time-averages, and the average magnetization, $\overline{\sigma^x}$,vanishes  (trajectory III in Fig.~\ref{figLMG}). 
 Indeed, the time-average 
\beq
\label{eq:neqop}
\overline{\sigma^x} = \lim_{T\to\infty} \frac{1}{T} \int_0^T dt \, \sigma^x(t),
\eeq
as a function of the post quench value, acts as a dynamical order parameter, and it vanishes abruptly at the dynamical critical value $g_{\text{dyn}}$. % , where $\Omega_0(g\to g_{cr})\to0$), 

This dynamical critical point therefore separates a \emph{dynamical ferromagnetic phase} with $\overline{\sigma^x}\ne0$ from a \emph{dynamical paramagnetic phase} with $\overline{\sigma^x}=0$.  %The dynamical critical point is determined by the initial condition and is given by $g_{cr} = (g_0 + g_{\text{cr}})/2 = \lambda + g_0/2$.
The   singularity of the equilibrium order parameter upon approaching the  critical point is algebraic, $\overline{\sigma^x}\sim(g_{\text{dyn}}-g)^\beta$, with  critical exponent $\beta=1/2$; on the contrary,   the non-equilibrium order parameter $\overline{\sigma^x}$ discussed here   displays a {logarithmic} singularity at the dynamical critical point.

Similar dynamical mean-field pictures hold also for   infinite dimensional   Bose-Hubbard systems~\cite{sciolla2011dynamical},   Jaynes-Cummings hamiltonians~\cite{sciolla2011dynamical},  or for the evolution of      classical $\phi^4$-field theories~\cite{Gambassi2011,sartori2015spin} (see  also the following Section~\ref{sec:ON}).  The DPTs occurring in these models can be all explained with a classical cartoon for their energy landscape after a quench, as done here for the LMG model. In   Section~\ref{sec:ON}, we review a model where the effects of Gaussian fluctuations on top of the evolution of a classical order parameter are instead crucial to   characterize the onset of the DPT.    

\subsection{Long-range interactions}\label{longrange}

 DPTs beyond the exact mean-field   behaviour of the LMG model can be also   inspected considering long-range couplings $J_{\lvert\mathbf{r}-\mathbf{r'}\rvert}\propto \lvert\mathbf{r}-\mathbf{r'}\rvert^{-\alpha}$  in  Eq.~\eqref{eq:modellmg} (see Refs.~\cite{vzunkovivc2018dynamical,halimeh2017prethermalization}). For $\alpha\leq3$, the model  describes, for instance, trapped ions simulators   (cf.  Sec.~\ref{sec:exp} for experimental observations of the DPT in a chain of trapped ions  with $0.8\lesssim\alpha\lesssim 1 $). The resulting non-equilibrium phase diagram can be derived using a MPS time-dependent variational principle~\cite{mps2016}. Equipped with long-range interactions, the quantum spin model in Eq.~\eqref{eq:modellmg}      interpolates  from the LMG limit ($\alpha=0$) to the short-range Ising model in transverse field ($\alpha\to\infty$), and it therefore offers a precious angle to  analyze the interplay of dimensionality and interactions  range in the formation of a dynamical critical point~\cite{PhysRevB.97.174401,PhysRevLett.125.040602,PhysRevResearch.2.012041,PhysRevB.104.115133,piccitto2019,PhysRevB.100.014434,PhysRevB.94.184403} (see also Sec.~\ref{sec:dmft} in this regard). In particular, the authors of Ref.~\cite{vzunkovivc2018dynamical}  have considered   dynamics in one dimension starting from a fully polarized state along the $\hat{x}$-direction,  unveiling a DPT for all values of $\alpha\leq2$. This critical value of $\alpha$ is the same for supporting a thermal phase transition in equilibrium  long-range interacting spin chains~\cite{dutta2001phase,campa2009statistical,defenu2021long,maghrebi2016causality}. An important difference with the LMG case, stands in the fact that the model is now not integrable, neither can its dynamics   be mapped to classical equations of motion. As a consequence,  the order parameter decays because of inhomogeneous dephasing~\cite{PolkovnikovRMP,Barthel2008,Rigol2008} towards a vanishing or non-vanishing expectation value depending on the dynamical phase  (paramagnetic or ferromagnetic), rather than exhibiting long-lived orbits as in the LMG model. This is a generic feature expected for  quantum many-body systems   whose dynamics cannot be simply described with the motion of   a collective mode. Once again in agreement with the statistical mechanics of equilibrium long-range systems~\cite{dutta2001phase,campa2009statistical,defenu2021long,maghrebi2016causality}, the order parameter     decays always to zero for $\alpha>2$, regardless of the post quench values of $g$, since the critical point flows into the universality class of short-range Ising chains.
 This is consistent with the expectation that a one dimensional model cannot support a phase transition at finite energy density, regardless where this results from a temperature or from the energy injected with a quantum quench~\cite{Sachdevbook}. 
 
 Furthermore, for $\alpha<1$, the order parameter acquires again the persistent oscillatory behaviour if the thermodynamic limit is taken before the long-time asymptotics. This is again explained by recalling that long-range Ising models with $\alpha<1$ fall into the universality class of the LMG model~\cite{dutta2001phase,campa2009statistical,defenu2021long,maghrebi2016causality}. It is therefore in the window $1<\alpha<2$ that a genuine quantum many body dynamical transition is observed. Upon changing the dimensionality, these two thresholds are expected to change with the qualitative picture remaining unaltered.  

Interestingly, dynamical transitions monitored by the cusps of the Loschmidt echo occur at any $\alpha$ (cf.\ Ref.~\cite{vzunkovivc2018dynamical}), showing that these two  notions of dynamical criticality describe qualitatively different physics. 
 
This section finds its natural experimental counterpart in Sec.~\ref{sec:exp} where realizations of DPTs in long-range interacting systems are discussed in detail.  
 
\subsection{ $O(\mathcal{N})$ models in the large $\mathcal{N}$ limit}
\label{sec:ON}

In this Section we discuss a prototypical example illustrating the richness of non-equilibrium phase transition of isolated systems beyond   mean-field. The model we review allows for an exact treatment of quantum fluctuations. It is furthermore iconical as it the deals with the quantum dynamics of a Landau-Ginzburg theory, and it might therefore offer a prototypical model for DPTs in more complex non-integrable quantum many-body systems.
We consider      a $\mathcal{N}$-component bosonic order parameter $\hat{\vecphi}=(\hat{\phi}_1, \hat{\phi}_2, ..., \hat{\phi}_\mathcal{N})$ in $d$ spatial dimensions, with an $O(\mathcal{N})$-invariant Hamiltonian~\cite{ZinnJustinbook}
\begin{equation}
\label{eq:hamiltonian}
\hat{H}(r,u)= \int d^d\mathbf{x} \left[\frac{1}{2}\hat{\mathbf{\Pi}}^2 + \frac{1}{2}(c\nabla\hat{\phi})^2 + \frac{r}{2}\hat{\phi}^2+\frac{u}{4!\mathcal{N}}(\hat{\phi}^2)^2\right],
\end{equation}
where   $\hat{\mathbf{\Pi}}=\partial_t\hat{\mathbf{\Phi}}$ is the $\mathcal{N}$-component momentum canonically conjugated to $\hat{\vecphi}$, and $r$ and $u$ parametrise respectively the mass term   and  the strength of the   non-linearity  (for   quantum quenches   of fermionic variants of this model in the context of DPTs, see Refs.~\cite{yin2021fermion,jian2019universal}). 
The requirement of invariance under rotations of the $O(\mathcal{N})$ group is the reason for the appearance of scalar products only, $\hat{\mathbf{\Pi}}^2$ and $\hat{\phi}^2=\hat{\mathbf{\Phi}}\cdot\hat{\mathbf{\Phi}}=\sum^\mathcal{N}_{i=1}\hat{\phi}^2_i$, in the Hamiltonian~\eqref{eq:hamiltonian}.
The field theory in Eq.~\eqref{eq:hamiltonian} is  customarily taken as an effective continuum description of a microscopic model~\cite{Sachdevbook} (with spin, fermionic or bosonic degrees of freedom) and it therefore requires   an ultraviolet momentum cutoff $\Lambda$, which is usually the inverse of   the lattice spacing.
Upon varying $\mathcal{N}$, Eq.~\eqref{eq:hamiltonian}   describes   the Ising ($\mathcal{N} = 1$) and Heisenberg ($\mathcal{N} = 3$) models, or the  Bose-Hubbard   at the particle-hole symmetric point ($\mathcal{N}=2$), to mention a few~\cite{Sachdevbook}.  

The success in modeling dynamical phase transitions with the field theory~\eqref{eq:hamiltonian} resides in a   closure at the Gaussian level of its hierarchy of equations of motion. Specifically, at $\mathcal{N}\to\infty$ the evolution of the state of the $O(\mathcal{N})$ model is fully characterized by the dynamics of the two point functions self-consistently coupled to the evolution of $\langle \hat{\mathbf{\Phi}} \rangle$, with higher order correlation functions parametrically suppressed in powers of $1/\mathcal{N}$. 
The large-$\mathcal{N}$ limit has    allowed     to extrapolate several qualitative features of the equilibrium thermal phase diagram of the $O(\mathcal{N})$ model to finite values of $\mathcal{N}$ and $d$ (see~\cite{Sachdevbook, ZinnJustinbook}). 
Specifically, in the limit  $\mathcal{N}\to\infty$,   quartic interactions in Eq.~\eqref{eq:hamiltonian} decouple at leading order which amounts to the formal substitution $(\hat{\phi}^2)^2\sim\langle \hat{\phi}^2\rangle \hat{\phi}^2$. In field theoretical language this corresponds   to a Hartree-Fock approximation~\cite{Chandran2013,Maraga2015}.
 Here, we have used the fact that, if the $O(\mathcal{N})$ symmetry of the initial state is not broken, the average  $\langle \hat{\phi}_a \hat{\phi}_b \rangle$ of two generic components of $\hat{\vecphi}$ vanishes unless $a=b$. Its non-vanishing value is independent of $a$ and equal to the fluctuation $\langle\hat{\phi}^2\rangle$ of a generic component of the field.  

 In diagrammatic language, the large-$\mathcal{N}$ limit amounts to retaining only tadpole corrections to $r$ and `candy' (RPA) diagram corrections to $u$~\cite{Berges2002}. The dynamics of the field components are therefore equivalent to a time-dependent quadratic Hamiltonian with an effective mass, $r_{\text{eff}}(t)$, to be self-consistently determined from the expectation value of $\langle \hat{\phi}^2 \rangle$ (see Eq.~\eqref{eq:effmass} below). 
The model becomes then exactly solvable via a time-dependent self-consistent Gaussian ansatz.   The large-$\mathcal{N}$ expansion allowed the authors of Refs.~\cite{Sotiriadis2010, Gambassi2011, Sciolla2013,Chandran2013,Smacchia2015,Maraga2015,Chiocchetta2016,halimeh2021quantum} to disregard thermalizing collisions that are effective at times parametrically large in $\mathcal{N}/u^2$, and which dictate equilibration. DPTs emerge in such a collisionless pre-thermal regime~\cite{Berges2004a,Aarts2000,gring2012relaxation,Langen2016} (cf. with Sec.~\ref{intro}),  as we will also emphasize later in Sec.~\ref{sec:RGviaLax}. 
%%%%%%%%%%%%%%%%%%%%
%   FIG 1  %%%%%%%%%%%%%%
%%%%%%%%%%%%%%%%%%%%%
\begin{figure}[t!]
\includegraphics[width=9cm]{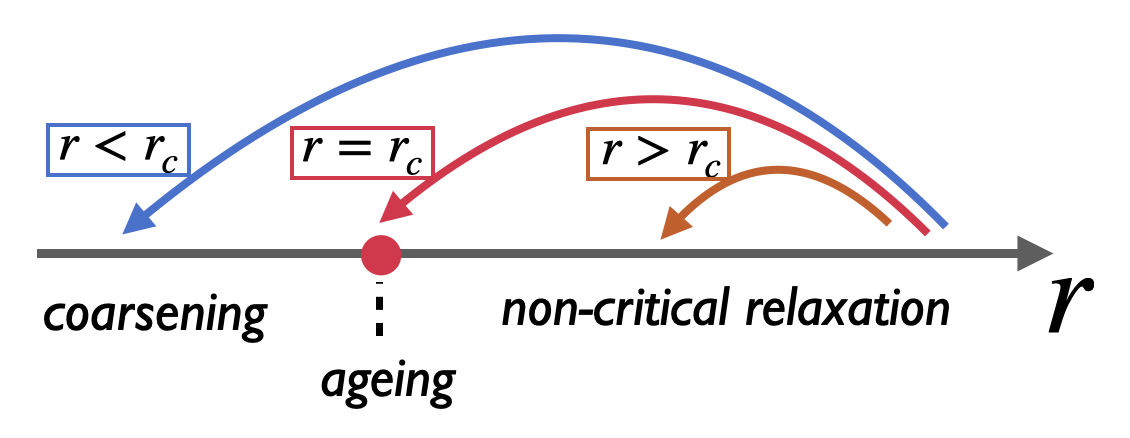}
\caption{Sketch of the dynamical phases    of the $O(\mathcal{N}\to\infty)$ model resulting from a quench with large initial mass ($r_0\gg \Lambda^2$). 
The system is quenched above ($r>r_c$), 
at ($r = r_c$), 
or below ($r<r_c$) the dynamical critical point;
 in the latter case dynamics display ageing. 
 This dynamical critical point separates non-critical relaxation (a dynamically disordered phase) from coarsening (a dynamically ordered phase). }
\label{fig:phase-diagram}
\end{figure}
%%%%%%%%%%%%%%%%%%%%%
%%%%%%%%%%%%%%%%%%%%%
% 

The typical quench protocol used to probe the dynamical phase diagram of this model consists in preparing the system in the ground state of the non-interacting Hamiltonian $\hat{H}(r_0,0)$, 
and then letting the dynamics evolve with $\hat{H}(r,u)$ at later times.  
Exact solvability in the $\mathcal{N}\to\infty$ limit allows us to elucidate the role of fluctuations in   the formation of the dynamical critical point.  The 
 time-dependent effective mass, $r_{\text{eff}}(t)$, dressed by Gaussian fluctuations reads 
\begin{equation}
\label{eq:effmass}
r_{\text{eff}}(t) = r + \frac{\mathcal{N}+2}{6\mathcal{N}} u\int \frac{\dd^dp}{(2\pi)^d}~\langle \hat{\phi}_{\pp}(t)\hat{\phi}_{-\pp}(t)  \rangle.
\end{equation}
with $\hat{\phi}_{\pp}$ the Fourier transform of the field.
The   right hand side of Eq.~\eqref{eq:effmass} reaches a steady state value as a result of inhomogeneous dephasing of the   oscillating factors with frequencies depending on the   momenta $\mathbf{p}$ in the integrand.    When the post-quench value of the mass $r$ is tuned to a critical value $r_c(r_0,u)$ such that it cancels the contribution from the fluctuations integral in~\eqref{eq:effmass},  the stationary value of the effective mass $r^*\equiv r_{\text{eff}}(t\to\infty)$ vanishes and, consequently, the spatial correlation length $\xi = (r^*)^{-1/2}$ diverges, thus signalling the onset of a dynamical critical  point.  Following Ref.~\cite{Smacchia2015}, we can evaluate the behaviour of the effective asymptotic mass $r^*$ for small deviations in the post-quench value, $\delta r$,   from the dynamical critical point  $r_c$, finding
\begin{equation}\label{eq:rstar}
    r^*=\delta r-\frac{u}{24} r^* \int^\Lambda_0 \frac{p^{d-3}dp}{(2\pi)^d} \frac{(p^2+r_0)^{1/2}}{(p^2+r^*)},
\end{equation}
with $\Lambda$ the ultraviolet momentum cutoff, typically related to inverse lattice spacing of the microscopic model from which  the field theory is derived.
 For $d>4$ the integral in Eq.~\eqref{eq:rstar}, which encodes the role of fluctuations, converges upon tuning $r^*\to0$  (close to criticality). 
For $2<d<4$ the integral is instead  the leading term, implying the scaling $r^*\sim (\delta r)^{2/(d-2)}$. This yields a correlation length  $1/\xi^*\sim \sqrt{r^*}$, with a static critical exponent $\nu=1/(d-2)$. For $d\geq4$ we find the mean field exponent $\nu=1/2$ since, as we just discussed, fluctuations are RG irrelevant. This upper and lower critical dimension for the DPT of the $O(\mathcal{N}\to\infty )$ model are the same for   thermal  equilibrium thermal phase transitions~\cite{Sachdevbook, *Sondhi, *Vojta}.     
Although Eqs.~\eqref{eq:effmass} and~\eqref{eq:rstar} are structurally analogous to   equilibrium thermal transitions,   the fluctuation corrections to the bare mass are quantitatively different, resulting in a different location for the critical point in the thermal and non-equilibrium cases.

% Furthermore, if only a quench in the mass is performed ($r_0=r$), one can easily derive that the lower and upper critical dimensions are respectively $d_l=1$ and $d_u=3$, with $\nu=1/(d-1)$. This again matches the equilibrium behaviour of zero temperature $O(N)$ models.
 %Although these quantum/classical pre-thermal critical points are analogous to the zero/high temperature ones, the dynamical phases appear radically different than the conventional ordered and disordered ones. \\

%
In order to explore the nature of the different dynamical phases, we assume to prepare the system with a given initial mass, $r_0$, and consider different post quench values of $r$. We will focus in the following on the asymptotic $t\to\infty$ value of the effective mass, $r^*$, on its real-time scaling features and on  momentum-resolved correlation functions.   Fig.~\ref{fig:phase-diagram} provides a graphical summary of the three dynamical phases and of their properties. We will be primary following Ref.~\cite{Maraga2015} in our exposition. 

(i) For $r>r_c(r_0,u)$, the system undergoes a non-critical relaxation characterized by a finite value of the correlation length and correlation time, and by a vanishing value of the order parameter. This is the \emph{dynamically disordered phase}. The symmetric correlation function $\mathcal{C}_\pp(t,t')=-i\langle \{\hat{\phi}_\pp(t),\hat{\phi}_{-\pp}(t')\}\rangle$ displays oscillations with asymptotic period set by $\sim (r^*)^{-1/2}$. 

(ii) For quenches to the \emph{   critical point},  $r \to r_c(r_0,u)$,   slow modes are characterized by aging dynamics, where  the symmetric correlation function  acquires the scaling form  $\mathcal{C}_\pp(t,t')\propto (tt')^{\theta}$ for $t\gg t'$ and $p\to0$, with a  non-equilibrium universal exponent $\theta=d/4$ in $2<d<4$ (see Refs. \cite{Chiocchetta2015, Maraga2015}). This is the analog  of the ageing exponent occurring in non-equilibrium classical systems      quenched close to   critical temperature~\cite{Janssen1989, Calabrese2005}; in contrast, in the case discussed here, the system acts as its own bath because isolated from the environment. The phenomenon is called ageing  because   the characteristic
  relaxation time-scale is the age of the system itself, i.e. the time, $t$, elapsed after the quench. 
As in its classical counterpart, ageing denotes   a dynamical regime where correlation functions break time-translational invariance  in a universal fashion~\cite{Chiocchetta2016}. Indeed, the exponent $\theta$ does not depend on microscopic details, rather it is dictated by dimensionality, symmetries and conservation laws, as in conventional   theory of critical phenomena. Remarkably, the value of the exponent $\theta$  cannot be related to  the equilibrium ones $\nu$, $\eta$, $z$, etc. The absence of a gap and therefore of a characteristic timescale at criticality, implies that each mode has a typical relaxation time scale of order $\sim 1/p$. Therefore the ageing window is expected to extend  for each mode within $1/\Lambda\lesssim t \lesssim 1/p$, and for well separated times $t\ll t'$. For $t\lesssim 1/\Lambda$, non-universal effects due to the microscopic details of the model will set in.

Concerning the steady state forming after deep quenches $r_0\gg\Lambda^2$, the infrared modes equilibrate at the effective temperature $T_{\text{eff}}\simeq \sqrt{r_0}$. Notice however that this does not imply that the system has thermalized  to a Gibbs state~\cite{Chandran2013}; on the contrary, higher momenta have occupations that violate      equipartition of energy. The  fact that  in the  presence of a finite energy density   low momenta modes effectively thermalize, while higher momentum tails do not satisfy detailed balance, is  a general feature of   non-equilibrium phase transitions, present also in driven-open  systems~\cite{Sieberer2015}.  

For small initial  masses $r_0\ll\Lambda^2$, the system undergoes a dynamical crossover around a time scale $\sim 1/r_0$, from a  regime where dynamical scaling is characterized by a `quantum' ageing exponent $\theta'$ into a regime ruled by the effectively thermal exponent $\theta$   we have just discussed.  This additional ageing exponent $\theta'$ corresponds to an unstable pre-thermal fixed point~\cite{chiocchetta2017dynamical} whose canonical scaling properties are akin to a zero-temperature quantum   critical point (the lower and upper critical dimensions  are   the same in the equilibrium and non-equilibrium case, $1<d<3$). % This fixed point is   destablized by any infinitesimal value of pre-quench mass, $r_0$, which dictates the crossover into the dynamical scaling regime ruled by   $\theta$.  

(iii) Finally, for $r\leq r_c(r_0,u)$, the time-dependent effective mass decays with a universal $\sim 1/t^2$ behaviour on long time scales, as it also happens  for quenches at the dynamical critical point (below the upper critical dimension $d<4$).  For $d\geq4$, this decay turns   into $\sim t^{-(d-2)}$ as it can be predicted  from dimensional analysis. 
However, in this regime with $r<r_c(c_0,r_0,u)$,   the system exhibits    coarsening ~\cite{Sciolla2013,Chandran2013,Maraga2015}, and correlation functions are characterized by a form of   dynamical scaling distinct from the ageing typical of the critical point. Coarsening is   caused by the formation of growing domains with different values of the order parameter, similarly to the phenomenon of   spinodal decomposition~\cite{Bray1994}. It occurs for quenches starting from a symmetric ground state, $r_0>r_c$. Physically, one expects that the system globally remains in a symmetric state with vanishing order parameter, however, the symmetry can be broken within spatial domains which appear separated by    walls. Within each of them,   the order parameter, $\hat{\phi}$, acquires the value of one of the two different  $Z_2$-symmetry broken  phases~\cite{Bray1994,Biroli2015,Cugliandolo2015}. The average
linear extension  of the ordered domains increases
with time, until a specific domain possibly prevails over
the others, establishing the equilibrium state. However,
   the size of domains grows algebraically   and actual equilibrium
is reached only asymptotically. This
lack of an intrinsic length scale in the system affects the
equal-time two-point correlation functions which scale as $\mathcal{C}_{p=0}(t,t)\sim t^\gamma$ with $\gamma=d-1$.  %(here $d$ is above the upper critical dimension $d>d_u=4$). 

We finally   remark that   the dynamical phases  of the $O(\mathcal{N}\to\infty)$ model remain qualitatively unaltered~\cite{maraga2016linear}  when the value of the mass is continuously ramped at a sufficiently fast speed  from $r_0$ to $r$, rather than suddenly quenched between the same two values. More precisely,    when the duration of the ramp is finite, the critical properties associated to the dynamical transition are qualitatively the same as in the sudden quench scenario, while as the ramp is infinitely slow the equilibrium quantum phase transition at zero temperature is eventually recovered. 
%

%%%%%%%%%%%%%%%%%%%%%%%%%%%%%%%%%%%%%%%%%%%%%%%%%%%%%%%%%%%%%%%%%%%%%%%%%%%%%%%%%%%%%%%%%%%%%%%%%%%%%%%%%
%%%%%%%%%%%%%%%%%%%%%%%%%%%%%%%%%%%%%%%%%%%%%%%%%%%%%%%%%%%%%%%%%%%%%%%%%%%%%%%%%%%%%%%%%%%%%%%%%%%%%%%%%
%%%%%%%%%%%%%%%%%%%%%%%%%%%%%%%%%%%%%%%%%%%%%%%%%%%%%%%%%%%%%%%%%%%%%%%%%%%%%%%%%%%%%%%%%%%%%%%%%%%%%%%%%
%%%%%%%%%%%%%%%%%%%%%%%%%%%%%%%%%%%%%%%%%%%%%%%%%%%%%%%%%%%%%%%%%%%%%%%%%%%%%%%%%%%%%%%%%%%%%%%%%%%%%%%%%
%%%%%%%%%%%%%%%%%%%%%%%%%%%%%%%%%%%%%%%%%%%%%%%%%%%%%%%%%%%%%%%%%%%%%%%%%%%%%%%%%%%%%%%%%%%%%%%%%%%%%%%%%
%%%%%%%%%%%%%%%%%%%%%%%%%%%%%%%%%%%%%%%%%%%%%%%%%%%%%%%%%%%%%%%%%%%%%%%%%%%%%%%%%%%%%%%%%%%%%%%%%%%%%%%%%
%%%%%%%%%%%%%%%%%%%%%%%%%%%%%%%%%%%%%%%%%%%%%%%%%%%%%%%%%%%%%%%%%%%%%%%%%%%%%%%%%%%%%%%%%%%%%%%%%%%%%%%%%
%%%%%%%%%%%%%%%%%%%%%%%%%%%%%%%%%%%%%%%%%%%%%%%%%%%%%%%%%%%%%%%%%%%%%%%%%%%%%%%%%%%%%%%%%%%%%%%%%%%%%%%%%
%%%%%%%%%%%%%%%%%%%%%%%%%%%%%%%%%%%%%%%%%%%%%%%%%%%%%%%%%%%%%%%%%%%%%%%%%%%%%%%%%%%%%%%%%%%%%%%%%%%%%%%%%
%%%%%%%%%%%%%%%%%%%%%%%%%%%%%%%%%%%%%%%%%%%%%%%%%%%%%%%%%%%%%%%%%%%%%%%%%%%%%%%%%%%%%%%%%%%%%%%%%%%%%%%%%
%%%%%%%%%%%%%%%%%%%%%%%%%%%%%%%%%%%%%%%%%%%%%%%%%%%%%%%%%%%%%%%%%%%%%%%%%%%%%%%%%%%%%%%%%%%%%%%%%%%%%%%%%
%%%%%%%%%%%%%%%%%%%%%%%%%%%%%%%%%%%%%%%%%%%%%%%%%%%%%%%%%%%%%%%%%%%%%%%%%%%%%%%%%%%%%%%%%%%%%%%%%%%%%%%%%

\section{Exactly solved dynamics in Richardson-Gaudin models via the Lax spectral method \label{sec:RGviaLax}}

In this Section, we review a class of all-to-all interacting spin exchange models of the Richardson type whose evolution under a quantum quench is solved exactly by 
a version of self-consistent mean field theory \cite{Barankov2004}. The solution reveals different dynamical phases of an 
order parameter, which can decay to zero (``phase I'') \cite{Barankov2006,Yuz}, 
assume a non-equilibrium steady-state value (``phase II'') \cite{Yuz2}, 
or even exhibit persistent oscillations (``phase III,'' a \emph{self-generated} Floquet phase \cite{Barankov2006}).
Although these dynamical phases of matter may carry some resemblance to those resulting from quenches of LMG-type models (cf. Sec.~\ref{sec:lmg} above), this Section aims at emphasising their markedly different origin which results from the classical and quantum integrability of Richardson magnets. 

The order parameter for this class of models can be defined via
\begin{align}\label{DeltaDef}
	\Delta
	\equiv
	- G \sum_{i} s_{i}^-.
\end{align}
This is an equal-weight sum of spin-1/2 lowering operators
in an $N$-spin system. It plays the role of the BCS
pairing gap in applications to quenched superfluids
and superconductors (see below).

\begin{figure}[b!]
\centering
\includegraphics[width=0.4\textwidth]{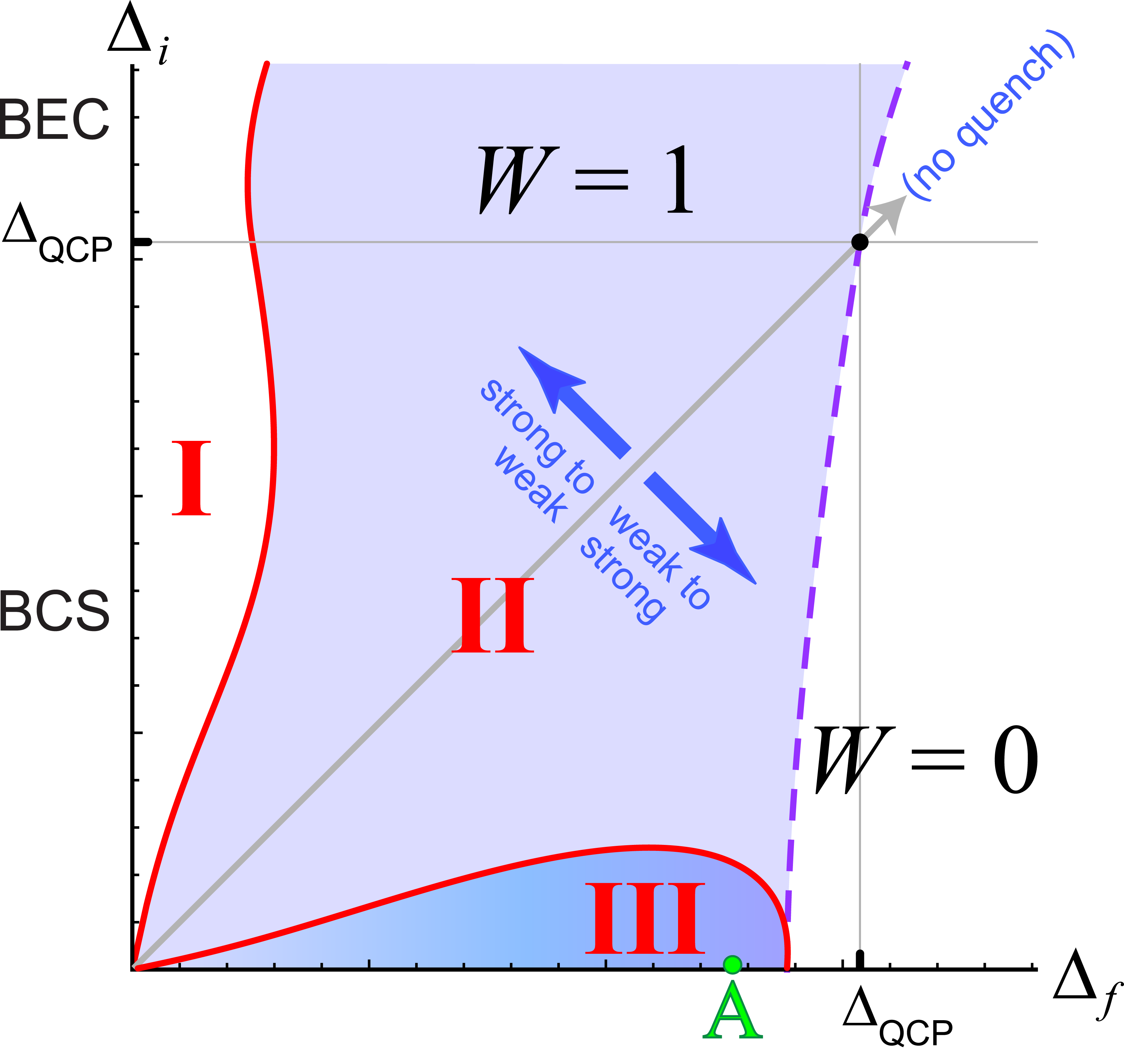}
\caption{Quench phase diagram for a 2D $p+ip$ fermion superfluid, from Refs.~\cite{Foster2013,Foster2014}.
Here $\Delta_i$ ($\Delta_f$) denotes the ground-state BCS order parameter in the pre- (post-)quench 
Hamiltonian. Any point with $\Delta_i \neq \Delta_f$ corresponds to a quench, while the line $\Delta_i = \Delta_f$ 
denotes the ground state. $\Delta_i$ ($\Delta_f$) is associated to an attractive BCS pairing interaction strength
$G_i$ ($G_f$), assumed to change instantaneously at the time of the quench.
The subsequent dynamics of $\Delta(t)$ fall into three different 
phases I, II, III, described in the text. Phase III is a self-generated Floquet phase,
which in the case of a $p+ip$ superfluid exhibits topological edge states, see Fig.~\ref{Fig--PwaveFloquet}. 
In this figure, $W = 1$ ($W = 0$) indicates topological (trivial) domains of phase II. 
The region with $W = 1$ hosts Majorana edge states, even for quenches that start
from the trivial BEC phase ($\Delta_i > \Delta_{\mathsf{QCP}}$).
$\Delta_{\mathsf{QCP}}$ denotes the ground-state topological quantum phase transition between
BCS and BEC phases \cite{Alicea2012,Levin2005}; the dashed purple line is its nonequilibrium extension.
}
\label{Fig--PwavePD}
\end{figure}

Fig.~\ref{Fig--PwavePD} shows the quench phase diagram for a system exhibiting phases I, II, and III.
In this figure, ground-state order-parameter values for the pre- ($\Delta_i$) and post-quench ($\Delta_f$) Hamiltonian label the vertical and horizontal axes, so that any point in the phase diagram away from the diagonal line is a quench. In this particular system (a quenched $p+ip$ superfluid), the ground state exhibits a topological transition at a certain value of the order parameter $\Delta_i = \Delta_f = \Delta_{\mathsf{QCP}}$. This extends to a line in the quench phase diagram, separating nonequilibrium phases characterized by different winding numbers $W$.

The models described in this section can be realized in a collisionless (prethermalization) regime 
for a range of physical systems, as will be discussed in Sec.~\ref{sec:exp}, 
including  
\begin{itemize}[leftmargin=10pt]
\item{
Far-from equilibrium dynamics in solid-state superconductors, following excitation by an intense subgap (terahertz) pulse 
\cite{Matsunaga2013,Shimano2019,Papenkort2007,Papenkort2008,KrullSchnyder2014,Chou2017,Papenkort2009},} 
\item{
Quenches in central-spin type problems (Gaudin magnets), or synthetic 
spin-1/2 magnets with infinite-range interactions realized in ultracold fermionic  atoms in a trap \cite{smale2019} and cavity QED systems \cite{Lewisswan2021}},
\item{Ultracold fermionic superfluids, with different pairing symmetries (such as $s$-wave or $p$-wave) where a  quench corresponds (e.g.)\ to a change in the strength of the attractive interactions responsible for pairing,
as could be accomplished by tuning a Feshbach resonance \cite{Barankov2006,Yuz2,Yuz,Foster2013,Yuzbashyan2015,Gurarie2007,Gurarie2009}.
The approach has been extended to multicomponent superconductors with competing orders \cite{Dzero2015}.}
\item{
Quenches in topological superfluids, such as a chiral $p+ip$ system that can host circulating Majorana edge modes \cite{Foster2013,Foster2014,Liao2015}.}
\end{itemize}

All of these systems can be approximately described by a Hamiltonian of Richardson-Gaudin type \cite{Richardson1964A,Richardson1964B,Gaudin}, 
which is both classically and quantum mechanically integrable. Quantum integrability implies that the many-body
spectrum can be obtained from the Bethe ansatz \cite{Richardson1964A,Richardson1964B,Gaudin,Dukelsky2004,Richardson2002,Skrypnyk2009,Ibanez2009,Dunning2010,Ortiz2010}.
Instead, here we focus on the thermodynamic (infinite-system-size) limit, where self-consistent
mean field theory can become exact \cite{Yuzbashyan2005Lax,Yuzbashyan2005,Yuz2,Barankov2006}. 
In this case, the system resides in a pure BCS-type state at all times 
(with time-dependent coherence factors), and the dynamics of generic observables (Green's functions) can 
be computed exactly by exploiting the \emph{classical integrability}. 

This section is organized as follows. 
First, we will describe how the Richardson-Gaudin spin model emerges from the description of a BCS superfluid. 
We will then illustrate how the dynamical phase diagram can be obtained using the ``Lax spectral method'' 
\cite{Yuzbashyan2005Lax,Yuzbashyan2005,Yuz2,Foster2013,Yuzbashyan2015}, 
which exploits the classical integrability and the simplifications that occur in the thermodynamic limit. 
For a particularly simple class of initial conditions (different from the usual assumption of a BCS-ground initial state), 
we show explicitly how the solution method works,
giving the dynamical phases I, II, III described above, and we will outline how to compute generic observables.
Sec.~\ref{sec:flucts} gives an overview of fluctuation phenomena beyond the scope of mean-field dynamics, including
quenches from an initial normal Fermi liquid state 
\cite{Barankov2006Replace,YuzbashyanTsyplyatyev2009,YuzbashyanDzero2009,Yuzbashyan2019,Mitra2017,Mitra2018A,Mitra2018B,Mitra2019}. 
Finally, we summarize a few key features of quench-induced dynamical topological phase transitions \cite{Foster2013,Foster2014,Liao2015}.

We limit the overview in this section to the idealized, thermodynamic limit for integrable Richardson-Gaudin models. 
Time and length scales for observing this physics in ultracold Fermi gases and THz-driven superconductors are
discussed later in Sec.~\ref{CMI}.

\subsection{BCS superfluids and Richardson-Gaudin}

Consider a spin-1/2 Fermi gas in $d$ spatial dimensions, with local, attractive interactions. The Hamiltonian is 
\begin{align}\label{HBCSmicro}
	H
	=&\,
	\sum_{\sigma \in \uparrow,\downarrow}
	\int d^d \vex{r}
	\,
	c_{\sigma}^\dagger(\vex{r})
	\left(- \frac{1}{2m} \nabla^2 \right)
	c_{\sigma}(\vex{r})
\nonumber\\
	&\,
	\quad
	-
	G
	\int d^d\vex{r}
	\,
	c_{\uparrow}^\dagger(\vex{r})
	\,
	c_{\downarrow}^\dagger(\vex{r})
	\,
	c_{\downarrow}(\vex{r})
	\,
	c_{\uparrow}(\vex{r})
\nonumber\\
	=&\,
	\sum_{\vex{k},\sigma}
	\e_{k}
	\,
	c_{\vex{k} \sigma}^\dagger
	\,	
	c_{\vex{k} \sigma}
\nonumber\\
	&\,
	\quad
	-
	G
	\sum_{\vex{k},\vex{k'},\vex{q}}
	c_{\vex{k} + \frac{\vex{q}}{2}  \uparrow}^\dagger
	\,
	c_{-\vex{k} + \frac{\vex{q}}{2}  \downarrow}^\dagger
	\,
	c_{-\vex{k'} + \frac{\vex{q}}{2} \downarrow}
	\,
	c_{\vex{k'} + \frac{\vex{q}}{2}  \uparrow},
\end{align}
where the second equality expresses the Hamiltonian in terms
of momentum modes in a finite volume. Here $\e_k = k^2/2m$ is the
kinetic energy, and $G > 0$ denotes a local, spin-singlet pairing interaction strength. 
The fermion creation and annihilation operators satisfy canonical
anticommutation relations, 
$
	c_{\vex{k} \sigma}
	\,
	c_{\vex{k'} \sigma'}^\dagger
	+
	c_{\vex{k'} \sigma'}^\dagger
	\,
	c_{\vex{k} \sigma}
	=
	\delta_{\vex{k},\vex{k'}}
	\delta_{\sigma,\sigma'},
$
where $\sigma,\sigma' \in \{\uparrow,\downarrow\}$ label the spin.

\begin{figure}[t!]
\centering
\includegraphics[width=0.4\textwidth]{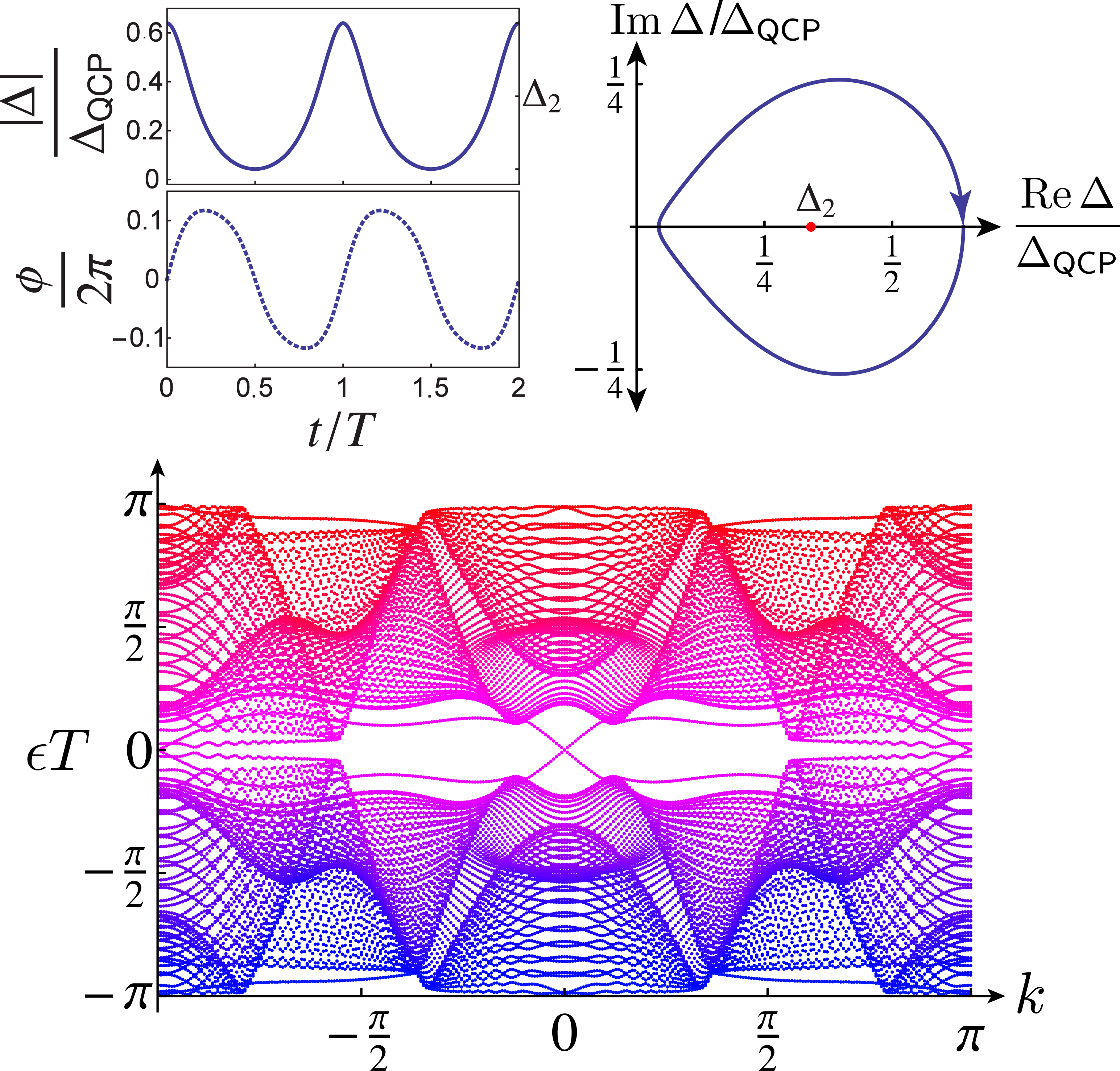}
\caption{Self-generated Floquet phase III for a sufficiently large weak-to-strong quench in the model
described in Fig.~\ref{Fig--PwavePD} \cite{Foster2014,Liao2015}. 
A Floquet phase is a dynamical state of matter characterized by parameters that oscillate periodically in time \cite{Oka2019}. Floquet systems can exhibit topological features \cite{QiZhang2011,Alicea2012} such as protected edge states at the sample boundary, even when the undriven system is topologically trivial \cite{Rudner2020}. The top left shows the periodic (but strongly anharmonic) amplitude and phase of the order
parameter $\Delta(t)$, while the top right depicts its orbit in the complex plane. 
On the bottom, diagonalization of a lattice-regularized strip shows Floquet-Majorana edge states (crossing in the center gap) 
for this particular quench, marked ``A'' in 
Fig.~\ref{Fig--PwavePD}.
Majorana edge modes will appear only in a fermionic superfluid or superconductor with a physical boundary.
}
\label{Fig--PwaveFloquet}
\end{figure}

Inspired by the solution to the Cooper problem that gives a minimal
bound-state energy for a two-electron pair with zero center-of-mass momentum,
we drop interaction terms in Eq.~(\ref{HBCSmicro}) with nonzero $\vex{q}$. 
The resulting ``reduced BCS Hamiltonian'' \cite{Schrieffer} can be expressed in 
terms of Anderson pseudospins \cite{Anderson1958}
\begin{align}
\begin{gathered}		
	S_{\vex{k}}^+ 
	\equiv
	c_{\vex{k}\uparrow}^\dagger
	\,
	c_{-\vex{k}\downarrow}^\dagger,
\qquad
	S_{\vex{k}}^- 
	\equiv 
	c_{-\vex{k}\downarrow}
	\,
	c_{\vex{k}\uparrow},
\\
	S_{\vex{k}}^z
	\equiv
	\frac{1}{2}\left(n_{\vex{k}\uparrow} + n_{-\vex{k}\downarrow} - 1\right),
\end{gathered}
\end{align}
where $n_{\vex{k}\sigma} \equiv c_{\vex{k}\sigma}^\dagger c_{\vex{k}\sigma}$. 
Simplifying notation by labeling single-particle levels with $i,j \in \{1,\cdots,N\}$ instead
of momenta, the reduced $s$-wave BCS Hamiltonian is 
\begin{align}\label{HRed}
	H
	=
	\sum_{i = 1}^N
	2 \e_i S_i^z
	-
	G
	\sum_{i,j = 1}^N
	S_i^+
	S_j^-.
\end{align}
This ``Richardson'' Hamiltonian \cite{Richardson1964A,Richardson1964B} 
is similar to the Dicke model, which features all-to-all coupling of the in-plane spin components,
but here each spin is subject to a potentially different Zeeman field $2 \e_i$. 
The all-to-all coupling is a necessary, but not sufficient condition for integrability \cite{Yuzbashyan2019};
integrable variants can describe $p$-wave and other pairings \cite{Richardson2002,Dukelsky2004,Skrypnyk2009,Ibanez2009,Dunning2010,Ortiz2010}.

Using the SU(2) algebra of the pseudospins, the Heisenberg equations of motion are 
\bsub
\begin{align}
	\dot{\vec{S}}_i 
	=&\,
	\vec{S}_i
	\times
	\vec{B}(\e_i),
\label{SEOM}
\\
	\vec{B}(\e_i)
	\equiv&\,
	-
	2 \e_i \hat{z}
	-
	\Delta 
	(\hat{x} + i \hat{y}) 
	-
	\Delta^*
	(\hat{x} - i \hat{y}),
\label{BEOM}
\end{align}
\esub
where the BCS order parameter was defined in Eq.~(\ref{DeltaDef}).

In the thermodynamic limit $N \rightarrow \infty$, the expectation of $\Delta$
in an initial BCS pure state becomes a classical observable (Ehrenfest's theorem).
Then, $\left\langle \vec{S}(t)\right\rangle \equiv \vec{s}(t)$ satisfies the
equation of motion for a classical spin in an external field. 
The usual BCS ground-state equation \cite{Schrieffer} obtains by aligning each spin to its own magnetic field \cite{Anderson1958}. 
The total particle number is encoded in the conserved $S^z \equiv \sum_{i} s_i^z$; its value
can be fixed by adding this to the field $\vec{B}$ with a chemical potential.

\subsection{The Lax spectral method \label{sec:LaxMethod}}

\subsubsection{Isolated roots and phases I, II, III}

The Lax spectral method \cite{Yuzbashyan2005Lax,Yuzbashyan2005,Yuz2,Foster2013,Yuzbashyan2015}
simplifies the dynamics for the classical spins,
generated by the BCS Hamiltonian in Eq.~(\ref{HRed}). The idea is that the
all-to-all coupling in this model leads to a reduction of 
the collective dynamics, such that the time evolution of $\Delta(t)$ can 
be determined by solving an emergent few-spin problem. This can be understood
as a sort of exact renormalization group: the effective few-spin problem
is governed by a ``Lax reduced'' Hamiltonian of precisely the same form as Eq.~(\ref{HRed}),
but with $M < N$ spins and renormalized parameters. 
The parameters of the reduced model can be determined from the $N$-spin quench
via the integrability. Specifically, these obtain from the roots of a certain
spectral polynomial, defined below. 

Here, we detail the Lax construction for the $s$-wave Hamiltonian in Eq.~(\ref{HRed}) 
\cite{Yuzbashyan2005Lax,Yuzbashyan2005,Yuz2,Yuzbashyan2015}.
A similar construction for a $p+ip$ superconductor can be found in \cite{Foster2013}.
See Refs.~\cite{Gaudin,Dukelsky2004,Richardson2002,Skrypnyk2009,Ibanez2009,Dunning2010,Ortiz2010} for 
the general classification of Richardson-Gaudin models. 

We define the Lax vector and its norm,
\begin{align}\label{LaxDef}
	\vec{L}(u)
	\equiv
	\sum_{i = 1}^N
	\frac{\vec{s}_i}{u - \e_i} 
	-
	\frac{\hat{z}}{G},
\quad
	L_2(u) \equiv \vec{L}\cdot\vec{L}(u),
\end{align}
where $u$ is an arbitrary parameter.
The spins can be taken to satisfy classical Poisson brackets $\{s_i^a,s_j^b\} = \delta_{i j} \epsilon^{a b c} s_i^c$. 
Then Eq.~(\ref{HRed}) implies that the Lax vector satisfies the same equation of motion as the spins, 
\begin{align}
	\dot{\vec{L}}(u) = \vec{L}(u) \times \vec{B}(u).
\end{align}
The Lax norm is conserved. 
Moreoever, it is easy to check that $\{L_2(u),L_2(v)\} = 0$
for any $u,v$. Using this fact and the explicit expression for the Lax norm, one can identify a set of $N$ independent
conserved quantities. These are Hamiltonians of central-spin type (``Gaudin magnets''), 
\begin{align}
	H_i \equiv \vec{s}_i \cdot \left[-\frac{\hat{z}}{G} + \sum_{j \neq i}\frac{\vec{s}_j}{\e_i - \e_j}\right],
	\quad
	\{H_i,H_j\} = 0 \;\; \forall \;\; i,j. 
\end{align}
Up to an additive constant, the BCS Hamiltonian is the particular combination
\begin{align}
	H = - G \sum_{i}2 \e_i H_i + G (S^z)^2.
\end{align}

From the conserved Lax norm, we define an order $2N$ spectral polynomial in the parameter $u$,
\begin{align}\label{SpecPoly}
	Q_{2N}(u) \equiv G^2 \prod_{i = 1}^N (u - \e_j)^2 L_2(u). 
\end{align}

To see how a small number of degrees of freedom can govern the dynamics, 
we trade the spins for an alternative set of ``separation variables,'' 
defined as follows. Eqs.~(\ref{LaxDef}) and (\ref{DeltaDef}) imply that 
\begin{align}
	L^-(u) \equiv L^x(u) - i L^y(u) = - \frac{\Delta}{G} \frac{\prod_{\alpha = 1}^{N-1}(u - u_\alpha)}{\prod_{j =1}^N(u - \e_j)}.
\end{align}
The $N-1$ zeroes of this equation $\{u_\alpha\}$ are complicated functions of the $\{s_i^-\}$, the precise form of which we do not need. 
By considering the $d/dt \left[L^-(u(t))\right]$ for a possibly time-dependent $u(t)$, we can derive the equations of motion
for these separation variables \cite{Yuzbashyan2005Lax},
\begin{align}\label{SepVarDyn}
	\dot{u}_\alpha
	=
	2 i 
	\frac{\sqrt{Q_{2N}(u_\alpha)}}{\prod_{\beta\neq\alpha} (u_\alpha - u_\beta)}.
\end{align}

The reduction to fewer degrees of freedom is best illustrated by an example.
Using the spin configuration that solves the BCS equations in the ground state, 
one can explicitly show that the spectral polynomial in Eq.~(\ref{SpecPoly})
possesses $N-1$ doubly-degenerate zeroes along the real $u$-axis. 
Since $L_2(u) = L^+(u) L^-(u) + \left[L^z(u)\right]^2$, any such zero $u_0$
also satisfies $L^-(u_0) = 0$. In other words, the $N-1$ separation variables $\{u_\alpha\}$ 
are static, and confined to the real axis [Eq.~(\ref{SepVarDyn})]. In addition, $Q_{2N}(u)$ possesses a complex-conjugate
pair of zeroes $\{u_1,u_1^*\}$ that encode the BCS ground-state order parameter $\Delta_0$ (which we take to be real). 
For a particle-hole symmetric $\e_i$ spectrum, one simply finds that $u_1 = i \Delta_0$.

Consider the case of a quench, wherein some initial pure state undergoes 
evolution according to the BCS Hamiltonian in Eq.~(\ref{HRed}). 
The initial state must be expressible as a configuration of the $N$ pseudospins 
(a pure BCS state), but is otherwise arbitrary \footnote{Instead of an $N$-fold spin product (spin coherent) state, one can also consider a state with $P \leq N$ ``blocked'' levels. These are states that possess a single fermion occupying one state of a Cooper pair. For the reduced BCS Hamiltonian in Eq.~(\ref{HRed}), blocked levels completely decouple from the time-evolution of the pseudospins (which are superpositions of doubly empty and occupied states of a Cooper pair).}. 
Feeding this initial spin configuration into the spectral polynomial 
[using the Lax vector in Eq.~(\ref{LaxDef})], one can characterize the state of the system in terms of 
the pattern of zeroes for $Q_{2N}(u)$, which is conserved. 
For finite $N$ and a generic initial state, one finds $N$ complex conjugate pairs of zeroes, implying that 
all separation variables $u_\alpha$ evolve nontrivially according to Eq.~(\ref{SepVarDyn}).
Despite the integrability of these equations, explicit results are only available for small $N$
in terms of hyperelliptic functions \cite{Yuzbashyan2005Lax,Yuzbashyan2005}.  
\begin{figure}[t!]
\centering
\includegraphics[width=0.4\textwidth]{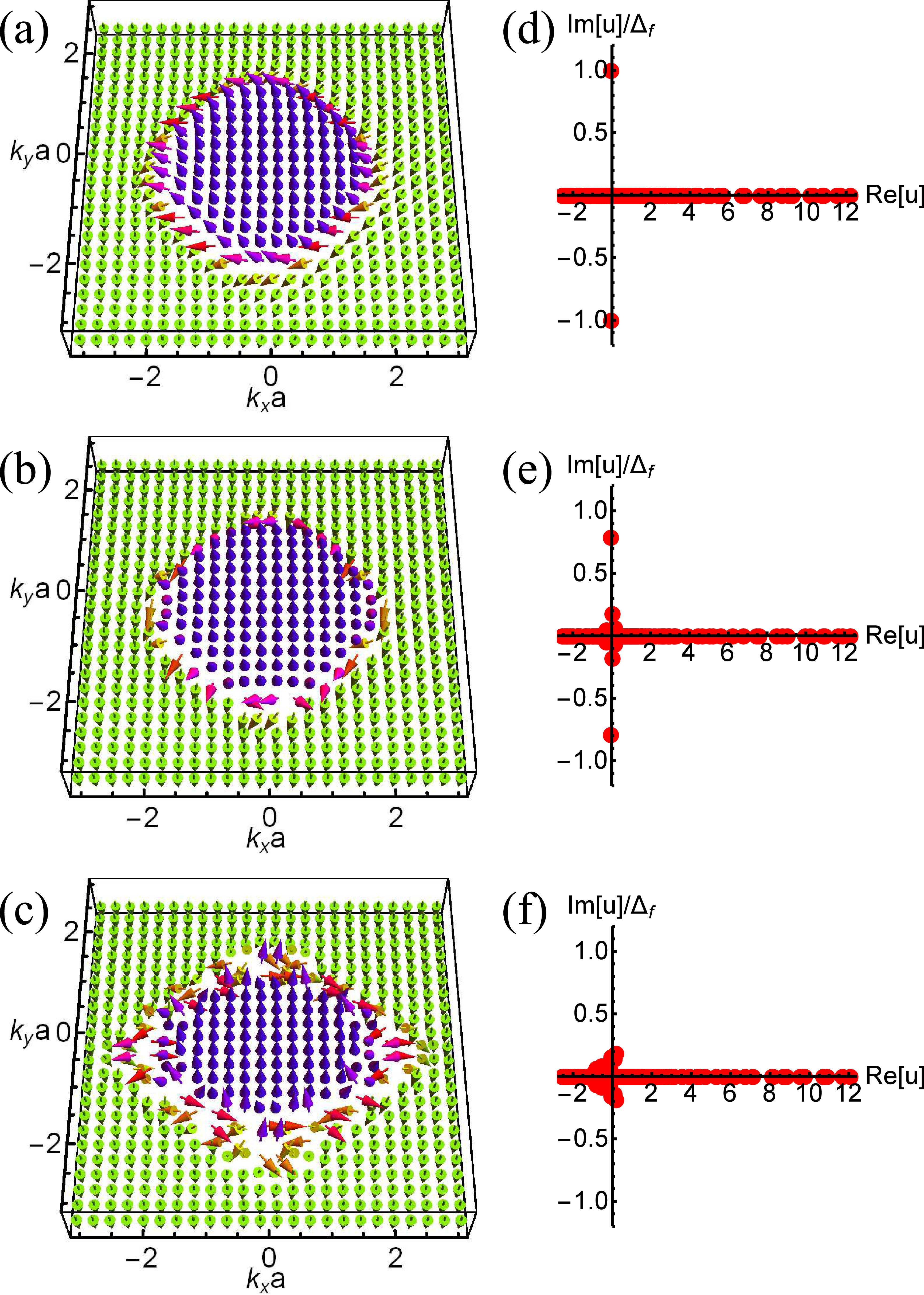}
\caption{This figure shows the Anderson pseudospin texture for a model 2D $s$-wave superconductor,
following a ``quench'' induced by a subgap electromagnetic pulse \cite{Chou2017}, 
see also Refs.~\cite{Papenkort2007,Papenkort2008,KrullSchnyder2014,Papenkort2009}. 
This work was inspired by the results of the experiment by Matsunaga et al.\ \cite{Matsunaga2013}, 
depicted in Fig.~\ref{Fig--Shimano} and reviewed in Sec.~\ref{CMI}, below.
The Anderson pseudospin textures and the Lax roots of the spectral polynomial are depicted immediately after 
the application of an intense THz pulse. 
(a), (b), and (c) are the spin textures after light exposure, corresponding to the
roots shown in (d), (e), and (f) respectively.
The pair of roots away from the real axis are called isolated roots, and these encode the key properties 
of the BCS state.  
The system is a quarter-filled square lattice tight-binding model with 24-by-24 sites
(the system is chosen to be small for the purpose of this illustration).  
Panels (a) and (d) correspond to a very weak pump energy, (b) and (e) to an intermediate pump energy, and 
(c) and (f) to a strong pump. 
The isolated roots $u^{\pm}_1 \simeq \pm i\Delta_{\infty}$
for the deformed spin textures encode
the asymptotic value of $\Delta(t \!\! \rightarrow\!\! \infty) = \Delta_\infty$ 
in the phase II pre-thermalization plateau. 
The main effect of the THz field quench is to twist Anderson pseudospins near the 
Fermi energy in the $x$-$y$ plane; more intense pulses produce more disordered twist patterns, 
and push the isolated roots towards the real axis.
For strong pulses (c,f), the isolated roots merge with the real axis and the 
system enters phase I \cite{Barankov2006,Yuz} with $\Delta_\infty = 0$, see Fig.~\ref{Fig--THz_Phase_Diagram}.}
\label{Fig--THz_Texture_Roots}
\end{figure}

\begin{figure}
\centering
\includegraphics[width=0.4\textwidth]{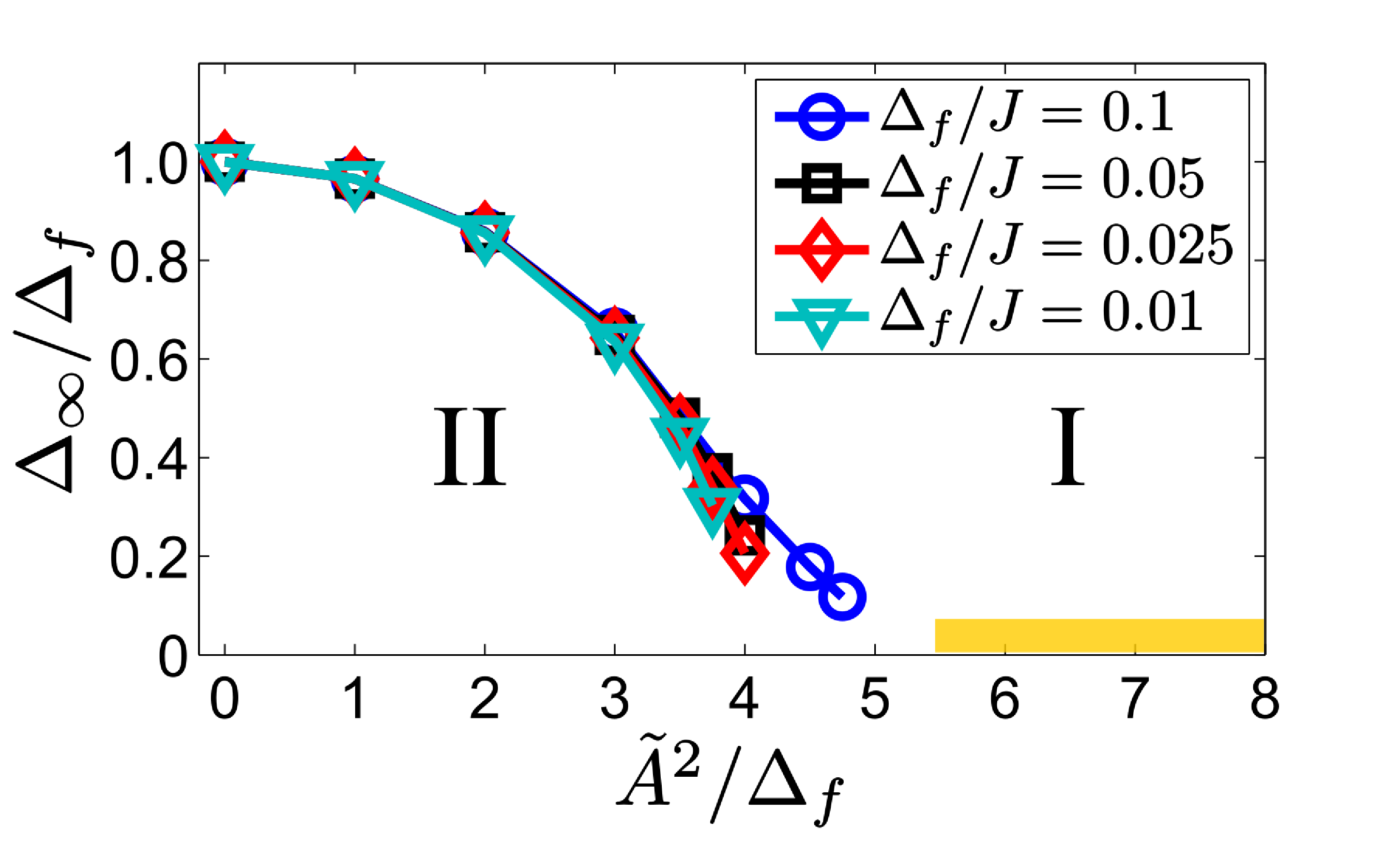}
\caption{Theoretical prediction for the nonequilibrium phase diagram of an $s$-wave BCS superconductor 
subject to an ultrashort ($\simeq$ monocycle) THz pulse with center frequency $\omega \simeq \Delta$,
from Ref.~\cite{Chou2017}, described in Fig.~\ref{Fig--THz_Texture_Roots}. 
The 2D square lattice has linear size $L=1000$ at quarter filling. 
Values of $\Delta_{\infty}$ are extracted from the isolated Lax roots corresponding 
to the spin configuration immediately after cessation of the pump pulse. 
We plot $\Delta_{\infty}$ as a function of the peak pulse intensity $\tilde{A}^2$, 
with different values of $\Delta/J$ ($J$ is the hopping strength). 
As illustrated in Fig.~\ref{Fig--THz_Texture_Roots}, more intense pulses produce larger deformations 
of the pseudospins along the Fermi surface, leading to a suppression of the asymptotic BCS gap $\Delta_\infty$. 
%Phase I ($\Delta_\infty = 0$) can be achieved with $\tilde{A}^2/\Delta > 5.5$ (yellow shaded region) in all cases. 
%The phase boundary for $\Delta/J=0.01$ is close to $\tilde{A}^2/\Delta = 4.5$.
}
\label{Fig--THz_Phase_Diagram}
\end{figure}

In the thermodynamic $N \rightarrow \infty$ limit, however, for certain classes of initial conditions the
situation simplifies dramatically. In particular, for quenches that evolve an initial weakly paired BCS or strongly paired BEC 
ground state
(characterized by coupling $G_i$) according to a post-quench Hamiltonian with $G \equiv G_f \neq G_i$, one
finds that only three possible dynamical phases appear. These are termed phase I, II, and III, 
and correspond respectively to zero, one, and two isolated pairs of roots of the spectral polynomial
\cite{Yuz2,Barankov2006,Yuz}.  
The remaining roots merge into a branch cut along the real axis, and thus correspond
to a continuum of ``frozen'' separation variables. The separation between isolated and remaining roots 
is illustrated for a finite system in Fig.~\ref{Fig--THz_Texture_Roots}. 
The physics of phases I--III are as follows (see also the example phase diagrams in Fig.~\ref{Fig--PwavePD} and \ref{Fig--THz_Phase_Diagram}.)
\begin{itemize}[leftmargin=10pt]
\item{In phase I, there are no isolated roots, and $\Delta(t) \rightarrow 0$ in the long-time limit. 
This occurs for large strong-to-weak
quenches, $G_f \ll G_i$ \cite{Yuz}, and corresponds to an effective zero-spin problem. 
In a superconductor quenched to phase I,
the optical conductivity or Meissner response would appear indistinguishable from a normal metal \cite{Chou2017}. 
Anomalous coherences persist however, encoded in the dephased precession of Anderson pseudospins; schemes for detecting these are discussed in \cite{Altman2004,Folling2005,Rom2006,Stahl2019}, see Sec.~\ref{MESymmetrybreaking}. 
Phase I could be accessed in a solid-state superconductor subject to a sufficiently strong subgap THz pulse, 
as illustrated in Figs.~\ref{Fig--THz_Texture_Roots} and \ref{Fig--THz_Phase_Diagram}.
} 
\item{In phase II there is one pair of isolated roots, corresponding to an effective one-spin problem.  
Phase II describes small quenches, and smoothly connects
to the ground-state configuration (no quench). $\Delta(t)$ asymptotes to a nonequilibrium, steady-state value $\Delta_\infty$.
For quenches entirely in the weak-pairing BCS regime, the approach to this prethermalized value takes a universal form \cite{VolkovKogan1973,Yuz2},
\begin{align}\label{PhaseIIApproach}
	\Delta(t) \simeq \frac{\alpha}{\sqrt{t}}\cos\left(2 \Delta_\infty t + \phi_0\right) + \Delta_\infty,
\end{align}
for some constants $\alpha,\phi_0$.
The value of $\Delta_\infty$ is precisely determined by the isolated roots of the spectral polynomial for the
$N \rightarrow \infty$ spin system.
Decaying oscillations at frequency $2 \Delta_\infty$ were observed in the THz pump-probe experiment Ref.~\cite{Matsunaga2013},
see Fig.~\ref{Fig--Shimano} and Sec.~\ref{CMI}.
} 
\item{In phase III, there are two pairs of isolated roots, corresponding to an effective two-spin problem. 
This occurs for sufficiently large weak-to-strong quenches, and connects in a somewhat intricate way to a
quench starting from a normal (unpaired) Fermi liquid state \cite{Barankov2004,Barankov2006Replace,YuzbashyanTsyplyatyev2009},
(see Sec.~\ref{sec:flucts}, below).
The isolated roots of the spectral polynomial in the thermodynamic limit determine the parameters 
of this two-spin problem, leading to an effective equation of motion for $\Delta(t)$. 
The solution can always be expressed in terms of elliptic functions, and thus corresponds
to a \emph{quench-generated Floquet phase} \cite{Barankov2006,Yuzbashyan2015}, as illustrated in Fig.~\ref{Fig--PwaveFloquet}
for a $p+ip$ superfluid quench \cite{Foster2013,Foster2014,Liao2015}.
Heating is a potential difficulty with driven Floquet systems \cite{Oka2019}, but the quench-induced phase III dynamics could evade this problem.
By contrast, a numerical study of quenches in a nodal $d$-wave superconductor 
did not exhibit phase III, which is damped out due to excited gapless quasiparticles \cite{Peronaci2015}.
This $d$-wave reduced BCS model is not of Richardson-Gaudin type, due to the form of the interaction coupling. 
}
\end{itemize}

\subsection{A simple example}\label{simple}

\subsubsection{Phases I and II}

In this subsection, we illustrate the Lax spectral method for the model in Eq.~(\ref{HRed}).
In order to keep the analysis as simple as possible, instead of a BCS (or BEC) initial Anderson 
pseudospin configuration, we consider a fully $x$-polarized initial state \cite{ZhangGurarieFoster}. In fact, this initial
state  was used in a recent ultracold fermionic atoms experiment \cite{smale2019} to probe the phase I to phase II DPT (see Sec.~\ref{sec:exp}). 

%\blue{[*** Ana: say something about the relevance to clock experiments?]}
\begin{align}\label{FullyxPol}
	\vec{s}_i = (1/2)\hat{x}, \quad \forall \;\; i \in \{1,\ldots,N\}.
\end{align}
The spectral polynomial [Eq.~(\ref{SpecPoly})] is 
\bsub
\begin{align}
	Q_{2N}(u) 
	=&\, 
	\left[
		\prod_{i = 1}^N
		(u - \e_i)^2
		+
		\left(\frac{G}{2}\right)^2
		P_{N-1}^2(u)
	\right],
\end{align}
where
\begin{align}
	P_{N-1}(u) 
	\equiv&\,
	\prod_{i = 1}^N
	(u - \e_i)
	\left[
	\sum_{j = 1}^N
	\frac{1}{u - \e_j}
	\right].
\end{align}
\esub
Assume that the single-particle energies reside in the bounded interval $-2J \leq \e_i \leq 2 J$.
Then, for $G \ll 2 J$, $Q_{2N}(u)$ possesses $2N$ zeroes that appear in complex conjugate pairs
close to each of the $N$ bare real energies $\{\e_i\}$. 
On the other hand, for $G \gg 2 J$, $2(N-1)$ zeroes appear close to the $N-1$ real zeroes of $P_{N-1}(u)$.
Thus there must be an additional, isolated pair.

To take the thermodynamic $N \rightarrow \infty$ limit, we consider a simple 1D cosine band $\e \rightarrow 2 J \cos(k)$,
with $|k| \leq \pi$. The Lax norm evaluates to 
\begin{align}\label{LaxNormIandII}
	L_2(u) = (L \nu_0)^2\left[\frac{\pi^2}{(u/J)^2 - 4} + \frac{1}{g^2}\right], \quad g = \nu_0 L G,
\end{align}
where $\nu_0 = 1/(2 \pi J)$ is the density of states at the band center and $L$ is the linear system size.
The roots are $u_1^{\pm} = \pm J\sqrt{4 - (\pi g)^2}$. 
There is a dynamical phase boundary at $g_c \equiv 2/\pi$: 
for $g < g_c$, the isolated roots are real and merge with the continuum branch cut---this is phase I.
For $g > g_c$, the isolated roots are purely imaginary, and we have phase II. 

For phase II, the Lax reduced problem corresponds to a single collective pseudospin $\vec{\sigma}$,
with a one-spin BCS Hamiltonian [compare to Eq.~(\ref{HRed})]
\begin{align}
	H_1 = 2 \chi \sigma^z - \frac{|\Delta_\infty|^2}{G}, \quad |\Delta_\infty| = - G |\sigma^-|. 
\end{align}
Here $\chi$, $\Delta_\infty$, and $\sigma^z$ are all constants that must be determined by 
the roots of the effective one-spin spectral polynomial,
\begin{align}
	Q_2(u) = (u - u_1^+)(u - u_1^-). 
\end{align}
Using the isolated roots from the many-body Lax norm in Eq.~(\ref{LaxNormIandII}), we find that
\begin{align}
	\Delta_\infty 
	= 
	|u_1^{\pm}|
	=
	2 J
	\,
	\theta(g - g_c)
	\,
	\sqrt{\left(\frac{g}{g_c}\right)^2 - 1},
\end{align}
where $\theta(x)$ is the unit step function.

It is important to note that not all information about the long-time evolution 
of the post-quench dynamics is encoded by the isolated roots. 
In a BCS ground state, each Anderson pseudospin $\vec{s}_i$ is aligned
to its magnetic field $\vec{B}(\e_i)$ [Eq.~(\ref{BEOM})]
(possibly modified by a chemical potential term to fix the density). 
In phase II, the effective magnetic field is determined by the 
asymptotic value of the gap $\Delta_\infty$,
\[
	\vec{B}_\infty(\e_i) = -2 \e_i \hat{z} - 2 \Delta_\infty \hat{x}.
\]
However, the steady-state pseudospins will in general exhibit precession
around this field at a finite canting angle,
\begin{align}\label{PrecSol}
	\lim_{t \rightarrow \infty}
	\vec{s}_i(t)
	=&\,
	\frac{1}{2}
	\sqrt{1 - \gamma_i^2}
	\left\{
	\begin{aligned}
	&\,
		\cos\left[2 E_\infty(\e_i) t\right]
		\hat{B}_\infty(\e_i) 
		\times
		\hat{y}
	\\&\,
		+
		\sin\left[2 E_\infty(\e_i) t\right]
		\hat{y}
	\end{aligned}
	\right\}
\nonumber\\
	&\,
	-
	\frac{\gamma_i}{2}
	\hat{B}_\infty(\e_i),
\end{align}
where $E_\infty(\e_i) \equiv \sqrt{\e_i^2 + \Delta_\infty^2}$ is the 
asymptotic steady-state BCS quasiparticle energy. 
The function $\gamma_i$ can be computed exactly by evaluating the 
Lax norm for the precessing solution in Eq.~(\ref{PrecSol}),
discarding terms that oscillate as $t \rightarrow \infty$, and
equating this with Lax norm for the pre-quench initial condition;
we omit details here, but see Refs.~\cite{Yuz,Foster2013,Yuzbashyan2015}. 
In phase II, the function $\gamma_i$ plays the role of a nonequilibrium distribution function for
the quasiparticle spectrum in the asymptotic steady state:
\begin{align}\label{gammaDef}
	-\gamma_i = 1 - 2 f_i,
\end{align}
where $f_i$ would take the Fermi-Dirac form in terms of the quasiparticle energies for
a system in thermal equilibrium. 

Once the gap $\Delta_\infty$ and the distribution function $\gamma_i$ are known,
generic $n$-body steady-state Green's functions (retarded, Keldysh, etc.) can be computed for the evolving pure BCS state.
See e.g.\ Refs.~\cite{Dzero2007,Foster2013,Liao2015} for examples of one-body functions and observables such as rf spectroscopy, 
tunneling, and time-resolved ARPES, 
and Ref.~\cite{Chou2017,Millis2017} for the current-current correlator that determines the optical conductivity and Meissner 
response of a quenched solid-state superconductor.

\subsubsection{Quench-generated Floquet phase III}

Phase III arises for sufficiently large weak-to-strong pairing BCS quenches, see e.g.\ Figs.~\ref{Fig--PwavePD} and \ref{Fig--PwaveFloquet}.
As we review below in Sec.~\ref{sec:flucts}, a special case corresponds to turning on attractive interactions
for an initially unpaired, noninteracting Fermi gas ground state. Fluctuations (either quantum or thermal)
play a crucial role in the real dynamics of such a quench, because the noninteracting ground state 
(a fully $\pm s^z$-polarized
Anderson pseudospin magnet with a sharp domain wall at the Fermi surface) is a metastable stationary state.

Nevertheless, one can use the Lax spectral method to consider the formal limit of $\Delta_i / \Delta_f \rightarrow 0$, 
where $\Delta_i$ ($\Delta_f$) denotes the prequench (postquench) ground state gap (associated to a pairing strength $G_i$ or $G_f$, respectively). 
In this limit, $\Delta(t)$ possesses a soliton solution that grows from zero, reaches a peak, and decays back to zero \cite{Barankov2004}.
The undamped oscillations of $\Delta(t)$ in phase III can be understood as a train of such solitons \cite{Barankov2004,Yuz2}, and are thus
in some sense adiabatically connected to the quench from an initial state with a sharp Fermi step. 

Instead of the metastable normal state, here we consider the fully $s^x$-polarized initial state 
in Eq.~(\ref{FullyxPol}), but now with a pair of sharp domain walls: \cite{ZhangGurarieFoster}
\begin{align}\label{FullyxPolDW}
	\vec{s}_i = (1/2)\hat{x} \sgn(\e_i).
\end{align}
As proposed in Ref.~\cite{Lewisswan2021} such an  initial state can be prepared in a cavity QED  system. In that work it is discussed how this state  can  used to probe  phases I, II and III in a controllable setting. See Sec.~\ref{sec:exp}.
For the 1D cosine band employed above, the Lax norm in the continuum limit evaluates to
[cf.\ Eq.~(\ref{LaxNormIandII})]
\begin{align}\label{LaxNormIII}
	L_2(u) 
	= 
	(L \nu_0)^2
	\left\{
		\frac{\left[\csc^{-1}(u/2J)\right]^2}{(u/2J)^2 - 1} + \frac{1}{g^2}
	\right\}.
\end{align}
For $g = 0$, this has only real roots at $u = \pm 2 J$.
For any $g > 0$, two pairs of conjugate isolated roots nucleate into the complex-$u$ plane. 
Denote these as $\{u_1,u_1^*,u_2,u_2^*\}$. For this particle-hole symmetric example, 
$u_2 = - u_1^*$. In the limit of large $g$, one has 
\begin{align}\label{u1Asym}
	\left(\frac{u_1}{2 J}\right)^2 
	\simeq 
	\frac{2 g^2}{1 - i \sqrt{4 g^2 -1 }}. 
\end{align}
 
Two pairs of isolated roots corresponds to an effective two-spin problem, with 
spectral polynomial 
\begin{align}
	Q_4(u) = (u - u_1)(u - u_1^*)(u - u_2)(u - u_2^*).
\end{align}
Parameterizing the order parameter $\Delta(t)$ in terms of an amplitude and phase via
\begin{align}
	\Delta \equiv \sqrt{\Rh} \, \exp(-i \phi),
\end{align}
one can solve the two-spin problem to obtain equations of motion in terms of the roots.
Details of this calculation can be found elsewhere \cite{Barankov2004,Foster2013,Yuzbashyan2015}. 
The results are
\begin{align}
\begin{aligned}
	\dot{\Rh}^2 
	=&\, 
	4(\Rh_+ - \Rh)(\Rh - \Rh_-)(\Rh + \tilde{\Rh}),
\\
	\dot{\phi}
	=&\,
	c_1
	+
	\frac{c_2}{\Rh},
\end{aligned}
\end{align}
where
\begin{align}
\begin{aligned}
	\Rh_{\pm}
	\equiv&\,
	\left(
		\left|\im \, u_1\right| \pm \left|\im \, u_2\right|
	\right)^2,
\\
	\tilde{\Rh} 
	\equiv&\,
	\left[\re \left(u_1 - u_2\right)\right]^2,
\\
	c_1 
	\equiv&\,
	\re(u_1 + u_2),
\\
	c_2
	\equiv&\,
	\left[\re(u_1 - u_2)\right]
	\left[(\im \, u_1)^2 - (\im \, u_2)^2\right].
\end{aligned}
\end{align}
In our example [Eqs.~(\ref{LaxNormIII}) and (\ref{u1Asym})],
we define $u_r \equiv \re \, u_1$, $u_i \equiv \im \, u_1$,
and $u_2 = - u_1^*$.
Then the amplitude $\Delta(t)$ is given by 
\begin{align}\label{JacobiDelta}
	\Delta(t)
	=
	2 u_i 
	\dn\left(2 u_i t \Big| M\right)
	\simeq
	2 u_i 
	\cos\left(\frac{2 \pi t}{\mathcal{T}}\right),
\end{align}
where 
\begin{align}\label{JacobiDeltaParams}
\begin{aligned}
	M 
	=&\,
	1 + {u_r^2}/{u_i^2},
	\\
	\mathcal{T}
	=&\,
	\frac{1}{2 ui}
	\left[4 K(M) + 4 i K(1-M)\right].
\end{aligned}
\end{align}
In Eq.~(\ref{JacobiDelta}), $\dn(z|M)$ is the Jacobi elliptic function, with $M$ the ``modulo'';
the approximate form of $\Delta(t)$ given by the second equality is a very good approximation,
except for very large $|u_{r,i}|$. 
In Eq.~(\ref{JacobiDeltaParams}), 
$K(M)$ is the complete elliptic integral of the first kind, and $M = k^2$, with $k$ the elliptic modulus parameter. 
%Using the asymptotic formula for the isolated roots in Eq.~(\ref{u1Asym}),
%we plot the Floquet period $\mathcal{T}$ versus $g$ in Fig.~\ref{Fig--PeriodPlot}.

We conclude that the $x$-polarized ``domain wall'' initial condition in 
Eq.~(\ref{FullyxPolDW}) gives rise to a self-generated Floquet phase III,
where the order parameter is given by the elliptic solution in Eq.~(\ref{JacobiDelta}).
The parameters of this solution are determined by the four isolated roots $\{u_1,u_1^*,-u_1,-u_1^*\}$,
where $u_1$ is given by Eq.~(\ref{u1Asym}) for sufficiently large $g$. 

Finally, we note that more complicated initial conditions with additional discontinuities 
can excite three or more pairs of isolated roots. These typically give rise to a self-generated, 
\emph{quasiperiodic Floquet phase} for $\Delta(t)$ in the prethermalized regime \cite{Yuz2,Yuzbashyan2008}.
These quasiperiodic phases have so far received little attention, and remain an attractive
avenue for future work.

%%%%%%%%%%%%%%%%%%%%%%%%%%%%%%%%%%%%%%%%%%%%%%%%%%%%%%%%%%%%%%%%%%%%%%%%%%%%
%%%%%%%%%%%%%%%%%%%%%%%%%%%%%%%%%%%%%%%%%%%%%%%%%%%%%%%%%%%%%%%%%%%%%%%%%%%%
%%%%%%%%%%%%%%%%%%%%%%%%%%%%%%%%%%%%%%%%%%%%%%%%%%%%%%%%%%%%%%%%%%%%%%%%%%%%
%%%%%%%%%%%%%%%%%%%%%%%%%%%%%%%%%%%%%%%%%%%%%%%%%%%%%%%%%%%%%%%%%%%%%%%%%%%%
%%%%%%%%%%%%%%%%%%%%%%%%%%%%%%%%%%%%%%%%%%%%%%%%%%%%%%%%%%%%%%%%%%%%%%%%%%%%

\subsection{Fluctuation phenomena \label{sec:flucts}}

The reduced BCS Hamiltonian in Eq.~(\ref{HRed}) is an approximation to the idealized
model for a Fermi gas with short-ranged interactions in Eq.~(\ref{HBCSmicro}). 
Only Cooper pairs with zero center-of-mass (COM) momentum are retained. This has several consequences.
Formally, when this model is coupled to an electric field \cite{Chou2017}, appropriate
for THz driving of a solid-state superconductor, the projection to zero-COM pairs breaks gauge invariance. 
This can be restored by incorporating the vector potential into the pairs, and still gives an 
effective classical system that can be efficiently simulated numerically \cite{Perakis2020}. 

Two more intricate problems are of particular theoretical and experimental interest. 
These are (1) a quench from the normal Fermi liquid state, either above $T_c$ or at zero temperature
with the pairing interactions turned off, and (2) a quench in a system of size $L \gg \xi$, where
$\xi$ is the coherence length. We discuss these in turn. 

Quenches in which attractive pairing interactions are switched on from an equilibrium normal state above $T_c$ 
have been studied in Refs.~\cite{Mitra2017,Mitra2018A,Mitra2018B,Mitra2019}. 
In this case, the fluctuation dynamics of the pairing amplitude are overdamped and can be treated classically,
while the Keldysh method can be employed to evaluate one- and two-fermion observables (such as the optical conductivity \cite{Millis2017,Mitra2018A,Mitra2018B}). 

By contrast, a quench from a normal system with energy density below $T_c$ is an extreme limit of a
weak-to-strong pairing quench, and thus would be expected to induce the self-generated Floquet phase III (see e.g.\ Fig.~\ref{Fig--PwavePD}). 
\begin{comment}
The period of Floquet oscillations for a phase III quench from a pre- (post-)quench system with ground state pairing gap $\Delta_i$ ($\Delta_f$)
is given by \cite{Barankov2004,Barankov2006,Foster2014}
\begin{align}\label{FloquetPeriod}
	\mathcal{T} 
	= 
	\frac{2}{\Delta_f} 
	\ln\left(\frac{\Delta_f}{\Delta_i}\right).
\end{align}
Quenching from a normal initial state instead, thermal fluctuations give 
$\Delta_i \sim \sqrt{\delta_\e T}$, where $T$ is the temperature and $\delta_\e$ is the single-particle level spacing \cite{YuzbashyanTsyplyatyev2009}. 
Even for a quench from a paired initial state, there is a variation in the Floquet period due to thermal fluctuations of order 
$\delta \mathcal{T} = \sqrt{\delta_\e T}/(\Delta_f \Delta_i)$. 
Thus, for a phase III quench from a finite-temperature BCS initial state with $\Delta_i > 0$, one expects
\begin{align}
	\frac{\mathcal{T}}{\delta \mathcal{T}}
	\sim
	\frac{\Delta_i}{\sqrt{\delta_\e T}}\ln\left(\frac{\Delta_f}{\Delta_i}\right),
\end{align} 
while for a quench from the finite-temperature normal state, one instead gets
\begin{align}
	\frac{\mathcal{T}}{\delta \mathcal{T}}
	\sim
	\ln\left(\frac{\Delta_f}{\sqrt{\delta_\e T}}\right).
\end{align} 
Since $\delta_\e \sim 1/N$, where $N$ is the number of particles, we see that
\end{comment}
A weak-to-strong BCS quench from a finite-temperature initial paired state,
in a system with $N$ pseudospins, can exhibit $\ord{\sqrt{N}}$ oscillation cycles, 
which can make phase III easily observable for $N \gg 1$. This follows from
the central-limit theorem \cite{YuzbashyanTsyplyatyev2009}.
On the other hand, the quench from the normal state would exhibit only $\ord{\ln N}$ oscillations 
\cite{YuzbashyanTsyplyatyev2009}. 
This would hardly be distinguishable from phase II [Eq.~(\ref{PhaseIIApproach})]
in an ultracold atom experiment (where $N$ is not exponentially large, unlike in solid state systems).

More problematic for ultrafast THz quench experiments in solid-state superconductors
(discussed below in Sec.~\ref{CMI}) is the fact that practical sample sizes typically have linear
dimension $L \gg \xi$, where $\xi = \hbar v_F/\pi \Delta$ is the coherence length. 
It can be shown that phase III is unstable in this case to the spontaneous generation
of spatial fluctuations (``Cooper pair turbulence'') \cite{YuzbashyanDzero2009,Chern2019}.
The mechanism is parametric resonance.

Finally, it must be remembered that even Eq.~(\ref{HBCSmicro}) is a significant simplification for both real ultracold Fermi gases and
solid-state superconductors. In the case of gases, it neglects the internal degrees of freedom that can lead to losses 
\cite{Salomon2004,Castin2008,Gurarie2008}, while for a superconductor, it neglects the retarded character
of the dynamical pairing interaction, as typically mediated by phonons \cite{Schrieffer,Tinkham}.

\subsection{Topological features \label{sec:topo}}

In an ordinary, $s$-wave fermionic superfluid or superconductor, the BCS
and BEC regimes respectively correspond to weak and strong pairing limits.
One speaks of the BCS-to-BEC crossover in the zero-temperature quantum phase diagram, 
as a function of the pairing interaction strength [$G$ in Eq.~(\ref{HRed})] \cite{Levin2005}. 
The BCS regime can be 
pictured as a gas of weakly bound, strongly overlapping Cooper pairs, while the cartoon 
for the BEC is a bosonic condensate of tightly bound two-fermion molecules. Since 
the zero temperature system is fully gapped for 
any nonzero pairing strength, the BCS and BEC regimes are limits
of the same phase. 

By contrast, for non-$s$-wave pairing, there is typically a genuine quantum phase
transition separating the BCS and BEC regimes. In particular, for 2D $p+ip$ 
superfluid, the weak-pairing BCS phase is topologically nontrivial. 
It means there exists a quantized integer-value winding number $W$, which characterizes
the phase, that takes a nonzero value. The most important physical consequence 
of $W \neq 0$ is the presence of gapless, chiral edge states that circulate around
a 2D sample of finite extent \cite{Alicea2012}. This is a superconducting analog
of the quantum Hall effect, except that the circulating edge fermions 
(called ``Majorana fermions'') carry only energy instead of charge. 
A $p+ip$ BCS superconductor can also host so-called ``Majorana zero modes''
in vortex cores; these are non-abelian anyons that could be exploited
for topological quantum computation \cite{Nayak2008,Alicea2012}. 

On the other hand, the strongly paired BEC regime of a 2D $p+ip$ superfluid is topologically 
trivial; there are no gapless edge states, and it cannot host Majorana zero modes. 
This is because topology is encoded in the effective bandstructure for the 
fermionic quasiparticles of the superfluid, but in the strongly paired BEC, the only low-lying excitations
are bosonic molecules. The quantum phase transition separating the BCS and BEC regimes
involves a closing of the excitation gap, such that gapless bulk quasiparticles
appear precisely at the transition. 

Other topologically nontrivial superconductors and superfluids are possible in three dimensions,
analogous to the distinct $p$-wave paired $^3$He $B$ and $A$ phases \cite{Volovik,Mizushima2016}.
The former (latter) is a fully gapped, time-reversal invariant (gapless Weyl, time-reversal breaking)
topological superfluid; both are predicted to host gapless two-dimensional Majorana fermion surface fluids.  

Because topology and its consequences (e.g., gapless edge and surface states) are \emph{global features}
of a quantum phase of matter \cite{QiZhang2011}, these are expected to be immune to weak 
symmetry-preserving perturbations. It is therefore natural to ask which features survive under
a \emph{global} quantum quench. 

Quench and Floquet dynamics in topological bandstructures is already a vast subject, 
see e.g.\ Refs.~\cite{Rigol2015,Bhaseen2015,Martin-Delgado2010,Perfetto2013,Sacramento2014,Vishveshwara2014,Refael2016,Budich2016,Pan2018,CooperRMP2019,Iyer2020,Oka2019,Rudner2020}.
Here we confine our attention to topological features that emerge in a self-consistent
quench, studied for a $p$-wave Richardson-Gaudin model in \cite{Foster2013,Foster2014,Liao2015},
see also \cite{Pu2015,Dzero2015B,Yuzbashyan2019}.

The dynamical phase diagram for quenches in the 2D $p+ip$ Richardson-Gaudin model is shown in 
Fig.~\ref{Fig--PwavePD}. It possesses the same three dynamical phases I, II, III for the order
parameter $\Delta(t)$ discussed above. In this case, the self-generated Floquet phase III
is topological, and features chiral Majorana edge states, shown in Fig.~\ref{Fig--PwaveFloquet}. 
These are similar to edge states obtained in equilibrium \cite{Alicea2012} or under external Floquet driving \cite{Rudner2020},
except that they are induced here by a sufficiently large weak-to-strong pairing quench. 
It was argued in Ref.~\cite{Foster2014} that phase III might be realizable in an ultracold Fermi gas,
by quenching from an undetectably small (but nonzero) initial gap strength $\Delta_i$ to 
an intermediate strength $\Delta_f \gg \Delta_i$. Although the latter regime is associated
to strong parasitic three-body losses \cite{Salomon2004,Castin2008,Gurarie2008}
that prevent the adiabatic cooling to an equilibrium
topological state, a parameter window with $\mathcal{T} \ll t_3$ could possibly 
allow experimental observation of phase III. Here $\mathcal{T}$ is the period of the Floquet phase, 
and $t_3$ is the three-body loss lifetime.

Another key feature of the phase diagram in Fig.~\ref{Fig--PwavePD} is the winding number $W$,
which takes different values in two regions of phase II. The $W = 1$ ($W = 0$) region
is a topological (trivial) phase II regime. The topological version would host chiral gapless
Majorana edge states for a realization in terms of paired fermions, with a sample boundary.

There are different notions of winding numbers that are equivalent in equilibrium,
but which must be distinguished for a quench. The winding number $W$ (Volovik invariant) 
is defined via \cite{Volovik,Foster2013}
\begin{widetext}
\begin{align}\label{WDef}
	W
	\equiv
	\frac{\epsilon_{\alpha \beta \gamma}}{3!}
	\int_{-\infty}^\infty
	d \omega
	\int \frac{d^2\vex{k}}{(2 \pi)^2}
	\Tr\left[
	\hat{\mathcal{G}}^{-1} \left(\partial_\alpha \hat{\mathcal{G}}\right)
	\hat{\mathcal{G}}^{-1} \left(\partial_\beta \hat{\mathcal{G}}\right)
	\hat{\mathcal{G}}^{-1} \left(\partial_\gamma \hat{\mathcal{G}}\right)
	\right].
\end{align}
\end{widetext}
Here $\alpha,\beta,\gamma \in \{\omega,k_x,k_y\}$ and repeated indices are summed. 
The matrix $\hat{\mathcal{G}}(\omega,\vex{k}) \equiv \hat{G}_R(i \omega,\vex{k})$;
the latter is the analytic continuation of the Fourier transform for the 
retarded, Bogoliubov-de Gennes Green's function $\hat{G}_R(t - t',\vex{k})$. 
In the long-time steady state of phase II, this Green's function is time-translationally
invariant and independent of the distribution function $\gamma_i$ [defined in Eqs.~(\ref{PrecSol}) and (\ref{gammaDef}), above].
Because the retarded Green's function encodes information about the \emph{spectrum of excitations} 
in the phase II steady-state, it is the appropriate topological index to determine the 
post-quench topology and its consequences (presence or absence of edge states).
In Fig.~\ref{Fig--PwavePD}, $\Delta_{\mathsf{QCP}}$ denotes the ground-state topological
phase transition between the BCS and BEC regimes; the dashed purple line is its nonequilibrium extension.

By contrast, a different winding number defined to characterize the many-body state of
the system does \emph{not} change following a quench in the collisionless, prethermalized regime \cite{Foster2013,Rigol2015,Bhaseen2015}.
This winding number can be defined as \cite{Foster2013}
\begin{align}\label{QDef}
	Q
	\equiv
	8 \pi \epsilon_{a b c}
	\int 
	\frac{d^2 \vex{k}}{(2 \pi)^2}
	\frac{1}{k}
	\,
	s_{\vex{k}}^a
	\,
	\partial_k
	s_{\vex{k}}^b
	\,
	\partial_{\phi_k}
	s_{\vex{k}}^c,
\end{align}
where $k,\phi_k$ are polar coordinates for the momentum plane, 
and $a,b,c\in \{1,2,3\}$ (again with repeated indices summed).
The winding number $Q$ characterizes the pseudospin texture of the many-body state.
For an equilibrium topological BCS state, this is a skyrmion texture with $Q = W = 1$ \cite{Alicea2012,Foster2013}.
In the BEC state $Q = W = 0$, and the texture can instead be deformed to a trivial ferromagnetic one. 
The unitary evolution under the Richardson-Gaudin Hamiltonian [which neglects center-of-mass spatial fluctuations 
for the order parameter $\Delta(t)$] prevents a change in $Q$ for any quench.

The most interesting consequence of the dual winding numbers $(W,Q)$ out-of-equilibrium appears for phase II
quenches across the topological quantum phase transition, such that $W \neq Q$. 
Then, one can show that consistency between Eqs.~(\ref{WDef}) and (\ref{QDef})
requires a \emph{population inversion} in the occupation of steady-state quasiparticle
states. I.e., the distribution function $-\gamma = 1 - 2 f$ [Eqs.~(\ref{PrecSol}) and (\ref{gammaDef})],
which has $\gamma_{\vex{k}} = -1$ for all quasiparticle states in either the BCS or BEC ground state, necessarily
``winds'' to $\gamma_{\vex{k}} = + 1$ for $\vex{k} \rightarrow 0$ (the bottom of the parabolic band
without pairing). This population inversion has detectable consequences for far-from-equilibrium phase
II observables, such as rf spectroscopy or time-resolved ARPES \cite{Foster2013,Liao2015}. 
The takeaway is that, in a far-from equilibrium situation, a topological change ($W \neq Q$) can be encoded
in real \emph{bulk} observables, due to the induced quasiparticle population inversion \cite{Foster2013,Gritsev2019}.

Finally, we note that the quenches studied of $p+ip$ superfluids in Refs.~\cite{Foster2013,Foster2014,Liao2015}
always assumed an initial nonzero $p+ip$ order ($\Delta_i \neq 0$ in Fig.~\ref{Fig--PwavePD}).
As discussed in Sec.~\ref{sec:flucts}, a quench from the normal state requires an analysis of thermal and or 
quantum fluctuations. However, even in the superconducting phase it is possible to have an initial
combination of both $p+ip$ (``$\Delta_+$'') and $p-ip$ (``$\Delta_-$'') orders, given that the natural $p$-wave pairing interaction
is time-reversal invariant. A very recent study \cite{Yuzbashyan2021} demonstrates that 
the coherent, topological phase III is replaced by a \emph{chaotic} non-Floquet phase III',
for a quench in such a $p$-wave system from any simultaneously initial nonzero combination of $\{\Delta_+,\Delta_-\}$. 
On the other hand, the topological Floquet phase III is recovered for arbitrarily weak 
time-reversal symmetry breaking in the Hamiltonian, which favors (e.g.) the development (suppression) of $p+ip$
($p-ip$) order \cite{Yuzbashyan2021}.

\section{Dynamical phase transitions \\ in infinite dimensions}
\label{sec:dmft}

For systems which follow a collisionless time evolution, non-thermal dynamical phases can be distinguished in terms of the asymptotic behavior at long times. Beyond this paradigm, a universal understanding of DPTs has yet to emerge. A generic interacting quantum system is expected to thermalize \cite{Polkovnikov2011}, in which case DPTs, such as between  phases I to III in the  collective spin models of Sec.~\ref{sec:RGviaLax}, turn into a crossover within the pre-thermal regime. To progress in an understanding of DPTs in non-integrable systems, we can therefore put forward the following questions: (i) How sharp are crossover phenomena which derive from exact mean-field DPTs under experimentally accessible conditions? Are they well-defined even in a realistic solid state setting, and can they be engineered on quantum simulation platforms? (ii) Can one identify different dynamical transitions which are entirely characterized  in terms of the short-time dynamics? (iii) Can thermalization be inhibited even in non-integrable systems, so that sharp dynamical transitions exist in the long time behavior? 

A  challenge in investigating these questions is that the time evolution in non-integrable quantum systems is typically  exponentially hard to compute.  One limit in which many-body systems can be studied in a numerically controlled way is the limit of infinite dimensions, or infinite lattice connectivity  $Z\to\infty$ \cite{Metzner1998}. In contrast to spin models  for which  mean-field theory becomes exact in this limit (cf. DPTs in Sec.~\ref{sec:lmg} and~\ref{sec:LaxMethod}), the dynamics of fermionic or bosonic lattice models generally remains non-integrable and ergodic. The limit of infinite dimensions therefore provides a useful setting to investigate the fate of DPTs when the dynamics is no longer collisionless. In the following section we will  briefly comment on the solution of lattice models,  in particular the Hubbard model, in this limit, and then review two types of DPTs:  (i) A  crossover in the pre-thermal evolution of symmetry-broken states  (Sec.~\ref{MESymmetrybreaking}), which is closely related to the mean-field DPTs discussed in the first part of this review, and (ii) DPTs in the short-time evolution (Sec.~\ref{MEQuenchHM}) which either separate distinct pre-thermal regimes or are related to Bloch oscillations in a time-independent potential gradient.

As a remark we emphasize that  the study of non-integrable quantum dynamics is a very active field of research also in the opposite numerically accessible limit, i.e., one-dimensional and finite systems. In particular, many-body localized phases in disordered systems show non-ergodic behavior \cite{Alet2018, Nandkishore2015, Abanin2019}, and the separation between non-ergodic many-body localized phases and thermal phases \cite{Parameswaran2017} can give rise to dynamical phase transitions. Weak ergodicity breaking  with unusually long relaxation times has also been observed in  translationally invariant systems due to dynamical bottlenecks and constraints, somewhat analogous to arrest in glassy dynamics \cite{Carleo2012, Smith2017a, Smith2017b,Yao2016, Michailidis2018, Lan2018, Horssen2015,Scherg2020}, and many-body dynamics can remain constrained to atypical eigenstates \cite{Iadecola2019, Vafek2017}, or quantum many-body scars \cite{Turner2018, Choi2019, Ho2019}. These settings are naturally interesting for a study of DPTs, but a detailed review of this field of research is beyond the scope of the present work (cf. discussion in Sec.~\ref{sec:exp}).

\subsection{The Hubbard model in infinite dimensions}
\label{MEDinf} 

A paradigmatic model for strongly interacting fermions is the Hubbard model, given by the Hamiltonian
\begin{align}
H_{\rm Hub}=-J\sum_{\langle i,j\rangle,\sigma} c_{i\sigma}^\dagger c_{j,\sigma}
+ U \sum_{j} n_{j\uparrow}  n_{j\downarrow}.
\end{align}
Here $c_{j,\sigma}$ ($c_{j,\sigma}^\dagger$) is the annihilation (creation) operator for a fermion with spin $\sigma\in\,\uparrow,\downarrow$ on a lattice site $j$, and $n_{j\sigma}=c_{j\sigma}^\dagger c_{j\sigma}$ is the onsite number operator. The model describes tunnelling of particles between neighbouring sites $\langle i,j\rangle$ on a lattice, with a local interaction $U$.  Depending on interaction and filling, the Hubbard model gives rise to Mott-insulating, magnetically-ordered, and superconducting phases in equilibrium, making it a suitable platform to explore DPTs. A nontrivial solvable limit of the Hubbard model is that of infinite lattice connectivity $Z\to\infty$. For example, the connectivity can be systematically varied on the Bethe lattice, or on the $D$-dimensional hypercubic lattice with $Z=2D$. When the tunnelling matrix elements are rescaled like $J=J_0/\sqrt{Z}$, with fixed $J_0$, interaction and kinetic energy for a system with given nonzero filling fraction (e.g.~half filling) remain of the same order, resulting in a meaningful competition of various phases \cite{Metzner1998}. At the same time, the many-body self-energy  $\Sigma(\kk,\omega)$ becomes local in space (independent of momentum $\kk$) \cite{MuellerHartmann1989}, and the model can be solved exactly within dynamical mean-field theory (DMFT) \cite{Georges1996}. Within DMFT, the local self-energy $\Sigma(\omega)$ is obtained from a quantum impurity model in which one site of the lattice is embedded in a self-consistently determined particle reservoir \cite{Georges1992}. Using Keldysh Green's functions, DMFT can be formulated to study lattice models in different non-equilibrium settings \cite{Schmidt2002, Freericks2006, Aoki2014}, including  the transient dynamics of isolated lattice systems, non-equilibrium steady states of dissipative driven systems \cite{Arrigoni2013, Joura2008, Scarlatella2021}, and periodically driven systems (Floquet DMFT) \cite{Tsuji2008}. The quantum impurity model for the Hubbard model can be solved numerically exactly for short times using real time Quantum Monte Carlo (QMC) \cite{Oka2009}, or matrix product state (MPS) simulations \cite{Wolf2014}. Long-time simulations are still exponentially hard due to the dynamical sign problem in case of QMC, and due to the unbounded growth of the entanglement typical for global quenches in many-body systems  in the  case of  MPS. Nevertheless, diagrammatic expansions on the level of the impurity model both at weak coupling \cite{Tsuji2013a} and at strong coupling \cite{Eckstein2010nca} allow for a solution at long times, and thus provide a unique possibility to achieve a non-perturbative description of the many-body dynamics in a high-dimensional system. 

\subsection{DPTs related to non-thermal symmetry breaking}
\label{MESymmetrybreaking}

The half-filled Hubbard model on a bipartite lattice supports antiferromagnetic (AFM) order in the repulsive case $U>0$, and superconductivity as well as charge density wave order in the attractive case $U<0$. In the following, we focus mainly on the repulsive model; the attractive model can be mapped onto the repulsive one by a particle hole transformation. At weak-coupling ($U\ll J_0$), the normal state is metallic and the AFM phase can be understood within mean-field theory. For $U\gg J_0$, on the other hand, the normal state is a Mott insulator, and the ordered phase is described by a low-energy Heisenberg model.  For $U<0$, the two limits correspond to BCS superconductivity at weak interactions, and a condensate of preformed pairs (BEC) in the Mott regime, respectively. The transition temperature is maximal at the crossover, where $U$ is comparable to the non-interacting bandwidth $W$ (which is proportional to the tunneling $J_0$).

The dynamics of the ordered phase after quenches of the interaction has been studied both at weak and at strong interactions. The weak-coupling limit is hereby closely linked to the mean-field models discussed in the first part of this review. To see this, one can choose a unit cell with two sites (corresponding to the two sub-lattices $A$ and $B$) and define momentum-space spinors ${\hat \psi}_{\kk\sigma} = (c_{\kk A\sigma},c_{\kk B\sigma})^T$; here $\kk$ is a quasi-momentum in the  Brillouin zone of the symmetry-broken state, which has a doubled unit cell with respect to the normal state due to the antiferromagnetic ordering. With this, one can introduce Anderson pseudo-spins $S^x_{\kk\sigma}=\frac{1}{2}\sum_\sigma  {\hat \psi}_{\kk\sigma}^\dagger \hat{\tau}_x {\hat \psi}_{\kk\sigma}$, $S^y_{\kk\sigma}=\frac{1}{2}\sum_\sigma \sigma {\hat \psi}_{\kk\sigma}^\dagger \hat{\tau}_y{\hat \psi}_{\kk\sigma}$, and $S^z_{\kk\sigma}=\frac{1}{2}\sum_\sigma\sigma  {\hat \psi}_{\kk\sigma}^\dagger \hat{\tau}_z{\hat \psi}_{\kk\sigma}$, with the Pauli matrices $\hat{\tau}_\alpha$. The Ne\'el order parameter is the sub-lattice magnetization, $m=\langle n_{A\uparrow}\rangle-\langle n_{B\uparrow}\rangle= \langle n_{B\downarrow}\rangle-\langle n_{A\downarrow}\rangle$, which becomes $m=\frac{1}{N_k}\sum_{\kk} \langle S_{\kk}^z \rangle$ in terms of the pseudo-spins. The mean-field Hamiltonian reads (up to terms which are absorbed in the chemical potential)
\begin{align}
\label{gwejke}
H_{\rm mf} = \sum_{\kk} \vec{B}_{\kk}(t)\cdot \vec{S}_{\kk},
\end{align}
with the self-consistent pseudo-magnetic field $\vec{B}_{\kk}(t) = (2\epsilon_{\kk}, 0, -Um(t))^T$; $\epsilon_{\kk}$ is the dispersion. Equation~\eqref{gwejke} is a spin model with all-to-all interaction as discussed in Sec.~\ref{sec:RGviaLax}. Upon a particle-hole transformation it is equivalent to the mean-field models studied for BCS superconductors \cite{Barankov2006}, see also Eq.~\eqref{HRed}. After an interaction quench starting from the symmetry-broken state, the order parameter $m$  will therefore either show damped collective oscillations (phase II), an exponential decay (phase I), or self-sustained oscillations (phase III). The solution of the Hubbard model beyond mean field theory allows to investigate the fate of these DPTs outside the collisionless regime. 

\begin{figure}[t]
\begin{center}
\includegraphics[width=0.99\columnwidth]{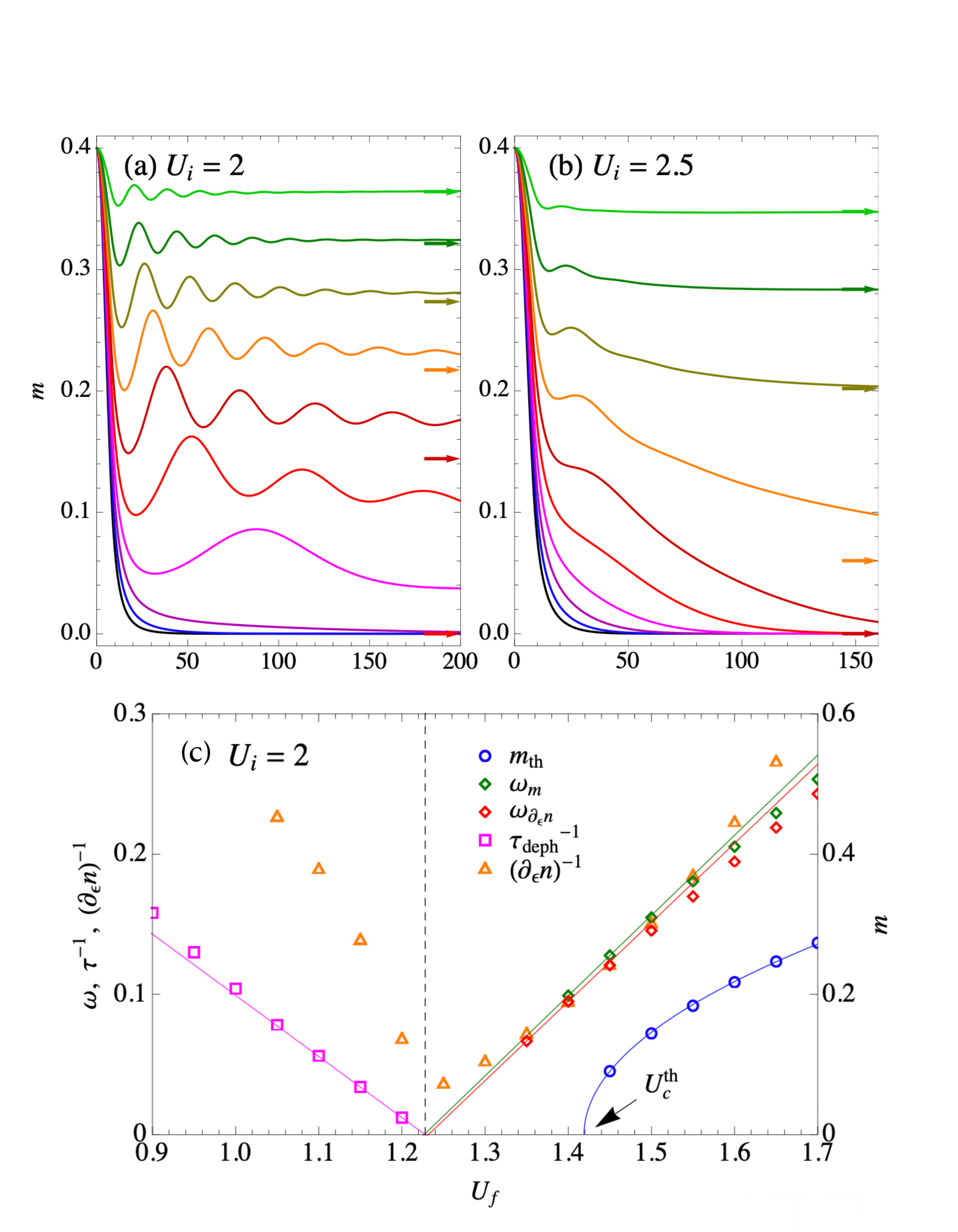}
\end{center}
\caption{
(a) Antiferromagnetic order $m$ after an interaction quench in the Hubbard model on the infinitely connected Bethe lattice. All energies (times) are measured in terms of the (inverse) tunneling $J_0$, the noninteracting bandwidth is $W=4$. Different curves correspond to different final interactions $U_{\rm f}=1.9,1.8,...,0.9$ from top to bottom. The arrow indicates the value of the order parameter after thermalization. (b) Analogous to (a), but for a larger initial value of the interaction, and $U_{\rm f}=2.4,2.3,...,1.4$. (c) Energy and time scales related to the transition (see text) for the initial interaction $U_{\rm i}=2$, as function of the quench amplitude $U_{\rm f}$. Figure adapted with permission from Ref.~\cite{Tsuji2013}.}
\label{figME01}
\end{figure}

To study quenches at weak coupling, one can prepare the system in the AFM phase at given interaction $U=U_{\rm i}$, and suddenly decrease $U$ to a smaller value $U_{\rm f}$. Figure~\ref{figME01} shows the results for a simulation on the infinitely coordinated Bethe lattice. For the  smaller value of the  initial interaction ($U_{\rm i}=2J_0$), one can clearly distinguish two dynamical regimes (Fig.~\ref{figME01}a): After weak quenches ($U_{\rm f}>1.2J_0$), the order parameter shows a damped oscillatory decay towards a finite value at the longest simulation times (phase II), while for larger quenches, it follows a monotonous exponential decay (phase I). Similar dynamical transitions related to symmetry broken states in infinite dimensional systems are found in a time-dependent Gutzwiller solution of the Hubbard model \cite{Sandri2013}, and for quenches in an O($\mathcal{N}$) model including fluctuations beyond mean field \cite{Sciolla2013}, though both approaches cannot describe thermalization. In the DMFT solution, interactions  lead to thermalization, and for quenches close to the dynamical critical point $U_c$ between the two regimes, the order parameter will eventually decay to zero both in phase I and in phase II. However, for sufficiently weak interactions the crossover between the two phases remains remarkably well-defined, and  one can even extract the critical behavior close to $U=U_c$: The decay time $\tau_{\rm def}$ of the order parameter in phase I diverges like $\tau_{\rm def}\sim |U-U_c|^{-1}$ and the oscillation frequency $\omega_{\rm m}$ in phase II vanishes like $\omega_{\rm m}\sim |U-U_c|$. Even for interactions $U$ which are half the bandwidth $W$, this critical behavior can be observed over an order of magnitude in time (Fig.~\ref{figME01}c). 

Because of thermalization, the dynamical crossover becomes less well-defined at larger interactions (Fig.~\ref{figME01}b).  A relevant question is therefore whether one can experimentally distinguish the decay of the order parameter related to phase I from thermalization, although both eventually lead to a vanishing of the order parameter. An indirect evidence would be a crossover in the relaxation time as the system proceeds from the dephasing dynamics to thermalization. More interesting would be a direct measure of the individual momentum-resolved pseudo-spins $S_{\kk}$: In the dephasing scenario (phase I), the magnitude of  $\langle \vec{S}_{\kk}\rangle$ remains nonzero for all $\kk$, but individual spins oscillate out of phase. In the thermal evolution, in contrast, the magnitude of $\langle \vec{S}_{\kk}\rangle$ for momenta at the Fermi surface decays to zero. Information on the individual pseudo-spins may be obtained by measuring correlations in the momentum occupations $\langle n_{\kk\sigma} n_{\kk+{\bm Q},\sigma'}\rangle-\langle n_{\kk\sigma} \rangle\langle n_{\kk+{\bm Q},\sigma'}\rangle$ between momenta $\kk$ and $\kk+{\bm Q}$ separated by the antiferromagnetic nesting vector ${\bm Q}$. For cold atoms, such quantities could be addressed using noise correlations in time of flight measurements \cite{Altman2004,Folling2005,Rom2006}, while in the solid-state setting intensity correlations in time-resolved photoemission spectroscopy may provide similar access to such higher order correlations in the momentum distribution and 
thus distinguish the phase I dephasing from thermalization \cite{Stahl2019}.

The findings for weak quenches in the Hubbard model are reminiscent of non-thermal fixed points in bosonic models \cite{Berges2008, Nowak2014,Oberthaler2018,Schmiedmayer2018,Langen2016}. Non-thermal critical behavior in this case has been linked to an emergent universal form of the momentum distribution function with a steady flow of energy between short and long length scales. To what extent DPTs between phases I and II in mean-field models give rise to non-thermal critical behavior when interactions beyond mean-field are taken into account  is still an open question. For the Fermi-Hubbard model, one could  look for universal power laws in the momentum-dependent spin structure factor (antiferromagnetic case) or pair distribution function (superconducting case). A close passage of the non-thermal critical point may moreover lead to a delay in thermalization. For the DMFT simulations, however,  the dynamical crossover and subsequent thermalization have so far been seen within the same simulation only for larger interactions, where  thermalization is fast, but at the same time the crossover is  already relatively broad. 

Recent simulations of the long-time dynamics using self-consistent second-order perturbation theory for the  $Z\to\infty$ Bethe lattice have focused on the thermalization of the order parameter in phase II after the decay of the collective oscillation \cite{Picano2021}. The system  remains in the non-thermal  symmetry broken state for a long period of time, with a slow decay of the order, and a gap separating a valence and conduction band. The transient state in this regime is approximately described by a quasi-steady state with separate chemical potentials in the valence and conduction band. At some point, however,  the dynamics speeds up, and the system rapidly approaches the disordered state. This highly non-monotonous decay contrasts the conventional evolution of pre-thermal states, which is expected to be  a single exponential relaxation of slow variables (almost conserved quantities) \cite{Langen2016, Moeckel2008, Stark2013, DAlessio2016, Mallayya2019}. The initial bottleneck against thermalization in the ordered phase may be linked to the gap in the electronic spectrum, so that the non-thermal symmetry-broken phase is somehow self-sustained.  Whether such self-sustained relaxation bottlenecks are robust beyond the second order perturbation theory and beyond the infinite dimensional limit is yet to be seen; if so, the prolonged pre-thermal regime would make the dynamical crossover between phase I and II more well-defined.

Next, one can ask for the existence of phase III, i.e.,  self-sustained oscillations of the order parameter, in the interacting Hubbard model. In the mean-field  model \eqref{gwejke}, phase III would require quenches to larger interaction, for which the mean-field description of the Hubbard model is probably no longer valid. It is nevertheless  interesting to see whether phase III can appear in large quenches of the Hubbard model, such as quenches from the Mott AFM to intermediate coupling, where the AFM order is strongest. Even within DMFT, there is at present  no approach which provides an unbiased solution to the long-time evolution in both the weak and strong coupling regime. However, one can obtain a numerically exact solution for short times within an MPS-based solution of the DMFT equations, when the initial state of the model is prepared as a perfect Ne\'el AFM, i.e., the mean-field state of the strong-coupling Heisenberg model \cite{Balzer2015}. In this case, no indications of phase III are found for quenches to arbitrary interactions \cite{Balzer2015}. For quenches to small $U$ one finds instead a prompt decay of the order parameter within few hopping times. The experimental realization of this scenario would be the analog of quantum simulation experiments that analyzed the decay of charge order in a Bose-Hubbard model in one dimension \cite{Trotzky2012}.

Quantum quenches in the symmetry broken states have also been performed at large $U$, within the Mott phase. In this case, the Mott gap itself provides a dynamical constraint which leads to slow thermalization of the interaction energy (double occupancy $d=\langle n_{j\uparrow}n_{j\downarrow}\rangle$) and the kinetic energy \cite{Kollath2007, Rosch2008}. Relaxation times are found to be exponentially long in the ratio $U/J_0$ \cite{Eckstein2011, Sensarma2010}, closely related to the slow decay of doublons in cold atom experiments \cite{Strohmaier2010,Morong2021}. A quenched state can therefore relax into non-thermal phases in which the density of doublons is fixed as a quasi-conserved quantity. In particular, the repulsive single-band Hubbard model favors superconductivity in the $\eta$-pairing channel \cite{Yang1989} if the doublon density is sufficiently increased with respect to the equilibrium state at the same effective temperature \cite{Rosch2008,Li2020, Murakami2021}. Such non-thermal phases in Mott insulators  have been discussed in connection to possible light-induced superconductivity \cite{Kaneko2019, Peronaci2020,Tindall2020, Li2020, Murakami2021}. Because of the exponentially slow thermalization, dynamical phases which correspond to a relaxation to different ordered or disordered non-thermal states can be well-defined, with a non-thermal critical behavior \cite{Werner2012}. However, as long as the non-thermal states are protected by a single or few quasi-conserved quantities,  dynamical transitions related to these non-thermal phases should be close to dynamical transitions related to the relaxation to different  equilibrium phases in a model which incorporates the constraints. For the Hubbard model at large $U$, where the doubly occupancy is the quasi-conserved quantity, the resulting model is a generalized $t-J$ model, which describes two species of fermions (holes and doubly occupied sites) moving on the background of quantum spins that interact via an antiferromagnetic Heisenberg exchange. The phase diagram of this model has been studied as an approximation of the non-equilibrium phase diagram of the photo-excited  Hubbard model for both the infinite-dimensional \cite{Li2020} and the one-dimensional case \cite{Murakami2021}.

\subsection{DPTs between pre-thermal phases at short-time}
\label{MEQuenchHM}

A non-perturbative solution of the short time dynamics after an interaction quench in the Hubbard model from the noninteracting Fermi sea has been obtained using DMFT and QMC \cite{Eckstein2009}. The system is initially prepared in the noninteracting Fermi sea, and at  time $t=0$, the interaction is suddenly quenched to a value $U_{\rm f}>0$. The subsequent dynamics can be monitored in terms of various observables, most importantly the momentum occupation $n_{\kk}(t) = \langle c_{\kk,\sigma}^\dagger  c_{\kk,\sigma}\rangle$ and the doubly occupancy $d(t)=\langle  n_{j\uparrow}  n_{j\downarrow} \rangle$. In the initial state, $n_{\kk}$ shows a unit size discontinuity $\Delta n_F = n_{\kk_F^-} - n_{\kk_F^+}$ across the Fermi surface; $\kk_F^\pm$ denotes a momentum infinitesimally above (below) the Fermi momentum. In thermal equilibrium, the discontinuity in the momentum distribution of a Fermi liquid exists only at zero temperature. Because a quenched system is excited with respect to the ground state, the existence of a finite jump $\Delta n_F$  indicates that the system is not yet thermalized. Moreover, one can show that the jump discontinuity is robust: While the magnitude $\Delta n_F$ can change at $t>0$, there remains an exact step-singularity in the momentum distribution, and the location of the discontinuity is precisely given by the noninteracting Fermi surface \cite{Uhrig2009}. Hence, $\Delta n_F$ can be taken as an order parameter to distinguish different dynamical regimes. 

\begin{figure}
\includegraphics[width=\columnwidth]{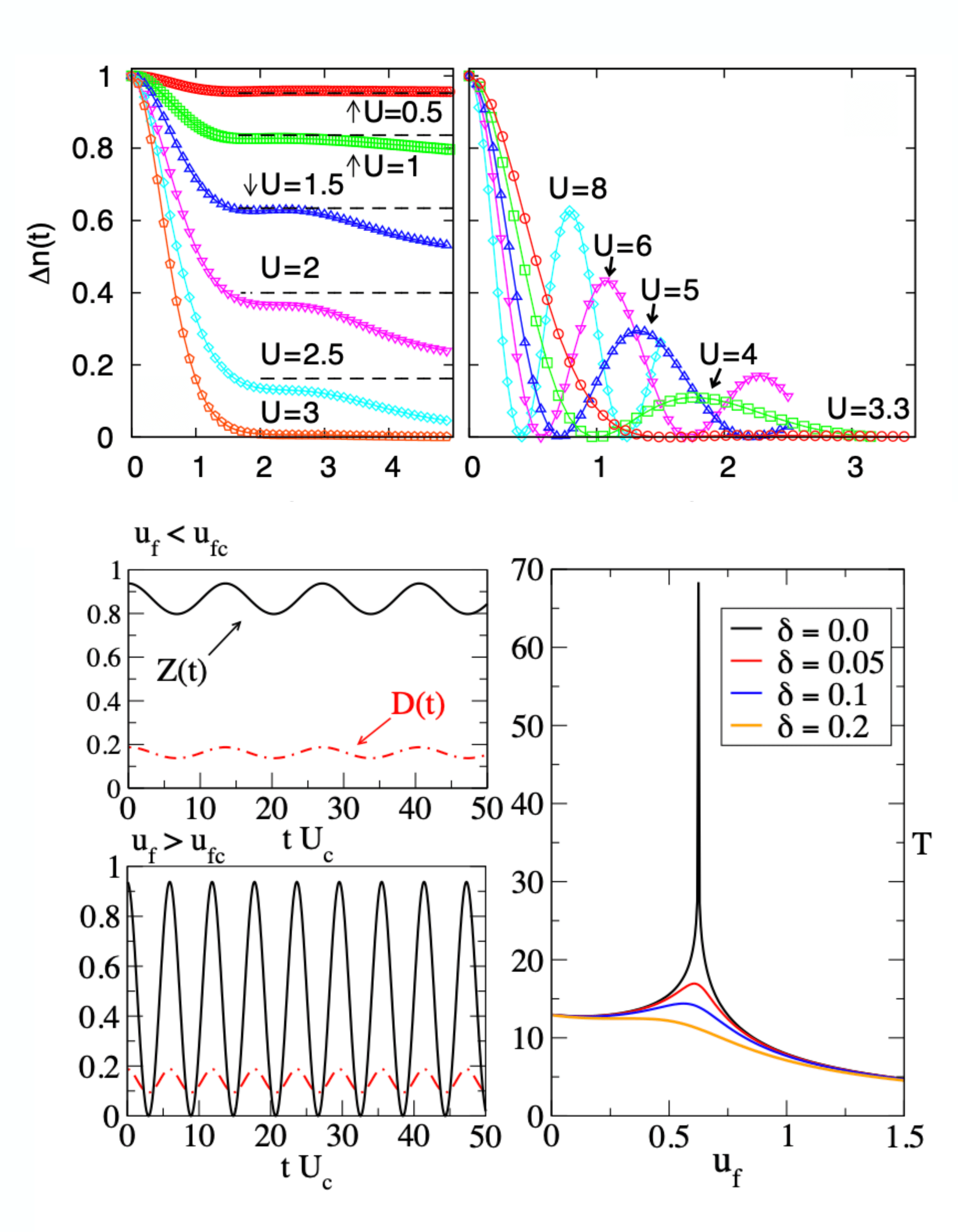}
\caption{
Upper panels: Fermi surface discontinuity $\Delta n_F (t)$ for an interaction quench in the Hubbard model on the Bethe lattice (noninteracting bandwidth $W=4$), from the noninteracting state to interactions $U$ as indicated. The left and right panel correspond to the different dynamical regimes below and above $U_{\rm fc}=3.3$. Adapted with permission from \cite{Eckstein2009}. Lower panels: Solution of the same model using the time-dependent Gutzwiller approach: Left is the quasiparticle weight ($Z$) and the double occupancy ($D$) for quenches below ($U_{\rm f}<U_{\rm fc}$) and above ($U_{\rm f}>U_{\rm fc}$) the DPT, right the period of oscillations for  different values of the doping $\delta$ away from half filling, as function of $U_{\rm f}$. A sharp DPT is observed for $\delta=0$ with a divergence of the period. Adapted with permission from \cite{Schiro2010}.
}
\label{figME02}
\end{figure}

After the quench, one observes two distinct dynamical regimes, see Fig.~\ref{figME02}: For small quenches below a value $U=U_{\rm fc}$ ($U_{\rm fc}=3.3J_0$ for the infinitely connected Bethe lattice), the Fermi surface singularity rapidly decreases to a value $\Delta n_F>0$. The subsequent slower decay towards the thermal value $\Delta n_F=0$  is not resolved within the short time simulations. For  quenches above $U_{\rm f}=U_{\rm fc}$, the dynamics features damped oscillations, with zero-crossings of $\Delta n_F$. For quenches to an interaction between the two dynamical regimes, the whole momentum distribution thermalizes rapidly. Thermalization is also verified by the fluctuation dissipation relation in  dynamical correlation functions \cite{Eckstein2010b}. 

The dynamical transition in the short time evolution is not directly related to the equilibrium Mott transition: $U_{\rm fc}$ is about a factor two smaller than the equilibrium Mott transition at temperature $T=0$, and the final state, after thermalization to a given temperature $T^*$, is in a bad metallic regime far from any known transition in the equilibrium phase diagram. Intriguingly, a singularity in the equilibrium two-particle vertex, which is related to multi-valuedness of self-consistent  perturbation theory \cite{Gunnarsson2017}, has been found right at the point $(U_{\rm cf}, T^*)$ \cite{Schaefer2013}. Whether this is coincidential, or whether such vertex singularities in general are related to DPTs is an open question.

It is useful to note that the two dynamical regimes reflect distinct pre-thermal behavior which can be understood in opposite perturbative limits (although these perturbative limits cannot give a description of the transition itself): Quenches from $U=0$ to  weak coupling have been studied using perturbative unitary transformations \cite{Moeckel2008} which are accurate for quenches $U\ll J_0$, on a timescale $t\ll J_0/U^2$. The unitary perturbation theory maps the Hubbard model on a system of non-interacting renormalized quasiparticles; the quasiparticle occupations $\tilde n_{\kk}$ are conserved, providing a constraint on the dynamics which prevents the system from reaching a thermal state. For quenches to $U\gg J_0$,  the effective model that describes the dynamics at least on timescales up to $U/J_0^2$ can be obtained by unitary perturbation theory, treating $J_0$ as a small parameter: It is the generalized $t-J$ model mentioned at the end of Sec.~\ref{MESymmetrybreaking}, where the number of doubly occupied sites and holes provides a constraint on the dynamics. The oscillations are well understood in the limit $J_0=0$, where the dynamics of the system would be $2\pi/U$-periodic, because the many-body spectrum is perfectly equidistant (analogous to the collapse and revival oscillations in the bosonic Hubbard model \cite{Greiner2002, Will2010}). 

While the two dynamical regimes can be understood in a respective perturbative limit, an analytical model of the dynamical transition itself has been observed within time-dependent Gutzwiller theory \cite{Schiro2010,Schiro2011}. The time-dependent Gutzwiller provides a variational approach to the dynamics; for the Hubbard model,  one starts from a variational ansatz wave function 
\begin{align}
|\Psi(t)\rangle = \prod_{j} e^{-iS_j(t)} P_j(t)|\Phi(t)\rangle,
\end{align}
where $|\Phi(t)\rangle$ is a time-dependent Slater determinant, $P_j(t)=\sum_{n=0,1,2} \lambda_{j,n}(t)P_{j,n}$, and $S_j(t)=\sum_{n} \phi_{j,n}(t)P_{j,n}$ are operators written in terms of variational parameters $\lambda_{j,n}(t)$ and $\phi_{j,n}(t)$ and the local projectors $P_{j,n}$ on the occupation $n=0,1,2$ on site $j$. The time-dependent variational principle $\delta \langle \Phi(t) | i\partial_t - H |\Phi(t)\rangle =0$ leads to a set of nonlinearly coupled differential equations for the variational parameters. Within the Gutzwiller approach, an exact dynamical transition is observed, which shares many features with the DMFT results (see Fig.~\ref{figME02}, lower panels): For small quenches, the quasiparticle weight $Z(t)$, which is proportional to the jump $\Delta n_F$ defined above, oscillates around a nonzero value. Its time average $\bar Z$ tends to zero as $U_{\rm f}$ approaches the dynamical transition. For quenches to the dynamical critical point, $Z(t)$ exponentially relaxes to its thermal value $Z=0$. The value $\bar Z$ follows the pre-thermal plateau obtained by DMFT with quantitative accuracy \cite{Schiro2010}. For quenches beyond $U_c$, $Z(t)$ oscillates with a zero crossing. In contrast to the full solution, the Gutzwiller approach does not describe thermalization, and the dynamical transition therefore distinguishes the dynamics at all times. 

Being computationally inexpensive, the Gutzwiller approach has allowed to study the DPT in a wider range of parameters: (i) By varying  the initial $U=U_{\rm i}$, one can map out a dynamical phase diagram which  connects the dynamical phase transition to the equilibrium Mott transition $U_c^{eq}$: For $U_{\rm i}=0$, the dynamical point $U_{\rm fc}$ is below $U_c^{eq}$, while it approaches $U_c^{eq}$ as $U_{\rm i}$ is increased \cite{Schiro2010}. (ii) Moreover, the dynamical transition persists if the interaction is ramped up with a nonzero ramp time $\tau$ instead of a quench \cite{Sandri2012, Hofmann2016}. The dynamics after the ramp still distinguishes different dynamical regimes, and the DPT evolves towards the equilibrium phase transition for larger $\tau$, as expected for an adiabatic dynamics.  (iii) When the system is doped away from half filling, where there is no Mott transition in equilibrium, the dynamical transition turns into a crossover. In particular, time averages of the double occupancy and of $Z$ evolve smoothly as a function of $U_{\rm f}$. Finally, (iv), time-dependent Gutzwiller simulations have been performed for a multi-orbital Hubbard model \cite{Behrmann2013}, which turns out to be different from single-band case due to inter-orbital fluctuations. The dynamical transition is replaced by a broad regime in which the different occupations evolve irregularly, which may be interpreted in terms of long-lived fluctuations between metallic and insulating states. 

Unfortunately, DMFT simulations have so far been performed neither away from half filling nor at nonzero $U_{\rm i}$, due to a more severe dynamical sign problem. Hence there are so far no numerically unbiased simulations to support these intriguing observations from the Gutzwiller approach, and a complete picture of the dynamical transition in the Hubbard model is yet to be developed. Another question is the extension of the transition to lower-dimensional systems. In one dimension, the dynamics has been calculated using a systematic solution of the equation of motion for the field operators $c_{i\sigma}$ in terms of higher order operator products \cite{Hamerla2013}. It is found that the non-equilibrium time-evolution after interaction quenches exhibits a similar dynamical transition as in the half-filled case, while the transition becomes a crossover upon doping. This may suggest that the dynamical transitions is  a general feature of quenches in such models \cite{Hamerla2013}. On the other hand  the prethermalization is typically less pronounced in lower dimensions \cite{Tsuji2014,Hamerla2013}.

Finally, a DPT has been found in the Bose-Hubbard model on an infinitely connected lattice \cite{Sciolla20102}, for quenches from the  superfluid phase towards to Mott phase. A solution of the Bose Hubbard within the non-equilibrium generalization of bosonic DMFT \cite{Strand2015} can be done at least for the short time evolution. In this case a single DPT is replaced by a richer dynamical phase diagram, including a non-thermal symmetry broken phase for weak quenches, an extended regime in which the system rapidly for intermediate quenches, and oscillations of the order parameter for quenches deep into the Mott phase.

Alternative to quenches of the interaction, dynamical phase transitions in  infinite-dimensional lattice models have also been observed after a sudden switch-on of a potential gradient. In the non-interacting case, this leads to perfect Bloch oscillations if the gradient is aligned with a high-symmetry direction of the lattice. In interacting models, these oscillations will be damped, as the system approaches an infinite temperature state (for ergodic systems) or a non-thermal steady state if the system does behaves ergodic. In infinite dimensions, the latter case is represented by the Falikov-Kimball model, which allows for an exact solution in DMFT \cite{Brandt1989,Freericks2003}. The Falikov-Kimball model describes a Hubbard model in which one species of fermions is frozen. Regarding the mobile species, the dynamics obtained by DMFT in the normal (non-symmetry broken) phase is identical to that of fermions with quenched binary alloy disorder. The model does not thermalize after a quench of the interaction, but instead reaches a non-equilibrium steady state at infinite time \cite{Eckstein2008}. The study of Bloch oscillations in the Hubbard and Falikov Kimball model in infinite dimensions gives  a rich dynamical phase diagram \cite{Freericks2006, Eckstein2011b, Fotso2014}, with transitions between oscillatory and non-oscillatory regimes.

%section 
 
\section{Experimental Observation }
\label{sec:exp}
This Section overviews a broad number of platforms    which are currently employed as quantum simulators for the evolution of  interacting many-particle systems where DPTs can be hosted. For continuity with the two previous theory sections, we start by overviewing the state of the art in experiments realizing dynamical phases in solid state platforms, and then move to cold atoms experiments. % some subsections to  

\subsection{DPTs in Condensed Matter Systems }\label{CMI}

The DPT in the integrable BCS model discussed in Section \ref{sec:RGviaLax} neglects the retarded character
of the dynamical pairing interaction, as typically mediated by phonons \cite{Schrieffer,Tinkham}. Nevertheless, under specific conditions  superconducting materials  could be   described by the integrable BCS model  and  display the 
three distinct dynamical phases  I, II, and III discussed  in Sec.~(\ref{sec:LaxMethod}). For that to happen, the asymptotically exact long-time results obtained from the Richardson-Gaudin model can be relevant  to experiments which  can access  a long prethermalization plateau. For a system of many particles $N \gg 1$, 
this requires that the minimum inelastic lifetime due e.g.\ to 
pair-breaking collisions, or inelastic electron-phonon scattering in a superconductor
must be much larger than $t_{\mathsf{dyn}}\simeq 1/\Delta_\infty.$ Here $t_{\mathsf{dyn}}$ is 
the timescale for transitioning to the asymptotic regime  and $\Delta_{\infty}$ is the steady-state value of the order parameter in the prethermalization plateau (phase II) \cite{Yuzbashyan2015}.

In the case of two-particle collisions for a quench entirely confined to the weak-pairing BCS regime,
the two-particle scattering time can be estimated using Fermi liquid theory \cite{Galaiko1972,Yuzbashyan2015}
$	t_{\mathsf{in}} \sim \left({\e_F}/{\Delta_\infty}\right) t_{\mathsf{dyn}} \gg t_{\mathsf{dyn}} ,$
where $\e_F$ is the ground-state Fermi energy. This is consistent with simulations for the Hubbard model in infinite dimensions (Sec. \ref{MESymmetrybreaking}), which 
find a well-defined separation of phase I and phase II even  when inelastic collisions are taken into 
account.

To observe phase III, the period $\mathcal{T}$ of the quench-induced Floquet oscillation should be much shorter than 
the minimum inelastic lifetime. For a quench from an initial BCS state with small (but nonzero) $\Delta_i$,
the Floquet period is of order 
$
	\mathcal{T} 
	= 
	\frac{2}{\Delta_f} 
	\ln\left(\frac{\Delta_f}{\Delta_i}\right) 
	\ll 
	\tau_{\mathsf{in}},
$
where now $\tau_{\mathsf{in}} \sim \left({\e_F}/{\Delta_f}^2\right)$,
and $\Delta_f$ denotes the ground-state pairing gap for the post-quench system \cite{Barankov2004,Barankov2006,Foster2014}.

In addition to the above requirement,  the quench that brings the superconductor out-of-equilibrium must occur on time-scales shorter than the inverse of the quasi-particle gap,
 to ensure the quasiparticle distribution does not adiabatically follow the variation of the system parameters.  At the same time the perturbation has to be  weak enough to avoid disrupting the system.  The latter  has been  the most important limitation.  
Near-visible femtosecond optical pulses, which are used as a non-adiabatic excitation of solid state systems are not compliant since excitations  with a frequency higher than the  BCS gap can break Cooper pairs into hot quasiparticles, and these can serve as an
efficient mechanism for rapid dissipation and thermalization.  Nevertheless  recent
developments in  THz technology are allowing now  the injection  of near monocycle pulses with center frequency close to the BCS gap, opening a window  to investigate  the coherent transient dynamics of superconductors in the nonadiabatic excitation regime over a window of
about 10 picoseconds (ps), well before thermalization occurs on a time scale of 100 ps.

\begin{figure}[t!]
\centering
\includegraphics[width=0.4\textwidth]{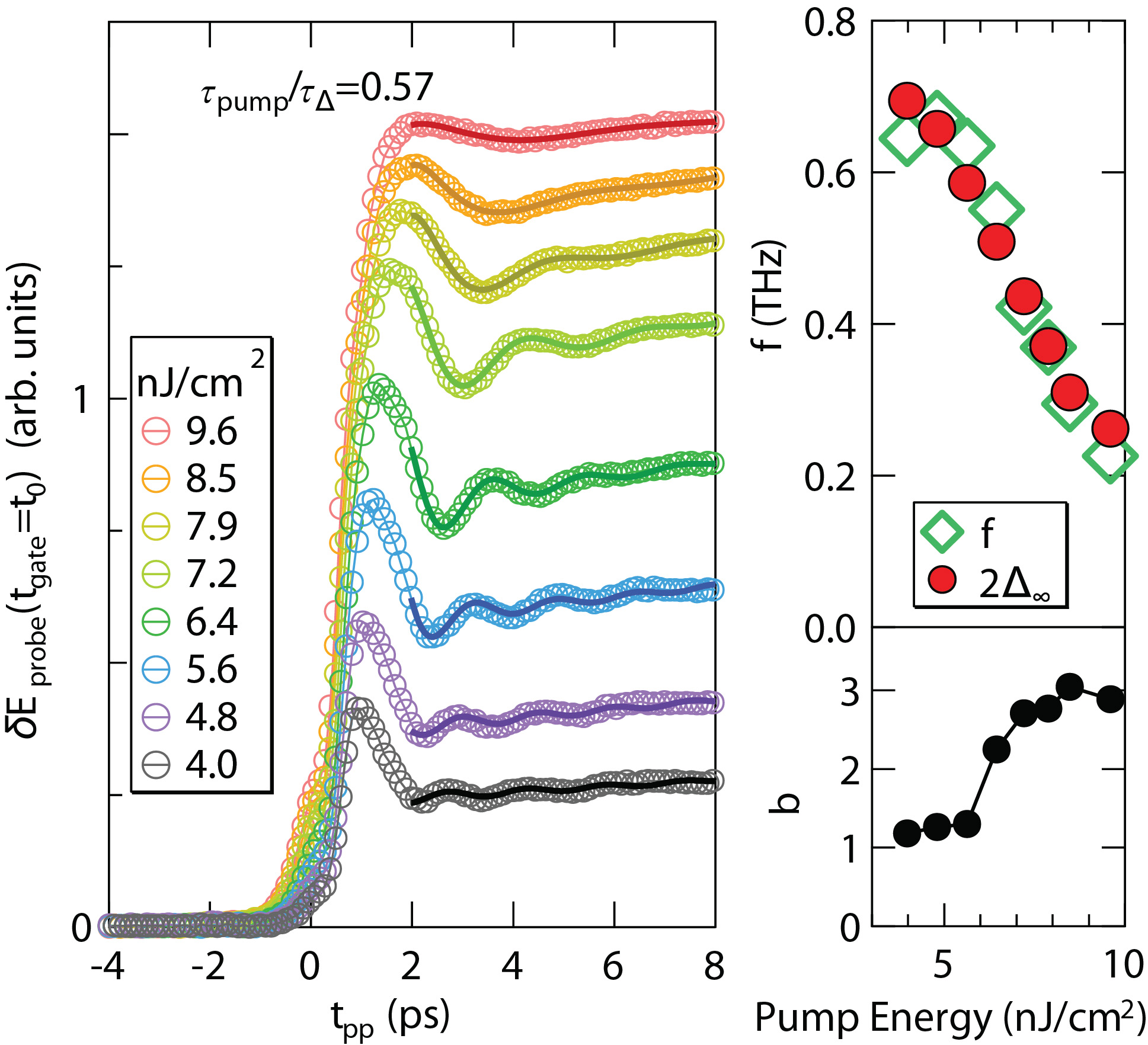}
\caption{A possible experimental realization of a phase II quench in a solid-state superconductor by Matsunaga et al.,
from Ref.~\cite{Matsunaga2013}, see also \cite{Shimano2019}.
This figure shows pump-probe THz spectroscopy data on thin film  Nb${}_{1-x}$Ti${}_x$N.
The left panel shows the probe signal oscillations of $\Delta(t)$, and 
implies the approach to a nonequilibrium value $\Delta_\infty < \Delta_i$
(weak phase II quench). The approach of $\Delta(t) \rightarrow \Delta_\infty$ is characterized
by decaying oscillations with frequency $2 \Delta_\infty$ (top right), consistent
with the theory [Eq.~(\ref{PhaseIIApproach})] \cite{VolkovKogan1973,Yuz2}. 
}
\label{Fig--Shimano}
\end{figure}

In particular the experiment done  by Matsunaga et al  injected an intense, short-duration THz pulse
with center frequency $\omega \simeq \Delta_0$ into a low-temperature   NbN thin film $s$-wave superconductor \cite{Matsunaga2013,Shimano2019}. 
Since most of the spectral weight of the pulse was below the optical gap edge $2 \Delta_0$, 
a weak pulse would not be expected to couple to the system \cite{Tinkham}. However, because of the   strong character of the injected  pulse, it coupled in a nonlinearly fashion and was able to  excite the ``Higgs'' (amplitude) mode of the superconductor. After  the application of the pulse,
the subsequent free evolution of $\Delta(t)$ was expected to   be described by the $s$-wave Richardson-Gaudin model.
The experiment \cite{Matsunaga2013} measured the system over a window  long enough
to observe decaying oscillations as predicted by the Lax method in Eq.~(\ref{PhaseIIApproach}), as shown in Fig.~\ref{Fig--Shimano}. Despite the different method of initial
state preparation (THz pulse versus instantaneous interaction quench), the phase II dynamics was
apparently observed.

A THz electric field pulse can be incorporated into the self-consistent solution for the BCS (Bogoliubov-de Gennes) equations.
In order to induce a nontrivial evolution of the system, it is necessary to either account for the finite (but very small) photon
momentum \cite{Papenkort2007,Papenkort2008,KrullSchnyder2014,Papenkort2009}, or to include non-quadratic corrections to the band 
dispersion \cite{Chou2017}. 
%The theoretical results depicted in Figs.~\ref{Fig--THz_Texture_Roots} and \ref{Fig--THz_Phase_Diagram} were obtained by a hybrid 
%numerical-Lax vector approach, incorporating the field in the temporal gauge with the $s$-wave Hamiltonian
%\begin{align}\label{HRedTHz}
%	H = \sum_{\vex{k}} 2 \e_{\vex{k} + \frac{e}{c}\vex{A}(t)} \, s^z_{\vex{k}} - G \sum_{\vex{k},\vex{k'}} s_{\vex{k}}^+ s_{\vex{k'}}^-,
%\end{align}
%where $\e_{\vex{k}}$ is the dispersion for nearest-neighbor hopping on a 2D square lattice, 
%and the electric field is $\vex{E}(t) = - (1/c) \, d \vex{A}(t) / d t$. 
Moreover, in  order to model the experiment in Ref.~\cite{Matsunaga2013}, a short-duration Gaussian electric pulse $\vex{E}(t)$, 
with frequency content mostly below the equilibrium optical gap $2\Delta_0$, 
needs to be  fed into the equations of motion for the Anderson pseudospins. The field twists the pseudospins in the $x$-$y$ plane, 
scrambling the ground state order 
as shown in Fig.~\ref{Fig--THz_Texture_Roots}. 
Importantly, the driving of the Higgs mode arises from nonlinear coupling to the field \cite{Chou2017,Shimano2019}. %expanding 
%$\e_{\vex{k} + \frac{e}{c}\vex{A}(t)}$ in powers of $\vex{A}(t)$, the linear coupling cancels, and the nontrivial
%drive arises from the $A^i A^j \partial_{k_i} \partial_{k_j} \e_{\vex{k}}$ term .
After the cessation of the pulse, the system  is expected to evolve freely according to the Richardson-Gaudin Hamiltonian. 
%in Eq.~(\ref{HRedTHz})
%(with $\vex{A} = 0$). The phase diagram in Fig.~\ref{Fig--THz_Phase_Diagram} was computed by evaluating the isolated roots of the Lax norm 
%(see Sec.~\ref{sec:LaxMethod}), and also computing the long-time spin dynamics numerically. The two methods gave identical results. 

The Richardson-Gaudin calculation \cite{Chou2017} predicts a transition to the gapless phase I for sufficiently strong pump energy, as shown in Fig.~\ref{Fig--THz_Phase_Diagram}. Indeed a putative gapless phase [with suppressed reactive component of the optical conductivity $\sigma_2(\omega)$]
was observed for sufficiently strong pump powers in Ref.~\cite{Wang2018}, potentially consistent with phase I. 
Strikingly, this experiment also observed extremely slow relaxational dynamics following the quench, on the order of 1000 ps. 
A subsequent experiment utilizing asymmetric, multicycle pulses exhibited an intervening ``gapless'' phase with $\Delta_\infty \neq 0$ \cite{Wang2019-A}.

A discrepancy between the collisionless phase II prediction in Eq.~(\ref{PhaseIIApproach}) and the data in Ref.~\cite{Matsunaga2013}
is the more rapid decay of the oscillating envelope than $t^{-1/2}$. This is also consistent with the simulations for the interacting Hubbard model in Sec.~\ref{MEDinf}. A better theoretical fit can be obtained by employing a phenomenological ``$T_2$'' 
dephasing time \cite{Orth2019}.   Another  key question is whether spatial fluctuations (``Cooper pair turbulence'' \cite{YuzbashyanDzero2009}) play an important
role in these solid-state experiments \cite{Matsunaga2013,Shimano2019,Wang2018,Wang2019-A}. See discussion in Sec. \ref{sec:flucts}.

%The Richardson-Gaudin calculation \cite{Chou2017} predicts a transition to the gapless phase I for %sufficiently strong pump energy $\propto A^2$,
%as shown in Fig.~\ref{Fig--THz_Phase_Diagram}. A putative gapless phase [with suppressed reactive %component of the optical conductivity $\sigma_2(\omega)$]
%was observed for sufficiently strong pump powers in Ref.~\cite{Wang2018}, potentially consistent with %phase I. 
%Strikingly, this experiment also observed extremely slow relaxational dynamics following the quench, %on the order of 1000 ps. 
%A subsequent experiment utilizing asymmetric, multicycle pulses exhibited an intervening ``gapless'' %phase with $\Delta_\infty \neq 0$ \cite{Wang2019-A}.
%Nonzero superfluid stiffness was inferred from a minimal change in $\sigma_2(\omega)$, whilst 
%the filling of the usual spectral depletion below $2 \Delta$ in the dissipative optical conductivity %component $\sigma_1(\omega)$ was interpreted as closing the quasiparticle gap.
%This gapless ``coherent'' phase (with nonzero $\Delta_\infty$) was argued to arise from Cooper pair %center-of-mass acceleration
%during the pulse application, an effect that is neglected in the pseudospin evolution %\cite{Wang2019-A,Perakis2020}. 
%A key question is whether spatial fluctuations (``Cooper pair turbulence'' \cite{YuzbashyanDzero2009}) play an important
%role in these solid-state experiments \cite{Matsunaga2013,Shimano2019,Wang2018,Wang2019-A}, 
%since the sample size needed for THz pump-probe is typically much larger than the coherence length $\xi$.
 
We finish this section by  noting that  we  have not survey here work involving the steady-state driving solid-state materials \cite{Oka2019}.
In particular, we do not review multipulse THz driving \cite{Orth2019} or continuous-wave mid-infrared laser excitation of superconductors. 
The latter has attracted significant 
interest as a possible mechanism for enhancing pairing or $T_c$ far from equilibrium
\cite{Cavelleri2011,Cavelleri2014,Cavelleri2016,Knap2016,Millis2016,Aleiner2018,Mitra2017,Mitra2018A,Mitra2018B,Mitra2019}. 
Continuous driving has also been discussed in the context of third-harmonic generation via the Higgs mode 
\cite{Matsunaga1145,Aoki2015,Cea2016,Shimano2019}. Although these are certainly interesting developments, our
focus here is on the collisionless dynamics of an effectively isolated system, free from external driving (following the quench).

\subsection{DPTs in Cold Atom Experiments }

\begin{figure*}[t]
    \includegraphics[width=2\columnwidth]{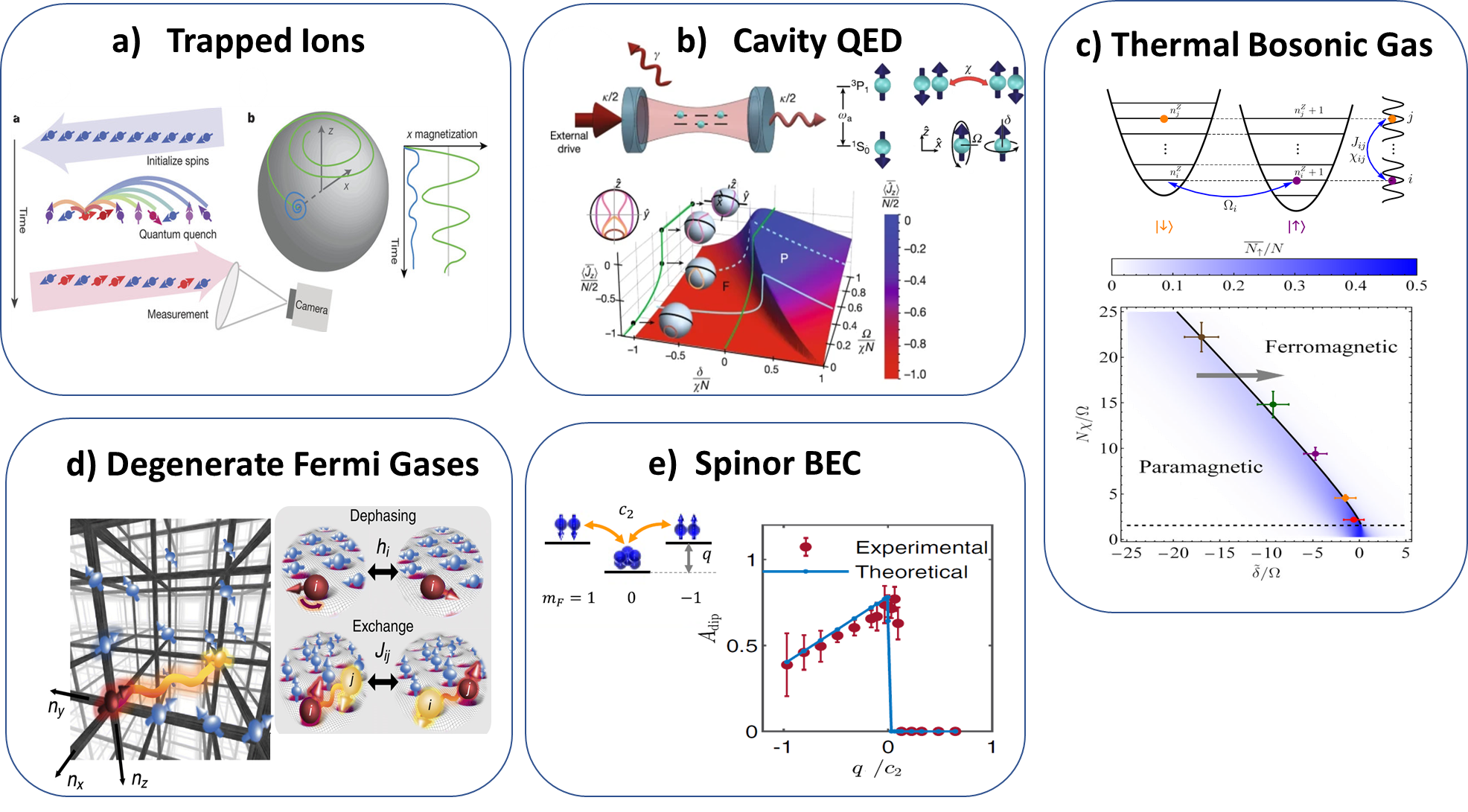}
    \caption{Dynamical phase transitions (DPTs) observed in ultra-cold atomic gases: a) The first DPT was observed in a trapped ion quantum simulator of the long range Ising model plus transverse field, with up to 53 ${}^{171}$Yb$^+$ ions \cite{Zhang2017}. b) A similar type of DPT, but  in the collective Ising limit plus an additional  longitudinal fields (so called LMG model), was also  observed in a cavity QED simulator with  with $N \approx 10^5$-$10^6$  ${}^{88}$Sr atoms \cite{Muniz2020}, followed up by the observation of a similar DPT in a thermal gas of $N=10^4-10^5$  bosonic  $^{87}$Rb atoms atoms using a sideband transition \cite{chu2020}. d) Using a quantum degenerate trapped Fermi gas of $N=10^4$ ${}^{40}$K  atoms the DPT between phase I and phase II predicted to exist in  BCS superconductors was observed in Ref.~\cite{smale2019}. e) Using three internal levels of ${}^{23}$Na atoms prepared   in a Bose Eisntein Condensate (BEC) with  $N\sim 10^5$ atoms  a DPT was observed  as the ratio between collective exchange interactions and an effective quadratic magnetic field was tuned from positive to negative \cite{yang2019,Tian2020}.   \label{Figexp} }
\end{figure*}

{\it \bf  DPT in the long-range Ising model.  } The first observation and quantitative characterization  of a DPT was done  in a trapped ion quantum simulator  consisting a chain of up to   $N=53$,  ${}^{171}$Yb${}^+$ ions  trapped  in   a linear radio-frequency Paul trap \cite{Zhang2017} (See Fig. \ref{Figexp}(a)). Two relevant   hyperfine levels were used  as a spin 1/2 degree of freedom. Note here we are not accounting for  experimental observations of DPTs described in terms of non-analytic behaviors of the Loschmidt echo amplitude which were observed at a similar time \cite{Flaschner2018,Jurcevic2017}. A discussion of these type of experiments can be found in Ref.~\cite{Heyl2019}. 

In Ref.~\cite{Zhang2017}  pairs of laser beams  were used to generate an optical dipole force that off resonantly  excited vibrational modes of the ion chain. The virtually excited phonon modes in turn   mediate tunable spin-spin interactions which lead to  long-range Ising couplings in the form of Eq.(\ref{eq:modellmg}). Explicitly, the engineered Hamiltonianis given by 
\begin{equation}
\frac{1}{2}\sum_{i>j} \chi_{i j} \hat{ \sigma}^x_i \hat{ \sigma}^x_j+\frac{B}{2} \sum_i\hat{ \sigma}^z_i ,
\end{equation} with coupling constants that fall off as as   $\chi_{i j} \approx J_0/ | i-j|^\alpha$, i.e, approximately algebraically with the distance between ions in the chain. The power-law exponent $\alpha$ was set to be  between 0.8 and 1.0 in the experiment.  Here $\hat{ \sigma}^{x,y,z}_i $ are Pauli matrices acting on the  $i$th ion.  The  competing transverse field proportional to  $ B$ was generated in the experiment by a controllable Stark shift on the spins  from the same laser field that generated the optical dipole force. The transverse field  $B$ was used  as the control parameter for crossing the DPT. At time $t=0$,  the system was initialized in the state with all the spins pointing along the $x$ direction of the Bloch sphere and the system was let to evolve under the combined Ising plus transverse field Hamiltonian for some time $t$ after which the collective magnetization 
\begin{equation}
\braket{ {\hat\sigma}^x(t)}=\frac{1}{ N}\sum_i \braket{ {\hat \sigma}^x_i (t)}, 
\end{equation}   and the corresponding time averaged were measured 
\begin{equation}
\overline{\braket{ {\hat\sigma}^x}}(T)=\frac{1}{T }\int_0^T\braket{ {\hat\sigma}^x(t)} dt,
\end{equation} equivalent to (Eq. \ref{eq:neqop}). 
At the mean field level the  time average magnetization  $\overline{\braket{ {\hat\sigma}^x}} $  is expected  to change, as discussed in Sec.  \ref{sec:lmg},  from a finite value to zero as the system crosses between  a dynamical  ferromagnetic phase  ($B< B^c$) to  a dynamical paramagnetic phase  ($B> B^c$). Nevertheless, as explained in Sec.\ref{longrange} the quantum  nature of the model resulted in a direct  decay of the order parameter towards a vanishing or non-vanishing expectation value depending on the dynamical phase considered, without the need of  long-time  averages  to  classify  phases.

To obtain further signatures of the DPT the experiment measured the  spatially averaged two-spin correlator
\begin{equation}
C_2 (t)=\frac{1}{N^2}\sum_{i,j}\braket{ {\hat \sigma}^x_i (t)  {\hat \sigma}^x_j (t)} 
\end {equation} This quantity was used  as a second   order parameters for the DPT.
It  should cross  from 1 (at small  $B$) to 1/2 (at large  $B$) with a dip at  $B^c$. However, given  the logarithmic scaling of  $C_2$   with $N$  at the critical point  and the limited  system size of the systems  under consideration,  the experiment  did not observe  sharp signatures in  $C_2$ at the   critical point. Nevertheless, measurements of the  distribution of domain sizes in the chain  (a  direct measurement of  higher-order correlations)   accessible in  the experiment allowed it to observe a sharp change of behavior at the critical point. \\

{\it \bf  DPT in  the Lipkin–Meshkov–Glick model.}
After this work a similar type of DPT was  observed in two different platforms, more specifically the one in the so called  Lipkin–Meshkov–Glick (LMG) model (Eq. \eqref{eq:MFH} in Sec.~\ref{sec:lmg}), which corresponds the  $\alpha=0$ limit of the system described above with an additional  longitudinal field, i.e 
\beq \hat{H} = \chi \hat{S}^+\hat{S}^- + B \hat{S}^x  - \delta \hat{S}^z \eeq . 
Here we  have introduced the collective spin operators $\hat{S}^{\alpha} = \sum_j \hat{\sigma}^{\alpha}_j/2$  and $\hat{S}^{\pm} = \hat{S}^x \pm i\hat{S}^y$. The summation runs over the individual spins $j=1,...,N$ and the  parameter $\chi$ sets the strength of an  infinite-range exchange interaction. To connect with the Hamiltonian,  Eq. \eqref{eq:MFH},  note that up to $1/N$ corrections  $\chi\hat{S}^+\hat{S}^- \approx \chi ({\boldsymbol{\hat S}}\cdot {\boldsymbol{ \hat S}} -(\hat{S}^z)^2 )$.  Since the first term ${\boldsymbol{\hat S}}\cdot {\boldsymbol{ \hat S}}$ is a constant when restricted to the fully symmetric spin manifold, which is the case of interest, the  Hamiltonian simplifies to $ \hat{H} \to - \chi \hat{S}_z^2 + B \hat{S}_x  - \delta \hat{S}_z $, which 
up to an overall  $\pi/2$ rotation along the $y$ axis of the Bloch sphere, coincides with the one realized in Ref.~\cite{Zhang2017} with an extra term proportional to $\delta$ which sets the longitudinal field (See Fig.~\ref{Figexp}(b)).

 The observation of  a DPT in the  Lipkin–Meshkov–Glick model was  first achieved in  a  cavity-QED simulator using   ensembles of $N \approx 10^5$-$10^6$   atoms \cite{Muniz2020}. We note that in the context of cavity QED great deal of  experimental progress has been done in the observation of  non-equilibrium phases characterized  by different steady states which therefore depend on  parameters such as pump or loss rates   but    are independent  of the initial conditions \cite{Landig2016,Landin2018,Kroeze2018,Kroeze2019,Baden2014,Baumann_2010, Ritsch_2013,Hemmerich_2015}. We will not further discuss these experiments  in the
following, as in this review we exclusively focus on
DPTs related to the unitary time evolution.

 In Ref.~\cite{Muniz2020}  the  internal spin degree of freedom was encoded in a long lived optical transition   (linewidth $\gamma/2\pi = 7.5$~kHz) between 
 the $\ket{\downarrow}$ [$^1$S$_0$ ($m_J=0$)] and $\ket{\uparrow}$ [$^3$P$_1$ ($m_J=0$)] states of $^{88}$Sr atoms.   The atoms were confined in a 1D optical lattice  and coupled to a single common  mode of the optical cavity far    detuned  from the atomic transition.  In this regime the photons can be adibatically eliminated. As a result their role is reduced  to mediate infinite range  elastic  spin exchange interactions between the atoms, with strength set by  $\chi$. The transverse field  was  engineered by pumping the cavity  with a  laser that  generated, at resonance,  Rabi flopping at frequency $B $. The longitudinal field was simultaneously  introduced by  detuning  the pump laser by a frequency $\delta$ from the atomic transition.

 For this system  the time-averaged collective magnetization along $\hat{z}$ in this case   $\overline{\langle \hat{S}^z \rangle} \equiv \lim_{T\to\infty} (1/T)\int_0^T \langle \hat{S}^z(t)\rangle dt$ serves as an ideal  order parameter as discussed in Sec.~\ref{sec:lmg}. This is equivalent to Eq.~(\ref{eq:neqop}) up to an overall unitary  rotation  that maps $S^x\leftrightarrow S^z$. However, in  the experiment additional  inhomogeneities  and other technical imperfections damped the oscillation in the paramagnetic phase. Under these conditions  the magnetization  $\langle \hat{S}^z \rangle$ after $4 \mu$s of  time evolution served as a proxy of the long-term time-averaged magnetization.  The experiment observed the DPT in  $\langle \hat{S}^z \rangle$ at $\delta=0$  using $B$ as the control parameter and   demonstrated the  expected  scaling with atom number.
The DPT was also  probed by varying the  longitudinal field for a fixed value of the drive strength $B$  and by observing   a sharp change of behavior of the order parameter  at the critical point. The robustness of the DPT was demonstrated by the symmetric response of the magnetization for  interaction shift $\chi N \leftrightarrow -\chi N$. The experiment also  explored the DPT as a function of the initial state. It tracked  the variation of the critical point as the system was initialized  in   a coherent spin state pointing along  different directions.

The same LMG Hamiltonian was implemented in an experiment operating with a  thermal gas of $N=10^4-10^5$ bosonic  $^{87}$Rb atoms  in a 3D harmonic trap using  two  hyperfine states  to set a spin-1/2 degree of freedom \cite{chu2020}.  The two internal states were coupled by laser fields tuned to one of the blue side transitions, i.e the laser not only generated a spin flip but at the same time increased one of the motional quantum numbers in the trap  by one unit. As a consequence the two coupled states had different motional eigenstates. 
Even though the interactions in this system are  contact interactions  during s-wave collisions, effective long range interactions can   due to  the delocalized nature of the single particle orbitals  when operating in the collisionless regime. This is  achieved when  the trapping potential is much larger than the interaction strength \cite{rey2014,Fuchs2002}. In this regime   it can be  assumed, to an excellent approximation, that each atom is fixed in a mode-space lattice with sites set by  the single-particle eigenstates of the 3D harmonic trap. The only relevant process between two colliding atoms is to either remain in the same internal states or to exchange them.  Even when dealing with bosonic  samples  restricting the Hilbert space to include only empty or singly-occupied lattice sites is appropriate when  the gas temperature  is  above quantum degeneracy. Under these conditions the contact interaction term  can be mapped to  a spin-$1/2$ long-range XXZ spin model where the indices $i,j$ run over the occupied  mode-space lattice sites,   \beq  H_{\mathrm{int}}=\frac{1}{4}\sum_{ij}J_{ij}\vec{ \hat{\sigma}}_i\cdot\vec{ \hat{\sigma}}_j+\frac{1}{4}\sum_{ij}\chi_{ij}{\hat \sigma}_i^z{\hat \sigma}_j^z+\sum_iB_i{\hat \sigma}_j^z.\eeq The XXZ couplings   depend on the  different scattering lengths  of the atoms and on  the overlap integral of the corresponding  3D harmonic oscillator wave functions.  The use of a blue sideband transition that generate mismatched motional states of the coupled internal levels allowed the experiment to generate a finite  $\chi_{ij}$. Otherwise it  would have been negligible for Rb if the experiment had used a carrier transition (when laser only flips the spin).  In addition to the interaction term, the interrogating laser  generated  transverse   and longitudinal fields with strengths set by the Rabi frequency, and  the laser detuning from the blue sideband transition.

To observe the DPT the atoms were  initialized in the $\ket{\downarrow}$  state.  For this initial state, the exchange interaction term not only becomes a constant of motion but also locks the atoms into the fully collective spin manifold reducing the XXZ Hamiltonian to a LMG model with coupling constants replaced by their averaged value.  Similar to the cavity experiment, instead  of direct measurements of the long-time-averaged excitation which is inevitably limited by technical issues, the order parameter was set to be  the excitation fraction at a probe time $0.5$s. The entire phase diagram was  obtained by scanning the two-photon detuning $\delta$ for fixed transverse field $B$  and by varying interactions using different atomic densities (See Fig.~\ref{Figexp}(c)).

We note that there is a direct connection between the  DPT in the LMG model  and the phenomena of macroscopic self-trapping and Josephson tunneling observed in coupled atomic condensates \cite{Albiez2005,Anker2005,Reinhard2013,Levy2007} and in solid state polariton condensates \cite{Abbarchi2013}. In this context the ferromagnetic and paramagnetic dynamical phases can be related to the self-trapped and tunneling phases respectively in the corresponding systems.  Under this correspondence one can say that  self-trapping   experiments done years back  did observe  indications of the  distinct dynamical behaviors. However,  they  did not provide a full  characterization of the DPTs.\\

{\it \bf  DPT in  
Richardson-Gaudin  models.   } A similar  mapping of the
single-particle eigenstates of the harmonic trap onto a lattice in mode
space, was used in a trapped gas of $10^4$ ultracold fermionic potassium atoms. In this  case the experiment observed a DPT predicted to exist in
quenched s-wave superconductors described by Richardson-Gaudin models as explained  in Sec. \ref{sec:RGviaLax}. As we will discuss  below this DPT 
has  only been   indirectly observed in real superconductors  (see Sec. \ref{CMI}).  The model that was simulated in the experiment \cite{smale2019} was a collective Heisenberg model,  in which the non-local spin-spin couplings $J_{ij}$ compete with an inhomogeneous axial field set by  $h_i$:
\beq  \hat{H}= \sum_i h_i \hat{S}^z_i - \sum_{i,j}J_{ij}  {\vec{\hat S}}_i \cdot {\vec {\hat S }}_j.\eeq The inhomogeneous axial field in mode space was generated by different harmonic confinement potentials experienced by the  two internal hyperfine  levels of  ${}^{40}$K. The strength of the inhomogeneity was tuned in two ways. Using the polarization of one of the laser beams forming the optical trap one  change the trapping frequency in a spin dependent manner. One can also change the temperature and therefore the distribution of the modes occupied by the atoms.  Interatomic collisions were  used to generate the exchange term. Due  to the extended nature of the motional wave functions, the $\mathcal J_{ij}$ were  again long-ranged. The    mode-dependent coupling factor  ${\mathcal J}_{ij}$ was controlled by  tuning 
 the s-wave scattering length  of the colliding atoms using   a magnetic Feshbach resonance \cite{RevModPhys2010}.  The expected critical  behavior was observed  in the experiment as a function  of the  mean and thermally averaged interaction strength $J = \left\langle\sum_{i,j} J_{ij}/N^{2}\right\rangle_T$  and  the  axial field inhomogeneity $\widetilde{h}= \left\langle\sqrt{\sum_i h_i^2/N - (\sum_i h_i/N)^2}\right\rangle_T$  where the indices $i,j$ run over $N$ populated modes (See Fig.~\ref{Figexp}d).

 In the  ``all-to-all'' limit, in which coupling constants are replaced by their mean value, $J_{ij}\to J$, the Hamiltonian   becomes integrable and maps to the Bardeen-Cooper-Schrieffer Hamiltonian for fermionic superconductors expressed in terms of Anderson pseudospin  discussed  in Eq.~(\ref {HRed}) with  $h_i \to 2\e_i$ and  $J \to G$.  Note that  at the  level of  mean field  the extra term $({\hat S}^{z})\to 2{\hat S}^{z}\langle {\hat S}^{z}\rangle$ acts as a simple collective rotation along $z$ which does not modify the dynamics since $\langle {\hat S}^{z}\rangle$ is conserved. Moreover this  term can be  removed  by choosing an initial condition  with $\langle {\hat S}^{z}\rangle=0$.
 
 Using the Lax analysis explained in Session \ref{sec:LaxMethod}, it is possible  to  obtain  the dynamical phases of this model, characterized by  the total transverse magnetization $ \mathcal{S}(t)=2\sqrt{\langle {\hat S}^{x}\rangle^2+\langle {\hat S}^{y}\rangle^2}/N$.  The initial conditions accessible in experiment, corresponding to a spin polarized state pointing along the $x$ direction of the Bloch sphere  is very similar to the simple example discussed in Sec.\ref{simple}, but in 3D and with a  different dispersion.  In this case there  is a DPT between   the so called  ``phase I''   below a critical coupling strength $J_c$  where the order parameter quick decays to zero 
 and  ``phase II''above $J_c$  where the order parameter  $\mathcal{S}(t)$ exhibits transient oscillations at the  frequency $ 2 |J| \mathcal{S}(\infty)$, which slowly damp as  $\mathcal{S}(t)$ reaches $\mathcal{S}(\infty)$ (See Eq. (\ref{PhaseIIApproach})). The oscillation  frequency goes to zero at $J_c$ in a non-analytic manner. 

The experimental work described in Ref.~\cite{smale2019} not only  fully characterized the phase diagram but in addition determined the parameter regime where the spin model was valid. The latter was explicitly demonstrated  via a many-body echo sequence  that  fully reversed the spin  Hamiltonian. Terms in the Hamiltonian such as the kinetic energy, assumed to play no role in the frozen mode approximation, were not reversed. The experiment indeed observed full  reversibility of the dynamics in the collisionless   regime where the spin model is expected to be valid, and non-reversibility otherwise.

Although the  phase diagram was characterized for the first time in Ref.~\cite{smale2019}, the understanding that  exchange interactions  can stabilize coherence in a very useful way had been demonstrated   experimentally years before. The experiments were   carried out using thermal Rb atoms, which feature a very similar Hamiltonian than the potassium fermionic gas.  Deep in phase II (dense sample)  a coherence time   up to 58 s owas observed,  while in phase I  (dilute conditions) the coherence time decreased to at most 3 s \cite{Deutsch:2010ky,Solaro:2016iv,Piechon:2009cr}.

 Although the Toronto experiment Ref.~\cite{smale2019}  was able to observe the DPT between  phase I and  phase II, by simulating a spin model, phase III has not been observed yet.

Fermionic ultracold atomic gas experiments   that can actually take  advantage of the Fermi statistics of the atoms instead of an indirect mapping to Anderson pseudospins, in principle are an ideal platform for the observation of the dynamical phase diagram. However, unfortunately it  has remained a challenge in cold atom experiments  to reach  conditions required to see  BCS pairing. This is  because both thermal and quantum fluctuations play a crucial role for quenches from the normal state for   systems with a finite number $N$ of Anderson pseudospins. 
Moreover,    the control and manipulation of  interactions in these systems,
specially for  the $p$-wave case, typically require Feshbach resonances. The latter unfortunately introduce strong three-body processes which make the gas unstable and destroy the desired  pairing processes \cite{Salomon2004,Castin2008,Gurarie2008,Foster2014}.

Future progress on cooling these systems might allow new opportunities  in that direction. An alternative  option  is the use of  an ensemble of cold atoms trapped in an optical cavity. In this case effective Cooper pairs can be  encoded in  internal states of the atoms and attractive interactions between the internal levels can be  realized via the exchange of virtual photons in the cavity. The control of the interaction strength via the parameters of the optical cavity  combined with the tunability of the  dispersion relation of the effective Cooper pairs via Stark shifts  should allow  for the near term exploration of the full dynamical phase diagram of the BCS model, as a function of system parameters and the prepared initial state \cite{Lewisswan2021}.   Ref.~\cite{Lewisswan2021} discusses a way to prepare  the  simple toy model discussed in Sec.\ref{simple} in a cavity setup. \\

{\it \bf  DPT in  
multi-level systems.    } Beyond two-level systems,  DPTs have been also observed in spinor Bose-Einstein  condensates (BEC) made of $N\sim 10^5$ ${}^{23}$Na atoms  using the three Zeeman states  of the lowest hyperfine level, $m_F=1,0,-1$ \cite{yang2019,Tian2020}(See Fig.~\ref{Figexp} e). The experiment operated in the    regime where  to an excellent approximation atoms remained frozen in their single particle motional levels.  Motional relaxation was only  an issue at long times. Because all atoms in the  BEC share the same motional level, in this limit  the bosonic statistics enforces atoms  to remain the in the fully symmetric spin  manifold. The Hamiltonian that describes the internal dynamics in the experiment  therefore becomes a fully collective spin model: \beq H=\frac{c_2}{2N}  {\boldsymbol{\hat S}}\cdot {\boldsymbol{ \hat S}} + q \sum_i  (\hat{S}^z_{i}) ^2.\eeq The first term encapsulates the spin dependent interactions,  which are infinite-range in mode space even though  the atomic collision interaction by itself is fully local. The second term describes a quadratic Zeeman field $q$ experience by the atoms from a real magnetic field plus a microwave dressing field. Here $ \hat{S}_{z,i}$ is the  spin-1 operator for the {\it ith} particle along the $z$  component of the Bloch sphere.

To observe the DPT, the experiment  prepared the majority of the atoms   in the $m_F=0$ level and then monitored the number of atoms  remaining in that level $N_0(t)$ as a function of time  for different values of the control parameter $q$. The order parameter used in the experiment was the  a
quantity $A_{\rm dip} \equiv 1 - N_0( t_{\rm dip} )$, with $N_0( t_{\rm dip} )$ being the
value of $N_0(t)$ at the first dip of the spin oscillations.  For positive $q>0$, $A_{\rm dip}$
was observed to  remain almost at 0. Across the DPT transition
point of $q_c = 0$, $A_{\rm dip}$ was seen to jump to finite value
which decreased linearly with $|q|$
when  $q<0$ (See Fig.~\ref{Figexp} e).  Further work  in this system was able to connect the  DPT  to an equilibrium  phase transition not for the ground state but  for the highest energy level in a
subspace with zero spin magnetization \cite{Tian2020}. 

Besides observations of DPT which happened at relatively short times,   a follow-up experiment  using Rb atoms  instead of Na atoms  observed  another complementary type of universal non-equilibrium behavior  associated with the emergence of  non-thermal fixed points \cite{Oberthaler2018}. This type of behavior was observed at intermediate times satisfying two conditions: i) Times that are long enough  to allow  the frozen mode approximation used to observe the DPT  to  become  invalid.   In this case  the system   loses the information about  the initial conditions  due to motional relaxation. ii) Times that are  not so  long that lead the system to reach a  quasi-stationary or an equilibrium situation.  In this intermediate regime it was observed that 
 the system develops a universal scaling behaviour in time and space.
Physically the emergence of  non-thermal fixed points have been  attributed to  the transport of an emergent collective conserved quantity
towards low momentum scales. A similar behavior has  observed by
another experiment  in a single-component Bose gas \cite{Schmiedmayer2018,Langen2016}. In both of these experiments it was experimentally confirmed  (by preparing different initial conditions and obtaining 
the same scaling behaviour) that
the observed non-thermal scaling phenomenon involved no fine-tuning of
parameters.

It is clear from all the discussion presented above that ultra-cold atomic system  are opening fantastic opportunities to probe  DPTs and even more general  non-equilibrium universal phenomena in controllable settings. While most of the observations so far have been guided by theory in conditions where  either a mean field analysis is sufficient, or when   exact calculations are possible,  soon   experiments will enter regimes intractable to theory. This may lead to the discovery of  new forms of non-equilibrium phenomena not yet predicted by theory, pushing the field  into even more  exciting  directions.

\section{Conclusions }
In this work we   reviewed DPTs occurring in the large $N$, $\mathcal{N}$, or $d$ limit of a broad variety of models ranging from statistical mechanics (Ising models, $\phi^4$ field theories) to condensed matter (Richardson-Gaudin magnets, strongly correlated fermions).
We have also discussed   numerous experimental platforms where the dynamics of these models can be realized (Sec.~\ref{sec:exp}). We have illustrated those regimes where  experiments are well fitted by the theory presented in Secs. II and III, as well as the limits where descriptions beyond collision-less regimes may  instead be required to describe experiments.  Such platforms include trapped ions, cavity QED systems, and ultracold Bose and Fermi systems.  They can be employed to demonstrate a broad variety of universality classes of DPTs in isolated quantum systems, and possibly study in the near future operational regimes where DPTs require a full quantum mechanical treatment and cannot be captured by an effective semi-classical   description.

There are  promising avenues for the future of   pre-thermal collisionless DPTs. 
For instance, Rydberg simulators have been recently shown to be capable to access long-lived athermal states beyond  the conventional pre-thermal paradigm~\cite{bernien2017probing,ebadi2021quantum},       referred as quantum scars~\cite{serbyn2021quantum} and  defying the  eigenstate thermalization hypothesis.  This could be potentially employed in the near future to study forms of DPTs which cannot be characterized within a simple semi-classical description. Recent implementations of models with a fragmented Hilbert space using ultracold fermions~\cite{kohlert2021experimental}   suggest similar lines of investigation in that context. 

Another exciting direction,  is the  understanding of the fate of the DPTs in systems where quantum fluctuations cannot be ignored.   At the theoretical level, it will be crucial to  derive new types of order parameters that can describe the distinct dynamical behavior even when quantum effects start to play a role. The order parameters  discussed in this review will decay in the ordered phase whenever quantum effects start to play a significant role. Can a properly defined order parameter,  genuinely characterize  the underlying phases associated with a DPT? It is likely there is an affirmative answer to this question.  Feasible candidates are observables that measure the   interaction energy of the evolving system and   therefore involve at the least two-body correlators, as illustrated by the case of the O($\mathcal{N}\to\infty$) model in Sec.~\ref{sec:ON}.  Closely connected with this question is the robustness of the topological order in non-trivial phases  in the presence of quantum effects.  Is there a redefinition of the winding number that remains nonzero even in the presence of strong quantum fluctuations? 

One of the most important goals of modern quantum science is to learn how to control and entangle many-body systems, and how to use it as a resource in quantum technologies, such as for the development of   improved quantum sensors, materials and technologies. A big limitation in this direction comes from the fact that entangled states are difficult to prepare and maintain, since noise and decoherence rapidly collapse them into classical statistical mixtures.  In the context of DPTs it is likely that the same parameter regime where interactions stabilize a finite order parameter is also promising for a generation of   robust entanglement. In fact, recent investigations have pointed  out the generation of robust spin squeezing in phase II ~\cite{He2019}.  Understanding the dynamics of entanglement across a DPT is  a fascinating new direction~\cite{Foss-Feig2017,Lewisswan20212}. Is entanglement maximum  at the critical point? What is the best entanglement witness for the state across the DPT? 

While all these questions are theoretically extremely challenging,  recent  experimental advances  in synthesizing, manipulating and detecting  quantum many-body systems are bringing  quantum control into a new  paradigm~\cite{Gross2017,ebadi2021quantum}. It is likely that near term experiments will  stimulate new theoretical methods, which may provide  unprecedented insight  into  novel classes of DPTs where strong quantum fluctuations  play a key role in shaping the dynamical phases.  We hope this review  will encourage future  work  and investigations in this direction. 

\section{Acknowledgements}

This project has been supported by the Deutsche Forschungsgemeinschaft (DFG, German Research Foundation) – Project-ID 429529648 – TRR 306 QuCoLiMa (”Quantum Cooperativity of Light and Matter”), ERC (starting grant No. 716648), AFOSR (FA9550-18-1-0319, FA9550-19-1-027), NSF QLCI OMA–2016244, NSF Phys-1734006, DOE National Quantum Information Science Research Centers (Quantum Systems Accelerator), and NIST.

\bibliography{Review2}

\begin{thebibliography}{100}

\bibitem{Henkel2008}
M.~Henkel, H~Hinrichsen, and S.~Lübeck.
\newblock {\em Non-Equilibrium Phase Transitions Vol. 1 and 2.}
\newblock Springer, Dordrecht, 2008.

\bibitem{Berges2004b}
J.~Berges.
\newblock Introduction to nonequilibrium quantum field theory.
\newblock {\em AIP Conf. Proc.}, 739(1):3--62, 2005.

\bibitem{gring2012relaxation}
Michael Gring, Maximilian Kuhnert, Tim Langen, Takuya Kitagawa, Bernhard Rauer,
  Matthias Schreitl, Igor Mazets, D~Adu Smith, Eugene Demler, and J{\"o}rg
  Schmiedmayer.
\newblock Relaxation and prethermalization in an isolated quantum system.
\newblock {\em Science}, 337(6100):1318--1322, 2012.

\bibitem{Langen2016}
Tim Langen, Thomas Gasenzer, and Jörg Schmiedmayer.
\newblock Prethermalization and universal dynamics in near-integrable quantum
  systems.
\newblock {\em J. Stat. Mech.}, 2016(6):064009, 2016.

\bibitem{Kollar2011}
Marcus Kollar, F.~Alexander Wolf, and Martin Eckstein.
\newblock Generalized gibbs ensemble prediction of prethermalization plateaus
  and their relation to nonthermal steady states in integrable systems.
\newblock {\em Phys. Rev. B}, 84:054304, Aug 2011.

\bibitem{Robinson2014}
F.~H.~L. Essler, S.~Kehrein, S.~R. Manmana, and N.~J. Robinson.
\newblock Quench dynamics in a model with tuneable integrability breaking.
\newblock {\em Phys. Rev. B}, 89:165104, Apr 2014.

\bibitem{Bertini2015}
Bruno Bertini, Fabian H.~L. Essler, Stefan Groha, and Neil~J. Robinson.
\newblock Prethermalization and thermalization in models with weak
  integrability breaking.
\newblock {\em Phys. Rev. Lett.}, 115:180601, Oct 2015.

\bibitem{marcuzzi2016prethermalization}
Matteo Marcuzzi, Jamir Marino, Andrea Gambassi, and Alessandro Silva.
\newblock Prethermalization from a low-density holstein-primakoff expansion.
\newblock {\em Physical Review B}, 94(21):214304, 2016.

\bibitem{PhysRevB.94.245117}
Bruno Bertini, Fabian H.~L. Essler, Stefan Groha, and Neil~J. Robinson.
\newblock Thermalization and light cones in a model with weak integrability
  breaking.
\newblock {\em Phys. Rev. B}, 94:245117, Dec 2016.

\bibitem{PhysRevLett.127.130601}
Joseph Durnin, M.~J. Bhaseen, and Benjamin Doyon.
\newblock Nonequilibrium dynamics and weakly broken integrability.
\newblock {\em Phys. Rev. Lett.}, 127:130601, Sep 2021.

\bibitem{PhysRevA.95.023621}
Cheng-Ju Lin and Olexei~I. Motrunich.
\newblock Quasiparticle explanation of the weak-thermalization regime under
  quench in a nonintegrable quantum spin chain.
\newblock {\em Phys. Rev. A}, 95:023621, Feb 2017.

\bibitem{heyl2013dynamical}
Markus Heyl, Anatoli Polkovnikov, and Stefan Kehrein.
\newblock Dynamical quantum phase transitions in the transverse-field ising
  model.
\newblock {\em Physical review letters}, 110(13):135704, 2013.

\bibitem{Jurcevic2017}
P.~Jurcevic, H.~Shen, P.~Hauke, C.~Maier, T.~Brydges, C.~Hempel, B.~P. Lanyon,
  M.~Heyl, R.~Blatt, and C.~F. Roos.
\newblock Direct observation of dynamical quantum phase transitions in an
  interacting many-body system.
\newblock {\em Phys. Rev. Lett.}, 119:080501, Aug 2017.

\bibitem{Heyl2019}
Markus Heyl.
\newblock Dynamical quantum phase transitions: A brief survey.
\newblock {\em EPL (Europhysics Letters)}, 125(2):26001, 2019.

\bibitem{ber2008}
Jürgen Berges, Alexander Rothkopf, and Jonas Schmidt.
\newblock Nonthermal fixed points: Effective weak coupling for strongly
  correlated systems far from equilibrium.
\newblock {\em Physical Review Letters}, 101(4), Jul 2008.

\bibitem{super2011}
Boris Nowak, Dénes Sexty, and Thomas Gasenzer.
\newblock Superfluid turbulence: Nonthermal fixed point in an ultracold bose
  gas.
\newblock {\em Physical Review B}, 84(2), Jul 2011.

\bibitem{pruf2018}
Maximilian Prüfer, Philipp Kunkel, Helmut Strobel, Stefan Lannig, Daniel
  Linnemann, Christian-Marcel Schmied, Jürgen Berges, Thomas Gasenzer, and
  Markus~K. Oberthaler.
\newblock Observation of universal dynamics in a spinor bose gas far from
  equilibrium.
\newblock {\em Nature}, 563(7730):217–220, Nov 2018.

\bibitem{erne2018}
Sebastian Erne, Robert Bücker, Thomas Gasenzer, Jürgen Berges, and Jörg
  Schmiedmayer.
\newblock Universal dynamics in an isolated one-dimensional bose gas far from
  equilibrium.
\newblock {\em Nature}, 563(7730):225–229, Nov 2018.

\bibitem{eigen2018}
Christoph Eigen, Jake A.~P. Glidden, Raphael Lopes, Eric~A. Cornell, Robert~P.
  Smith, and Zoran Hadzibabic.
\newblock Universal prethermal dynamics of bose gases quenched to unitarity.
\newblock {\em Nature}, 563(7730):221–224, Nov 2018.

\bibitem{gasen2019}
Christian-Marcel Schmied, Aleksandr~N. Mikheev, and Thomas Gasenzer.
\newblock Non-thermal fixed points: Universal dynamics far from equilibrium.
\newblock {\em International Journal of Modern Physics A}, 34(29):1941006, Oct
  2019.

\bibitem{Haken1975}
H.~Haken.
\newblock Cooperative phenomena in systems far from thermal equilibrium and in
  nonphysical systems.
\newblock {\em Rev. Mod. Phys.}, 47:67--121, Jan 1975.

\bibitem{Chandrasekhar1968}
S.~Chandrasekhar.
\newblock {\em Hydrodynamic and Hydromagnetic Stability}.
\newblock Clarendon, Oxford, 1968.

\bibitem{Chakrabarti1995}
C.~G. Chakrabarti, Sutapa Ghosh, and Syamali Bhadra.
\newblock Non-equilibrium thermodynamics of lotka-volterra ecosystems:
  Stability and evolution.
\newblock {\em Journal of Biological Physics}, 21(4):273--284, 1995.

\bibitem{Glansdorff1971}
P.~Glansdorff and I.~Prigogine.
\newblock {\em Thermodynamic Theory of Structure, Stability and Fluctuations}.
\newblock Wiley, New York, 1971.

\bibitem{landau1969statistical}
LD~Landau and EM~Lifshitz.
\newblock Statistical physics, 1969.
\newblock {\em Google Scholar}, pages 237--241, 1969.

\bibitem{hohenberg2015introduction}
PC~Hohenberg and AP~Krekhov.
\newblock An introduction to the ginzburg--landau theory of phase transitions
  and nonequilibrium patterns.
\newblock {\em Physics Reports}, 572:1--42, 2015.

\bibitem{Berges2004a}
J.~Berges, Sz. Bors\'anyi, and C.~Wetterich.
\newblock Prethermalization.
\newblock {\em Phys. Rev. Lett.}, 93:142002, Sep 2004.

\bibitem{ZinnJustinbook}
J~Zinn-Justin.
\newblock {\em Quantum Field Theory and Critical Phenomena}.
\newblock Oxford Clarendon Press, 1989.

\bibitem{Amit/Martin-Mayor}
Daniel~J. Amit and Victor Martin-Mayor.
\newblock {\em Field Theory, the Renormalization Group, and Critical
  Phenomena}.
\newblock World Scientific, Singapore, third edition, 2005.

\bibitem{Cardy1996}
John Cardy and Uwe~C. T\"auber.
\newblock Theory of branching and annihilating random walks.
\newblock {\em Phys. Rev. Lett.}, 77:4780--4783, Dec 1996.

\bibitem{Hohenberg1977}
P.~C. Hohenberg and B.~I. Halperin.
\newblock Theory of dynamic critical phenomena.
\newblock {\em Rev. Mod. Phys.}, 49:435--479, Jul 1977.

\bibitem{Tauberbook2014}
Uwe~C T{\"a}uber.
\newblock {\em Critical Dynamics: a Field Theory Approach to Equilibrium and
  Non-Equilibrium Scaling Behavior}.
\newblock Cambridge University Press, 2014.

\bibitem{Kamenevbook2011}
A.~Kamenev.
\newblock {\em Field Theory of Non-Equilibrium Systems}.
\newblock Cambridge University Press, 2011.

\bibitem{Gagel2014}
Pia Gagel, Peter~P. Orth, and J{\"o}rg Schmalian.
\newblock Universal postquench prethermalization at a quantum critical point.
\newblock {\em Phys. Rev. Lett.}, 113:220401, Nov 2014.

\bibitem{Sieberer2016review}
L~M Sieberer, M~Buchhold, and S~Diehl.
\newblock Keldysh field theory for driven open quantum systems.
\newblock {\em Reports on Progress in Physics}, 79(9):096001, 2016.

\bibitem{1965NucPh..62..211G}
A.~J. {Glick}, H.~J. {Lipkin}, and N.~{Meshkov}.
\newblock {Validity of many-body approximation methods for a solvable model.
  (III). Diagram summations}.
\newblock {\em Nuclear Physics}, 62(2):211--224, February 1965.

\bibitem{1965NucPh..62..188L}
H.~J. {Lipkin}, N.~{Meshkov}, and A.~J. {Glick}.
\newblock {Validity of many-body approximation methods for a solvable model.
  (I). Exact solutions and perturbation theory}.
\newblock {\em Nuclear Physics}, 62(2):188--198, February 1965.

\bibitem{1965NucPh..62..199M}
N.~{Meshkov}, A.~J. {Glick}, and H.~J. {Lipkin}.
\newblock {Validity of many-body approximation methods for a solvable model.
  (II). Linearization procedures}.
\newblock {\em Nuclear Physics}, 62(2):199--210, February 1965.

\bibitem{1999PhLB..451....1P}
Feng {Pan} and J.~P. {Draayer}.
\newblock {Analytical solutions for the LMG model}.
\newblock {\em Physics Letters B}, 451(1-2):1--10, April 1999.

\bibitem{2004RvMP...76..643D}
J.~{Dukelsky}, S.~{Pittel}, and G.~{Sierra}.
\newblock {Colloquium: Exactly solvable Richardson-Gaudin models for many-body
  quantum systems}.
\newblock {\em Reviews of Modern Physics}, 76(3):643--662, July 2004.

\bibitem{1983PhRvB..28.3955B}
R.~{Botet} and R.~{Jullien}.
\newblock {Large-size critical behavior of infinitely coordinated systems}.
\newblock {\em \prb}, 28(7):3955--3967, October 1983.

\bibitem{2005PhRvA..71f0304D}
S{\'e}bastien {Dusuel} and Julien {Vidal}.
\newblock {Finite-size scaling exponents and entanglement in the two-level BCS
  model}.
\newblock {\em \pra}, 71(6):060304, June 2005.

\bibitem{PhysRevLett.99.050402}
Pedro Ribeiro, Julien Vidal, and R\'emy Mosseri.
\newblock Thermodynamical limit of the lipkin-meshkov-glick model.
\newblock {\em Phys. Rev. Lett.}, 99:050402, Aug 2007.

\bibitem{lmg2008}
Pedro Ribeiro, Julien Vidal, and Rémy Mosseri.
\newblock Exact spectrum of the lipkin-meshkov-glick model in the thermodynamic
  limit and finite-size corrections.
\newblock {\em Physical Review E}, 78(2), Aug 2008.

\bibitem{defenu2021long}
Nicol{\`o} Defenu, Tobias Donner, Tommaso Macr{\`\i}, Guido Pagano, Stefano
  Ruffo, and Andrea Trombettoni.
\newblock Long-range interacting quantum systems.
\newblock {\em arXiv preprint arXiv:2109.01063}, 2021.

\bibitem{latorre2005entanglement}
Jos{\'e}~I Latorre, Rom{\'a}n Or{\'u}s, Enrique Rico, and Julien Vidal.
\newblock Entanglement entropy in the lipkin-meshkov-glick model.
\newblock {\em Physical Review A}, 71(6):064101, 2005.

\bibitem{orus2008equivalence}
Rom{\'a}n Or{\'u}s, S{\'e}bastien Dusuel, and Julien Vidal.
\newblock Equivalence of critical scaling laws for many-body entanglement in
  the lipkin-meshkov-glick model.
\newblock {\em Physical review letters}, 101(2):025701, 2008.

\bibitem{2020JPhA...53a3001M}
Somnath {Maity}, Utso {Bhattacharya}, and Amit {Dutta}.
\newblock {One-dimensional quantum many body systems with long-range
  interactions}.
\newblock {\em Journal of Physics A Mathematical General}, 53(1):013001,
  January 2020.

\bibitem{Moeckel2008}
Michael Moeckel and Stefan Kehrein.
\newblock Interaction quench in the hubbard model.
\newblock {\em Phys. Rev. Lett.}, 100:175702, May 2008.

\bibitem{marcuzzi2013prethermalization}
Matteo Marcuzzi, Jamir Marino, Andrea Gambassi, and Alessandro Silva.
\newblock Prethermalization in a nonintegrable quantum spin chain after a
  quench.
\newblock {\em Physical review letters}, 111(19):197203, 2013.

\bibitem{PhysRevLett.113.210402}
N.~Nessi, A.~Iucci, and M.~A. Cazalilla.
\newblock Quantum quench and prethermalization dynamics in a two-dimensional
  fermi gas with long-range interactions.
\newblock {\em Phys. Rev. Lett.}, 113:210402, Nov 2014.

\bibitem{PhysRevX.9.021027}
Krishnanand Mallayya, Marcos Rigol, and Wojciech De~Roeck.
\newblock Prethermalization and thermalization in isolated quantum systems.
\newblock {\em Phys. Rev. X}, 9:021027, May 2019.

\bibitem{lerose2019impact}
Alessio Lerose, Bojan {\v{Z}}unkovi{\v{c}}, Jamir Marino, Andrea Gambassi, and
  Alessandro Silva.
\newblock Impact of nonequilibrium fluctuations on prethermal dynamical phase
  transitions in long-range interacting spin chains.
\newblock {\em Physical Review B}, 99(4):045128, 2019.

\bibitem{Zhang2017}
J.~Zhang, G.~Pagano, P.~W. Hess, A.~Kyprianidis, P.~Becker, H.~Kaplan, A.~V.
  Gorshkov, Z.~X. Gong, and C.~Monroe.
\newblock Observation of a many-body dynamical phase transition with a 53-qubit
  quantum simulator.
\newblock {\em Nature}, 551(7682):601--604, 2017.

\bibitem{Muniz2020}
Juan~A. Muniz, Diego Barberena, Robert~J. Lewis-Swan, Dylan~J. Young, Julia
  R.~K. Cline, Ana~Maria Rey, and James~K. Thompson.
\newblock Exploring dynamical phase transitions with cold atoms in an
  optical  cavity.
\newblock {\em Nature}, 580(7805):602--607, Apr 2020.

\bibitem{chu2020}
Anjun Chu, Johannes Will, Jan Arlt, Carsten Klempt, and Ana~Maria Rey.
\newblock Simulation of $xxz$ spin models using sideband transitions in trapped
  bosonic gases.
\newblock {\em Phys. Rev. Lett.}, 125:240504, Dec 2020.

\bibitem{polkovnikov2010phase}
Anatoli Polkovnikov.
\newblock Phase space representation of quantum dynamics.
\newblock {\em Annals of Physics}, 325(8):1790--1852, 2010.

\bibitem{das06}
Arnab Das, K.~Sengupta, Diptiman Sen, and Bikas~K. Chakrabarti.
\newblock Infinite-range ising ferromagnet in a time-dependent transverse
  magnetic field: Quench and ac dynamics near the quantum critical point.
\newblock {\em Phys. Rev. B}, 74:144423, 2006.

\bibitem{kelly2019detecting}
Shane~P Kelly, Eddy Timmermans, and S-W Tsai.
\newblock Detecting macroscopic indefiniteness of cat states in bosonic
  interferometers.
\newblock {\em Physical Review A}, 100(3):032117, 2019.

\bibitem{kelly2020thermalization}
Shane~P Kelly, Eddy Timmermans, and S-W Tsai.
\newblock Thermalization and its breakdown for a large nonlinear spin.
\newblock {\em Physical Review A}, 102(5):052210, 2020.

\bibitem{sciolla2011dynamical}
Bruno Sciolla and Giulio Biroli.
\newblock Dynamical transitions and quantum quenches in mean-field models.
\newblock {\em Journal of Statistical Mechanics: Theory and Experiment},
  2011(11):P11003, 2011.

\bibitem{PhysRevLett.121.240403}
Nicol\`o Defenu, Tilman Enss, Michael Kastner, and Giovanna Morigi.
\newblock Dynamical critical scaling of long-range interacting quantum magnets.
\newblock {\em Phys. Rev. Lett.}, 121:240403, Dec 2018.

\bibitem{PhysRevB.78.104426}
Tommaso Caneva, Rosario Fazio, and Giuseppe~E. Santoro.
\newblock Adiabatic quantum dynamics of the lipkin-meshkov-glick model.
\newblock {\em Phys. Rev. B}, 78:104426, Sep 2008.

\bibitem{lerose2018chaotic}
Alessio Lerose, Jamir Marino, Bojan {\v{Z}}unkovi{\v{c}}, Andrea Gambassi, and
  Alessandro Silva.
\newblock Chaotic dynamical ferromagnetic phase induced by nonequilibrium
  quantum fluctuations.
\newblock {\em Physical review letters}, 120(13):130603, 2018.

\bibitem{lerose2019prethermal}
Alessio Lerose, Jamir Marino, Andrea Gambassi, and Alessandro Silva.
\newblock Prethermal quantum many-body kapitza phases of periodically driven
  spin systems.
\newblock {\em Physical Review B}, 100(10):104306, 2019.

\bibitem{Zunkovic}
B.~\v{Z}unkovi\v{c}, A.~Silva, and M.~Fabrizio.
\newblock {Dynamical phase transitions and Loschmidt echo in the infinite-range
  XY model}.
\newblock {\em Phil. Trans. R. Soc.}, A 374:20150160, 2016.

\bibitem{Gambassi2011}
A.~Gambassi and P.~Calabrese.
\newblock Quantum quenches as classical critical films.
\newblock {\em Europhys. Lett.}, 95(6):66007, 2011.

\bibitem{sartori2015spin}
Alberto Sartori, Jamir Marino, Sandro Stringari, and Alessio Recati.
\newblock Spin-dipole oscillation and relaxation of coherently coupled
  bose--einstein condensates.
\newblock {\em New Journal of Physics}, 17(9):093036, 2015.

\bibitem{vzunkovivc2018dynamical}
Bojan {\v{Z}}unkovi{\v{c}}, Markus Heyl, Michael Knap, and Alessandro Silva.
\newblock Dynamical quantum phase transitions in spin chains with long-range
  interactions: Merging different concepts of nonequilibrium criticality.
\newblock {\em Physical review letters}, 120(13):130601, 2018.

\bibitem{halimeh2017prethermalization}
Jad~C Halimeh, Valentin Zauner-Stauber, Ian~P McCulloch, Ines De~Vega, Ulrich
  Schollw{\"o}ck, and Michael Kastner.
\newblock Prethermalization and persistent order in the absence of a thermal
  phase transition.
\newblock {\em Physical Review B}, 95(2):024302, 2017.

\bibitem{mps2016}
Jutho Haegeman, Christian Lubich, Ivan Oseledets, Bart Vandereycken, and Frank
  Verstraete.
\newblock Unifying time evolution and optimization with matrix product states.
\newblock {\em Physical Review B}, 94(16), Oct 2016.

\bibitem{PhysRevB.97.174401}
Johannes Lang, Bernhard Frank, and Jad~C. Halimeh.
\newblock Concurrence of dynamical phase transitions at finite temperature in
  the fully connected transverse-field ising model.
\newblock {\em Phys. Rev. B}, 97:174401, May 2018.

\bibitem{PhysRevLett.125.040602}
Paraj Titum and Mohammad~F. Maghrebi.
\newblock Nonequilibrium criticality in quench dynamics of long-range spin
  models.
\newblock {\em Phys. Rev. Lett.}, 125:040602, Jul 2020.

\bibitem{PhysRevResearch.2.012041}
Alessio Lerose and Silvia Pappalardi.
\newblock Origin of the slow growth of entanglement entropy in long-range
  interacting spin systems.
\newblock {\em Phys. Rev. Research}, 2:012041, Feb 2020.

\bibitem{PhysRevB.104.115133}
Jad~C. Halimeh, Maarten Van~Damme, Lingzhen Guo, Johannes Lang, and Philipp
  Hauke.
\newblock Dynamical phase transitions in quantum spin models with
  antiferromagnetic long-range interactions.
\newblock {\em Phys. Rev. B}, 104:115133, Sep 2021.

\bibitem{piccitto2019}
Giulia Piccitto and Alessandro Silva.
\newblock Crossover from fast to slow dynamics in a long range interacting
  ising chain.
\newblock {\em Journal of Statistical Mechanics: Theory and Experiment},
  2019(9):094017, Sep 2019.

\bibitem{PhysRevB.100.014434}
Nicol\`o Defenu, Tilman Enss, and Jad~C. Halimeh.
\newblock Dynamical criticality and domain-wall coupling in long-range
  hamiltonians.
\newblock {\em Phys. Rev. B}, 100:014434, Jul 2019.

\bibitem{PhysRevB.94.184403}
Steve Campbell.
\newblock Criticality revealed through quench dynamics in the
  lipkin-meshkov-glick model.
\newblock {\em Phys. Rev. B}, 94:184403, Nov 2016.

\bibitem{dutta2001phase}
Amit Dutta and JK~Bhattacharjee.
\newblock Phase transitions in the quantum ising and rotor models with a
  long-range interaction.
\newblock {\em Physical Review B}, 64(18):184106, 2001.

\bibitem{campa2009statistical}
Alessandro Campa, Thierry Dauxois, and Stefano Ruffo.
\newblock Statistical mechanics and dynamics of solvable models with long-range
  interactions.
\newblock {\em Physics Reports}, 480(3-6):57--159, 2009.

\bibitem{maghrebi2016causality}
Mohammad~F Maghrebi, Zhe-Xuan Gong, Michael Foss-Feig, and Alexey~V Gorshkov.
\newblock Causality and quantum criticality in long-range lattice models.
\newblock {\em Physical Review B}, 93(12):125128, 2016.

\bibitem{PolkovnikovRMP}
Anatoli Polkovnikov, Krishnendu Sengupta, Alessandro Silva, and Mukund
  Vengalattore.
\newblock \textit{Colloquium} : Nonequilibrium dynamics of closed interacting
  quantum systems.
\newblock {\em Rev. Mod. Phys.}, 83:863--883, Aug 2011.

\bibitem{Barthel2008}
T.~Barthel and U.~Schollwoeck.
\newblock Dephasing and the steady state in quantum many-particle systems.
\newblock {\em Phys. Rev. Lett.}, 100(10):100601, Mar 2008.

\bibitem{Rigol2008}
Marcos Rigol, Vanja Dunjko, and Maxim Olshanii.
\newblock Thermalization and its mechanism for generic isolated quantum
  systems.
\newblock {\em Nature}, 452(7189):854--858, 2008.

\bibitem{Sachdevbook}
S.~Sachdev.
\newblock {\em Quantum Phase Transitions}.
\newblock Cambridge University Press, 2011.

\bibitem{yin2021fermion}
Shuai Yin and Shao-Kai Jian.
\newblock Fermion-induced dynamical critical point.
\newblock {\em Physical Review B}, 103(12):125116, 2021.

\bibitem{jian2019universal}
Shao-Kai Jian, Shuai Yin, and Brian Swingle.
\newblock Universal prethermal dynamics in gross-neveu-yukawa criticality.
\newblock {\em Physical review letters}, 123(17):170606, 2019.

\bibitem{Chandran2013}
Anushya Chandran, Arun Nanduri, S.~S. Gubser, and S.~L. Sondhi.
\newblock Equilibration and coarsening in the quantum $o(n)$ model at infinite
  $n$.
\newblock {\em Phys. Rev. B}, 88:024306, Jul 2013.

\bibitem{Maraga2015}
Anna Maraga, Alessio Chiocchetta, Aditi Mitra, and Andrea Gambassi.
\newblock Aging and coarsening in isolated quantum systems after a quench:
  Exact results for the quantum $\text{O}(n)$ model with $n$
  $\ensuremath{\rightarrow}$ $\ensuremath{\infty}$.
\newblock {\em Phys. Rev. E}, 92:042151, Oct 2015.

\bibitem{Berges2002}
J.~Berges, N.~Tetradis, and C.~Wetterich.
\newblock Non-perturbative renormalization flow in quantum field theory and
  statistical physics.
\newblock {\em Phys. Rep.}, 363:223--386, 2002.

\bibitem{Sotiriadis2010}
Spyros Sotiriadis and John Cardy.
\newblock Quantum quench in interacting field theory: A self-consistent
  approximation.
\newblock {\em Phys. Rev. B}, 81:134305, Apr 2010.

\bibitem{Sciolla2013}
Bruno Sciolla and Giulio Biroli.
\newblock Quantum quenches, dynamical transitions, and off-equilibrium quantum
  criticality.
\newblock {\em Phys. Rev. B}, 88:201110, Nov 2013.

\bibitem{Smacchia2015}
Pietro Smacchia, Michael Knap, Eugene Demler, and Alessandro Silva.
\newblock Exploring dynamical phase transitions and prethermalization with
  quantum noise of excitations.
\newblock {\em Phys. Rev. B}, 91:205136, May 2015.

\bibitem{Chiocchetta2016}
Alessio Chiocchetta, Marco Tavora, Andrea Gambassi, and Aditi Mitra.
\newblock Short-time universal scaling and light-cone dynamics after a quench
  in an isolated quantum system in $d$ spatial dimensions.
\newblock {\em Phys. Rev. B}, 94:134311, Oct 2016.

\bibitem{halimeh2021quantum}
Jad~C Halimeh and Mohammad~F Maghrebi.
\newblock Quantum aging and dynamical universality in the long-range o model.
\newblock {\em Physical Review E}, 103(5):052142, 2021.

\bibitem{Aarts2000}
Gert Aarts, Gian~Franco Bonini, and Christof Wetterich.
\newblock Exact and truncated dynamics in nonequilibrium field theory.
\newblock {\em Phys. Rev. D}, 63:025012, Dec 2000.

\bibitem{Sondhi}
S.~L. Sondhi, S.~M. Girvin, J.~P. Carini, and D.~Shahar.
\newblock Continuous quantum phase transitions.
\newblock {\em Rev. Mod. Phys.}, 69:315--333, Jan 1997.

\bibitem{Vojta}
M.~Vojta.
\newblock Quantum phase transitions.
\newblock {\em Rep. Prog. Phys.}, 66:2069, 2003.

\bibitem{Chiocchetta2015}
Alessio Chiocchetta, Marco Tavora, Andrea Gambassi, and Aditi Mitra.
\newblock Short-time universal scaling in an isolated quantum system after a
  quench.
\newblock {\em Phys. Rev. B}, 91:220302, Jun 2015.

\bibitem{Janssen1989}
H.-K. Janssen, B.~Schaub, and B.~Schmittmann.
\newblock New universal short-time scaling behaviour of critical relaxation
  processes.
\newblock {\em Z. Phys. B Cond. Mat.}, 73(4):539--549, 1989.

\bibitem{Calabrese2005}
Pasquale Calabrese and Andrea Gambassi.
\newblock Ageing properties of critical systems.
\newblock {\em J. Phys. A: Math. Gen.}, 38(18):R133, 2005.

\bibitem{Sieberer2015}
L.~M. {Sieberer}, M.~{Buchhold}, and S.~{Diehl}.
\newblock {Keldysh Field Theory for Driven Open Quantum Systems}.
\newblock {\em arXiv:1512.00637}, 2015.

\bibitem{chiocchetta2017dynamical}
Alessio Chiocchetta, Andrea Gambassi, Sebastian Diehl, and Jamir Marino.
\newblock Dynamical crossovers in prethermal critical states.
\newblock {\em Physical review letters}, 118(13):135701, 2017.

\bibitem{Bray1994}
Alan~J Bray.
\newblock Theory of phase-ordering kinetics.
\newblock {\em Adv. Phys.}, 43(3):357--459, 1994.

\bibitem{Biroli2015}
G.~{Biroli}.
\newblock {Slow Relaxations and Non-Equilibrium Dynamics in Classical and
  Quantum Systems}.
\newblock {\em arXiv:1507.05858}, 2015.

\bibitem{Cugliandolo2015}
Leticia~F. Cugliandolo.
\newblock Coarsening phenomena.
\newblock {\em C. R. Phys.}, 16(3):257 -- 266, 2015.

\bibitem{maraga2016linear}
Anna Maraga, Pietro Smacchia, and Alessandro Silva.
\newblock Linear ramps of the mass in the o (n) model: Dynamical transition and
  quantum noise of excitations.
\newblock {\em Physical Review B}, 94(24):245122, 2016.

\bibitem{Barankov2004}
R.~A. Barankov, L.~S. Levitov, and B.~Z. Spivak.
\newblock Collective rabi oscillations and solitons in a time-dependent bcs
  pairing problem.
\newblock {\em Phys. Rev. Lett.}, 93:160401, Oct 2004.

\bibitem{Barankov2006}
R.~A. Barankov and L.~S. Levitov.
\newblock Synchronization in the bcs pairing dynamics as a critical phenomenon.
\newblock {\em Phys. Rev. Lett.}, 96:230403, Jun 2006.

\bibitem{Yuz}
Emil~A. Yuzbashyan and Maxim Dzero.
\newblock Dynamical vanishing of the order parameter in a fermionic condensate.
\newblock {\em Phys. Rev. Lett.}, 96:230404, Jun 2006.

\bibitem{Yuz2}
Emil~A. Yuzbashyan, Oleksandr Tsyplyatyev, and Boris~L. Altshuler.
\newblock Relaxation and persistent oscillations of the order parameter in
  fermionic condensates.
\newblock {\em Phys. Rev. Lett.}, 96:097005, Mar 2006.

\bibitem{Foster2013}
Matthew~S. Foster, Maxim Dzero, Victor Gurarie, and Emil~A. Yuzbashyan.
\newblock Quantum quench in a $p+ip$ superfluid: Winding numbers and
  topological states far from equilibrium.
\newblock {\em Phys. Rev. B}, 88:104511, Sep 2013.

\bibitem{Foster2014}
Matthew~S. Foster, Victor Gurarie, Maxim Dzero, and Emil~A. Yuzbashyan.
\newblock Quench-induced floquet topological $p$-wave superfluids.
\newblock {\em Phys. Rev. Lett.}, 113:076403, Aug 2014.

\bibitem{Alicea2012}
Jason Alicea.
\newblock New directions in the pursuit of majorana fermions in solid state
  systems.
\newblock {\em Reports on Progress in Physics}, 75(7):076501, Jun 2012.

\bibitem{Levin2005}
Qijin Chen, Jelena Stajic, Shina Tan, and K.~Levin.
\newblock Bcs–bec crossover: From high temperature superconductors to
  ultracold superfluids.
\newblock {\em Physics Reports}, 412(1):1--88, 2005.

\bibitem{Matsunaga2013}
Ryusuke Matsunaga, Yuki~I. Hamada, Kazumasa Makise, Yoshinori Uzawa, Hirotaka
  Terai, Zhen Wang, and Ryo Shimano.
\newblock Higgs amplitude mode in the bcs superconductors
  ${\mathrm{nb}}_{1\mathrm{\text{\ensuremath{-}}}x}{\mathrm{ti}}_{x}\mathbf{N}$
  induced by terahertz pulse excitation.
\newblock {\em Phys. Rev. Lett.}, 111:057002, Jul 2013.

\bibitem{Shimano2019}
Ryo Shimano and Naoto Tsuji.
\newblock Higgs mode in superconductors.
\newblock {\em Annual Review of Condensed Matter Physics}, 11(1):103--124,
  2020.

\bibitem{Papenkort2007}
T.~Papenkort, V.~M. Axt, and T.~Kuhn.
\newblock Coherent dynamics and pump-probe spectra of bcs superconductors.
\newblock {\em Phys. Rev. B}, 76:224522, Dec 2007.

\bibitem{Papenkort2008}
T.~Papenkort, T.~Kuhn, and V.~M. Axt.
\newblock Coherent control of the gap dynamics of bcs superconductors in the
  nonadiabatic regime.
\newblock {\em Phys. Rev. B}, 78:132505, Oct 2008.

\bibitem{KrullSchnyder2014}
H.~Krull, D.~Manske, G.~S. Uhrig, and A.~P. Schnyder.
\newblock Signatures of nonadiabatic bcs state dynamics in pump-probe
  conductivity.
\newblock {\em Phys. Rev. B}, 90:014515, Jul 2014.

\bibitem{Chou2017}
Yang-Zhi Chou, Yunxiang Liao, and Matthew~S. Foster.
\newblock Twisting anderson pseudospins with light: Quench dynamics in
  terahertz-pumped bcs superconductors.
\newblock {\em Phys. Rev. B}, 95:104507, Mar 2017.

\bibitem{Papenkort2009}
T~Papenkort, T~Kuhn, and V~M Axt.
\newblock Nonequilibrium dynamics and coherent control of {BCS} superconductors
  driven by ultrashort {THz} pulses.
\newblock {\em Journal of Physics: Conference Series}, 193:012050, Nov 2009.

\bibitem{smale2019}
Scott Smale, Peiru He, Ben~A Olsen, Kenneth~G Jackson, Haille Sharum, Stefan
  Trotzky, Jamir Marino, Ana~Maria Rey, and Joseph~H Thywissen.
\newblock Observation of a transition between dynamical phases in a quantum
  degenerate fermi gas.
\newblock {\em Sci. Adv.}, 5(8):eaax1568, 2019.

\bibitem{Lewisswan2021}
Robert~J. Lewis-Swan, Diego Barberena, Julia R.~K. Cline, Dylan~J. Young,
  James~K. Thompson, and Ana~Maria Rey.
\newblock Cavity-qed quantum simulator of dynamical phases of a
  bardeen-cooper-schrieffer superconductor.
\newblock {\em Phys. Rev. Lett.}, 126:173601, Apr 2021.

\bibitem{Yuzbashyan2015}
E.~A. Yuzbashyan, M.~Dzero, V.~Gurarie, and M.~S. Foster.
\newblock Quantum quench phase diagrams of an $s$-wave bcs-bec condensate.
\newblock {\em Phys. Rev. A}, 91:033628, Mar 2015.

\bibitem{Gurarie2007}
V.~Gurarie and L.~Radzihovsky.
\newblock Resonantly paired fermionic superfluids.
\newblock {\em Annals of Physics}, 322(1):2--119, 2007.
\newblock January Special Issue 2007.

\bibitem{Gurarie2009}
V.~Gurarie.
\newblock Nonequilibrium dynamics of weakly and strongly paired
  superconductors.
\newblock {\em Phys. Rev. Lett.}, 103:075301, Aug 2009.

\bibitem{Dzero2015}
Maxim Dzero, Maxim Khodas, and Alex Levchenko.
\newblock Amplitude modes and dynamic coexistence of competing orders in
  multicomponent superconductors.
\newblock {\em Phys. Rev. B}, 91:214505, Jun 2015.

\bibitem{Liao2015}
Yunxiang Liao and Matthew~S. Foster.
\newblock Spectroscopic probes of isolated nonequilibrium quantum matter:
  Quantum quenches, floquet states, and distribution functions.
\newblock {\em Phys. Rev. A}, 92:053620, Nov 2015.

\bibitem{Richardson1964A}
R.W. Richardson and N.~Sherman.
\newblock Exact eigenstates of the pairing-force hamiltonian.
\newblock {\em Nuclear Physics}, 52:221--238, 1964.

\bibitem{Richardson1964B}
R.W. Richardson and N.~Sherman.
\newblock Pairing models of pb206, pb204 and pb202.
\newblock {\em Nuclear Physics}, 52:253--268, 1964.

\bibitem{Gaudin}
Michel Gaudin.
\newblock {\em The Bethe Wavefunction}.
\newblock Cambridge University Press, 2014.

\bibitem{Dukelsky2004}
J.~Dukelsky, S.~Pittel, and G.~Sierra.
\newblock {Colloquium: Exactly solvable Richardson-Gaudin models for many-body
  quantum systems}.
\newblock {\em Rev. Mod. Phys.}, 76:643--662, 2004.

\bibitem{Richardson2002}
R.W. Richardson.
\newblock New class of solvable and integrable many-body models, 2002.

\bibitem{Skrypnyk2009}
T.~Skrypnyk.
\newblock Non-skew-symmetric classical r-matrices, algebraic bethe ansatz, and
  bardeen–cooper–schrieffer–type integrable systems.
\newblock {\em Journal of Mathematical Physics}, 50(3):033504, 2009.

\bibitem{Ibanez2009}
Miguel Iba\~nez, Jon Links, Germ\'an Sierra, and Shao-You Zhao.
\newblock Exactly solvable pairing model for superconductors with
  ${p}_{x}+i{p}_{y}$-wave symmetry.
\newblock {\em Phys. Rev. B}, 79:180501, May 2009.

\bibitem{Dunning2010}
Clare Dunning, Miguel Iba{\~{n}}ez, Jon Links, Germ{\'{a}}n Sierra, and
  Shao-You Zhao.
\newblock Exact solution of the p $+$ ip pairing hamiltonian and a hierarchy of
  integrable models.
\newblock {\em Journal of Statistical Mechanics: Theory and Experiment},
  2010(08):P08025, Aug 2010.

\bibitem{Ortiz2010}
Stefan M.~A. Rombouts, Jorge Dukelsky, and Gerardo Ortiz.
\newblock Quantum phase diagram of the integrable ${p}_{x}+i{p}_{y}$ fermionic
  superfluid.
\newblock {\em Phys. Rev. B}, 82:224510, Dec 2010.

\bibitem{Yuzbashyan2005Lax}
Emil~A Yuzbashyan, Boris~L Altshuler, Vadim~B Kuznetsov, and Victor~Z Enolskii.
\newblock Solution for the dynamics of the {BCS} and central spin problems.
\newblock {\em Journal of Physics A: Mathematical and General},
  38(36):7831--7849, Aug 2005.

\bibitem{Yuzbashyan2005}
Emil~A. Yuzbashyan, Boris~L. Altshuler, Vadim~B. Kuznetsov, and Victor~Z.
  Enolskii.
\newblock Nonequilibrium cooper pairing in the nonadiabatic regime.
\newblock {\em Phys. Rev. B}, 72:220503, Dec 2005.

\bibitem{Barankov2006Replace}
R.~A. Barankov and L.~S. Levitov.
\newblock Excitation of the dissipationless higgs mode in a fermionic
  condensate, 2007.

\bibitem{YuzbashyanTsyplyatyev2009}
Emil~A. Yuzbashyan and Oleksandr Tsyplyatyev.
\newblock Dynamics of emergent cooper pairing at finite temperatures.
\newblock {\em Phys. Rev. B}, 79:132504, Apr 2009.

\bibitem{YuzbashyanDzero2009}
M.~Dzero, E.~A. Yuzbashyan, and B.~L. Altshuler.
\newblock Cooper pair turbulence in atomic fermi gases.
\newblock {\em {EPL} (Europhysics Letters)}, 85(2):20004, jan 2009.

\bibitem{Yuzbashyan2019}
Jasen~A. Scaramazza, Pietro Smacchia, and Emil~A. Yuzbashyan.
\newblock Consequences of integrability breaking in quench dynamics of pairing
  hamiltonians.
\newblock {\em Phys. Rev. B}, 99:054520, Feb 2019.

\bibitem{Mitra2017}
Yonah Lemonik and Aditi Mitra.
\newblock Time-resolved spectral density of interacting fermions following a
  quench to a superconducting critical point.
\newblock {\em Phys. Rev. B}, 96:104506, Sep 2017.

\bibitem{Mitra2018A}
Yonah Lemonik and Aditi Mitra.
\newblock Model predictions for time-resolved transport measurements made near
  the superfluid critical points of cold atoms and
  ${\mathrm{k}}_{3}{\mathrm{c}}_{60}$ films.
\newblock {\em Phys. Rev. Lett.}, 121:067001, Aug 2018.

\bibitem{Mitra2018B}
Yonah Lemonik and Aditi Mitra.
\newblock Quench dynamics of superconducting fluctuations and optical
  conductivity in a disordered system.
\newblock {\em Phys. Rev. B}, 98:214514, Dec 2018.

\bibitem{Mitra2019}
Yonah Lemonik and Aditi Mitra.
\newblock Transport and spectral signatures of transient fluctuating
  superfluids in the absence of long-range order.
\newblock {\em Phys. Rev. B}, 100:094503, Sep 2019.

\bibitem{Oka2019}
Takashi Oka and Sota Kitamura.
\newblock Floquet engineering of quantum materials.
\newblock {\em Annual Review of Condensed Matter Physics}, 10(1):387--408,
  2019.

\bibitem{QiZhang2011}
Xiao-Liang Qi and Shou-Cheng Zhang.
\newblock Topological insulators and superconductors.
\newblock {\em Rev. Mod. Phys.}, 83:1057--1110, Oct 2011.

\bibitem{Rudner2020}
Fenner Harper, Rahul Roy, Mark~S. Rudner, and S.L. Sondhi.
\newblock Topology and broken symmetry in floquet systems.
\newblock {\em Annual Review of Condensed Matter Physics}, 11(1):345--368,
  2020.

\bibitem{Schrieffer}
J.R. Schrieffer.
\newblock {\em Theory Of Superconductivity}.
\newblock CRC Press, 2018.

\bibitem{Anderson1958}
P.~W. Anderson.
\newblock Random-phase approximation in the theory of superconductivity.
\newblock {\em Phys. Rev.}, 112:1900, 1958.

\bibitem{Note1}
Instead of an $N$-fold spin product (spin coherent) state, one can also
  consider a state with $P \leq N$ ``blocked'' levels. These are states that
  possess a single fermion occupying one state of a Cooper pair. For the
  reduced BCS Hamiltonian in Eq.~(\ref {HRed}), blocked levels completely
  decouple from the time-evolution of the pseudospins (which are superpositions
  of doubly empty and occupied states of a Cooper pair).

\bibitem{Altman2004}
Ehud Altman, Eugene Demler, and Mikhail~D. Lukin.
\newblock Probing many-body states of ultracold atoms via noise correlations.
\newblock {\em Phys. Rev. A}, 70:013603, Jul 2004.

\bibitem{Folling2005}
Simon Fölling, Fabrice Gerbier, Artur Widera, Olaf Mandel, Tatjana Gericke,
  and Immanuel Bloch.
\newblock Spatial quantum noise interferometry in expanding ultracold atom
  clouds.
\newblock {\em Nature}, 434(7032):481--484, 2005.

\bibitem{Rom2006}
T.~Rom, Th~Best, D.~van Oosten, U.~Schneider, S.~Fölling, B.~Paredes, and
  I.~Bloch.
\newblock Free fermion antibunching in a degenerate atomic fermi gas released
  from an optical lattice.
\newblock {\em Nature}, 444(7120):733--736, 2006.

\bibitem{Stahl2019}
Christopher Stahl and Martin Eckstein.
\newblock Noise correlations in time- and angle-resolved photoemission
  spectroscopy.
\newblock {\em Phys. Rev. B}, 99:241111, Jun 2019.

\bibitem{VolkovKogan1973}
A.~F. Volkov and Sh.~M. Kogan.
\newblock Collisionless relaxation of the energy gap in superconductors.
\newblock {\em JETP}, 38:1018, 1974.
\newblock [Russian original - Zh. Eksp. Teor. Fiz. {\bf 65}, 2038 (1973)].

\bibitem{Peronaci2015}
Francesco Peronaci, Marco Schir\'o, and Massimo Capone.
\newblock Transient dynamics of $d$-wave superconductors after a sudden
  excitation.
\newblock {\em Phys. Rev. Lett.}, 115:257001, Dec 2015.

\bibitem{ZhangGurarieFoster}
X.~Zhang, V.~Gurarie, and M.~S. Foster.
\newblock {\em unpublished}.

\bibitem{Dzero2007}
M.~Dzero, E.~A. Yuzbashyan, B.~L. Altshuler, and P.~Coleman.
\newblock Spectroscopic signatures of nonequilibrium pairing in atomic fermi
  gases.
\newblock {\em Phys. Rev. Lett.}, 99:160402, Oct 2007.

\bibitem{Millis2017}
D.~M. Kennes, E.~Y. Wilner, D.~R. Reichman, and A.~J. Millis.
\newblock Nonequilibrium optical conductivity: General theory and application
  to transient phases.
\newblock {\em Phys. Rev. B}, 96:054506, Aug 2017.

\bibitem{Yuzbashyan2008}
Emil~A. Yuzbashyan.
\newblock Normal and anomalous solitons in the theory of dynamical cooper
  pairing.
\newblock {\em Phys. Rev. B}, 78:184507, Nov 2008.

\bibitem{Perakis2020}
M.~Mootz, J.~Wang, and I.~E. Perakis.
\newblock Lightwave terahertz quantum manipulation of nonequilibrium
  superconductor phases and their collective modes.
\newblock {\em Phys. Rev. B}, 102:054517, Aug 2020.

\bibitem{Chern2019}
Gia-Wei Chern and Kipton Barros.
\newblock Nonequilibrium dynamics of superconductivity in the attractive
  hubbard model.
\newblock {\em Phys. Rev. B}, 99:035162, Jan 2019.

\bibitem{Salomon2004}
J.~Zhang, E.~G.~M. van Kempen, T.~Bourdel, L.~Khaykovich, J.~Cubizolles,
  F.~Chevy, M.~Teichmann, L.~Tarruell, S.~J. J. M.~F. Kokkelmans, and
  C.~Salomon.
\newblock $p$-wave feshbach resonances of ultracold $^{6}\mathrm{Li}$.
\newblock {\em Phys. Rev. A}, 70:030702, Sep 2004.

\bibitem{Castin2008}
M.~Jona-Lasinio, L.~Pricoupenko, and Y.~Castin.
\newblock Three fully polarized fermions close to a $\mathit{p}$-wave feshbach
  resonance.
\newblock {\em Phys. Rev. A}, 77:043611, Apr 2008.

\bibitem{Gurarie2008}
J.~Levinsen, N.~R. Cooper, and V.~Gurarie.
\newblock Stability of fermionic gases close to a $p$-wave feshbach resonance.
\newblock {\em Phys. Rev. A}, 78:063616, Dec 2008.

\bibitem{Tinkham}
M.~Tinkham.
\newblock {\em Introduction to Superconductivity: Second Edition}.
\newblock Dover Books on Physics. Dover Publications, 2004.

\bibitem{Nayak2008}
Chetan Nayak, Steven~H. Simon, Ady Stern, Michael Freedman, and Sankar
  Das~Sarma.
\newblock Non-abelian anyons and topological quantum computation.
\newblock {\em Rev. Mod. Phys.}, 80:1083--1159, Sep 2008.

\bibitem{Volovik}
G.E Volovik.
\newblock {\em The Universe in a Helium Droplet}.
\newblock International Series of Monographs on Physics. Clarendon Press, 2003.

\bibitem{Mizushima2016}
Takeshi Mizushima, Yasumasa Tsutsumi, Takuto Kawakami, Masatoshi Sato, Masanori
  Ichioka, and Kazushige Machida.
\newblock Symmetry-protected topological superfluids and superconductors
  —from the basics to 3he—.
\newblock {\em Journal of the Physical Society of Japan}, 85(2):022001, 2016.

\bibitem{Rigol2015}
Luca D’Alessio and Marcos Rigol.
\newblock Dynamical preparation of floquet chern insulators.
\newblock {\em Nature Communications}, 6:8336, Oct 2015.

\bibitem{Bhaseen2015}
M.~D. Caio, N.~R. Cooper, and M.~J. Bhaseen.
\newblock Quantum quenches in chern insulators.
\newblock {\em Phys. Rev. Lett.}, 115:236403, Dec 2015.

\bibitem{Martin-Delgado2010}
A~Bermudez, L~Amico, and M~A Martin-Delgado.
\newblock Dynamical delocalization of majorana edge states by sweeping across a
  quantum critical point.
\newblock {\em New Journal of Physics}, 12(5):055014, may 2010.

\bibitem{Perfetto2013}
E.~Perfetto.
\newblock Dynamical formation and manipulation of majorana fermions in driven
  quantum wires in contact with a superconductor.
\newblock {\em Phys. Rev. Lett.}, 110:087001, Feb 2013.

\bibitem{Sacramento2014}
P.~D. Sacramento.
\newblock Fate of majorana fermions and chern numbers after a quantum quench.
\newblock {\em Phys. Rev. E}, 90:032138, Sep 2014.

\bibitem{Vishveshwara2014}
G.~Kells, D.~Sen, J.~K. Slingerland, and S.~Vishveshwara.
\newblock Topological blocking in quantum quench dynamics.
\newblock {\em Phys. Rev. B}, 89:235130, Jun 2014.

\bibitem{Refael2016}
Justin~H. Wilson, Justin C.~W. Song, and Gil Refael.
\newblock Remnant geometric hall response in a quantum quench.
\newblock {\em Phys. Rev. Lett.}, 117:235302, Nov 2016.

\bibitem{Budich2016}
Ying Hu, Peter Zoller, and Jan~Carl Budich.
\newblock Dynamical buildup of a quantized hall response from nontopological
  states.
\newblock {\em Phys. Rev. Lett.}, 117:126803, Sep 2016.

\bibitem{Pan2018}
Wei Sun, Chang-Rui Yi, Bao-Zong Wang, Wei-Wei Zhang, Barry~C. Sanders,
  Xiao-Tian Xu, Zong-Yao Wang, Joerg Schmiedmayer, Youjin Deng, Xiong-Jun Liu,
  Shuai Chen, and Jian-Wei Pan.
\newblock Uncover topology by quantum quench dynamics.
\newblock {\em Phys. Rev. Lett.}, 121:250403, Dec 2018.

\bibitem{CooperRMP2019}
N.~R. Cooper, J.~Dalibard, and I.~B. Spielman.
\newblock Topological bands for ultracold atoms.
\newblock {\em Rev. Mod. Phys.}, 91:015005, Mar 2019.

\bibitem{Iyer2020}
Deepak Iyer and Matthew~S. Foster.
\newblock Topological quantum control: Edge currents via floquet depinning of
  skyrmions in the $\ensuremath{\nu}=0$ graphene quantum hall antiferromagnet.
\newblock {\em Phys. Rev. B}, 101:241403, Jun 2020.

\bibitem{Pu2015}
Ying Dong, Lin Dong, Ming Gong, and Han Pu.
\newblock Dynamical phases in quenched spin–orbit-coupled degenerate fermi
  gas.
\newblock {\em Nature Communications}, 6:6103, Jan 2015.

\bibitem{Dzero2015B}
Maxim Dzero, Ammar~A. Kirmani, and Emil~A. Yuzbashyan.
\newblock Nonadiabatic dynamics of superfluid spin-orbit-coupled degenerate
  fermi gas.
\newblock {\em Phys. Rev. A}, 92:053626, Nov 2015.

\bibitem{Gritsev2019}
Eyzo Stouten, Pieter~W. Claeys, Jean-S\'ebastien Caux, and Vladimir Gritsev.
\newblock Integrability and duality in spin chains.
\newblock {\em Phys. Rev. B}, 99:075111, Feb 2019.

\bibitem{Yuzbashyan2021}
Aidan Zabalo and Emil~A. Yuzbashyan.
\newblock Time reversal symmetry protected chaotic fixed point in the quench
  dynamics of a topological $p$-wave superfluid, 2021.

\bibitem{Polkovnikov2011}
Anatoli Polkovnikov, Krishnendu Sengupta, Alessandro Silva, and Mukund
  Vengalattore.
\newblock Colloquium: Nonequilibrium dynamics of closed interacting quantum
  systems.
\newblock {\em Rev. Mod. Phys.}, 83:863--883, Aug 2011.

\bibitem{Metzner1998}
Walter Metzner and Dieter Vollhardt.
\newblock Correlated lattice fermions in $d=\ensuremath{\infty}$ dimensions.
\newblock {\em Phys. Rev. Lett.}, 62:324--327, Jan 1989.

\bibitem{Alet2018}
Fabien Alet and Nicolas Laflorencie.
\newblock Many-body localization: An introduction and selected topics.
\newblock {\em Comptes Rendus Physique}, 19(6):498--525, 2018.

\bibitem{Nandkishore2015}
Rahul Nandkishore and David~A. Huse.
\newblock Many-body localization and thermalization in quantum statistical
  mechanics.
\newblock {\em Annual Review of Condensed Matter Physics}, 6(1):15--38, 2015.

\bibitem{Abanin2019}
Dmitry~A. Abanin, Ehud Altman, Immanuel Bloch, and Maksym Serbyn.
\newblock Colloquium: Many-body localization, thermalization, and entanglement.
\newblock {\em Rev. Mod. Phys.}, 91:021001, May 2019.

\bibitem{Parameswaran2017}
S.~A. Parameswaran, A.~C. Potter, and R.~Vasseur.
\newblock Eigenstate phase transitions and the emergence of universal dynamics
  in highly excited states.
\newblock {\em Ann. Phys. (Berlin)}, 529:1600302, 2017.

\bibitem{Carleo2012}
Giuseppe Carleo, Federico Becca, Marco Schir{\'o}, and Michele Fabrizio.
\newblock Localization and glassy dynamics of many-body quantum systems.
\newblock {\em Scientific Reports}, 2(1):243, 2012.

\bibitem{Smith2017a}
A.~Smith, J.~Knolle, R.~Moessner, and D.~L. Kovrizhin.
\newblock Absence of ergodicity without quenched disorder: From quantum
  disentangled liquids to many-body localization.
\newblock {\em Phys. Rev. Lett.}, 119:176601, Oct 2017.

\bibitem{Smith2017b}
A.~Smith, J.~Knolle, D.~L. Kovrizhin, and R.~Moessner.
\newblock Disorder-free localization.
\newblock {\em Phys. Rev. Lett.}, 118:266601, Jun 2017.

\bibitem{Yao2016}
N.~Y. Yao, C.~R. Laumann, J.~I. Cirac, M.~D. Lukin, and J.~E. Moore.
\newblock Quasi-many-body localization in translation-invariant systems.
\newblock {\em Phys. Rev. Lett.}, 117:240601, Dec 2016.

\bibitem{Michailidis2018}
Alexios~A. Michailidis, Marko \ifmmode \check{Z}\else
  \v{Z}\fi{}nidari\ifmmode~\check{c}\else \v{c}\fi{}, Mariya Medvedyeva,
  Dmitry~A. Abanin, Toma\ifmmode \check{z}\else~\v{z}\fi{} Prosen, and
  Z.~Papi\ifmmode~\acute{c}\else \'{c}\fi{}.
\newblock Slow dynamics in translation-invariant quantum lattice models.
\newblock {\em Phys. Rev. B}, 97:104307, Mar 2018.

\bibitem{Lan2018}
Zhihao Lan, Merlijn van Horssen, Stephen Powell, and Juan~P. Garrahan.
\newblock Quantum slow relaxation and metastability due to dynamical
  constraints.
\newblock {\em Phys. Rev. Lett.}, 121:040603, Jul 2018.

\bibitem{Horssen2015}
Merlijn van Horssen, Emanuele Levi, and Juan~P. Garrahan.
\newblock Dynamics of many-body localization in a translation-invariant quantum
  glass model.
\newblock {\em Phys. Rev. B}, 92:100305, Sep 2015.

\bibitem{Scherg2020}
Sebastian Scherg, Thomas Kohlert, Pablo Sala, Frank Pollmann, Bharath
  Hebbe~Madhusudhana, Immanuel Bloch, and Monika Aidelsburger.
\newblock Observing non-ergodicity due to kinetic constraints in tilted
  fermi-hubbard chains.
\newblock {\em Nature Communications}, 12(1):4490, 2021.

\bibitem{Iadecola2019}
Thomas Iadecola and Marko \ifmmode \check{Z}\else
  \v{Z}\fi{}nidari\ifmmode~\check{c}\else \v{c}\fi{}.
\newblock Exact localized and ballistic eigenstates in disordered chaotic spin
  ladders and the fermi-hubbard model.
\newblock {\em Phys. Rev. Lett.}, 123:036403, Jul 2019.

\bibitem{Vafek2017}
Oskar Vafek, Nicolas Regnault, and B.~Andrei Bernevig.
\newblock {Entanglement of Exact Excited Eigenstates of the Hubbard Model in
  Arbitrary Dimension}.
\newblock {\em SciPost Phys.}, 3:043, 2017.

\bibitem{Turner2018}
C.~J. Turner, A.~A. Michailidis, D.~A. Abanin, M.~Serbyn, and Z.~Papi{\'c}.
\newblock Weak ergodicity breaking from quantum many-body scars.
\newblock {\em Nature Physics}, 14(7):745--749, 2018.

\bibitem{Choi2019}
Soonwon Choi, Christopher~J. Turner, Hannes Pichler, Wen~Wei Ho, Alexios~A.
  Michailidis, Zlatko Papi\ifmmode~\acute{c}\else \'{c}\fi{}, Maksym Serbyn,
  Mikhail~D. Lukin, and Dmitry~A. Abanin.
\newblock Emergent su(2) dynamics and perfect quantum many-body scars.
\newblock {\em Phys. Rev. Lett.}, 122:220603, Jun 2019.

\bibitem{Ho2019}
Wen~Wei Ho, Soonwon Choi, Hannes Pichler, and Mikhail~D. Lukin.
\newblock Periodic orbits, entanglement, and quantum many-body scars in
  constrained models: Matrix product state approach.
\newblock {\em Phys. Rev. Lett.}, 122:040603, Jan 2019.

\bibitem{MuellerHartmann1989}
E.~M{\"u}ller-Hartmann.
\newblock Correlated fermions on a lattice in high dimensions.
\newblock {\em Zeitschrift f{\"u}r Physik B Condensed Matter}, 74(4):507--512,
  1989.

\bibitem{Georges1996}
Antoine Georges, Gabriel Kotliar, Werner Krauth, and Marcelo~J. Rozenberg.
\newblock Dynamical mean-field theory of strongly correlated fermion systems
  and the limit of infinite dimensions.
\newblock {\em Rev. Mod. Phys.}, 68:13--125, Jan 1996.

\bibitem{Georges1992}
Antoine Georges and Gabriel Kotliar.
\newblock Hubbard model in infinite dimensions.
\newblock {\em Phys. Rev. B}, 45:6479--6483, Mar 1992.

\bibitem{Schmidt2002}
Schmidt P. and Monien H.
\newblock Nonequilibrium dynamical mean-field theory of a strongly correlated
  system.

\bibitem{Freericks2006}
J.~K. Freericks, V.~M. Turkowski, and V.~Zlati\ifmmode~\acute{c}\else
  \'{c}\fi{}.
\newblock Nonequilibrium dynamical mean-field theory.
\newblock {\em Phys. Rev. Lett.}, 97:266408, Dec 2006.

\bibitem{Aoki2014}
Hideo Aoki, Naoto Tsuji, Martin Eckstein, Marcus Kollar, Takashi Oka, and
  Philipp Werner.
\newblock Nonequilibrium dynamical mean-field theory and its applications.
\newblock {\em Rev. Mod. Phys.}, 86:779--837, Jun 2014.

\bibitem{Arrigoni2013}
Enrico Arrigoni, Michael Knap, and Wolfgang von~der Linden.
\newblock Nonequilibrium dynamical mean-field theory: An auxiliary quantum
  master equation approach.
\newblock {\em Phys. Rev. Lett.}, 110:086403, Feb 2013.

\bibitem{Joura2008}
A.~V. Joura, J.~K. Freericks, and Th. Pruschke.
\newblock Steady-state nonequilibrium density of states of driven strongly
  correlated lattice models in infinite dimensions.
\newblock {\em Phys. Rev. Lett.}, 101:196401, Nov 2008.

\bibitem{Scarlatella2021}
Orazio Scarlatella, Aashish~A. Clerk, Rosario Fazio, and Marco Schir\'o.
\newblock Dynamical mean-field theory for markovian open quantum many-body
  systems.
\newblock {\em Phys. Rev. X}, 11:031018, Jul 2021.

\bibitem{Tsuji2008}
Naoto Tsuji, Takashi Oka, and Hideo Aoki.
\newblock Correlated electron systems periodically driven out of equilibrium:
  $\text{Floquet}+\text{DMFT}$ formalism.
\newblock {\em Phys. Rev. B}, 78:235124, Dec 2008.

\bibitem{Oka2009}
Philipp Werner, Takashi Oka, and Andrew~J. Millis.
\newblock Diagrammatic monte carlo simulation of nonequilibrium systems.
\newblock {\em Phys. Rev. B}, 79:035320, Jan 2009.

\bibitem{Wolf2014}
F.~Alexander Wolf, Ian~P. McCulloch, and Ulrich Schollw\"ock.
\newblock Solving nonequilibrium dynamical mean-field theory using matrix
  product states.
\newblock {\em Phys. Rev. B}, 90:235131, Dec 2014.

\bibitem{Tsuji2013a}
Naoto Tsuji and Philipp Werner.
\newblock Nonequilibrium dynamical mean-field theory based on weak-coupling
  perturbation expansions: Application to dynamical symmetry breaking in the
  hubbard model.
\newblock {\em Phys. Rev. B}, 88:165115, Oct 2013.

\bibitem{Eckstein2010nca}
Martin Eckstein and Philipp Werner.
\newblock Nonequilibrium dynamical mean-field calculations based on the
  noncrossing approximation and its generalizations.
\newblock {\em Phys. Rev. B}, 82:115115, Sep 2010.

\bibitem{Tsuji2013}
Naoto Tsuji, Martin Eckstein, and Philipp Werner.
\newblock Nonthermal antiferromagnetic order and nonequilibrium criticality in
  the hubbard model.
\newblock {\em Phys. Rev. Lett.}, 110:136404, Mar 2013.

\bibitem{Sandri2013}
Matteo Sandri and Michele Fabrizio.
\newblock Nonequilibrium dynamics in the antiferromagnetic hubbard model.
\newblock {\em Phys. Rev. B}, 88:165113, Oct 2013.

\bibitem{Berges2008}
J\"urgen Berges, Alexander Rothkopf, and Jonas Schmidt.
\newblock Nonthermal fixed points: Effective weak coupling for strongly
  correlated systems far from equilibrium.
\newblock {\em Phys. Rev. Lett.}, 101:041603, Jul 2008.

\bibitem{Nowak2014}
Boris Nowak, Jan Schole, and Thomas Gasenzer.
\newblock Universal dynamics on the way to thermalization.
\newblock {\em New Journal of Physics}, 16(9):093052, 2014.

\bibitem{Oberthaler2018}
Maximilian Prüfer, Philipp Kunkel, Helmut Strobel, Stefan Lannig, Daniel
  Linnemann, Christian-Marcel Schmied, Jürgen Berges, Thomas Gasenzer, and
  Markus~K. Oberthaler.
\newblock Observation of universal dynamics in a spinor bose gas far from
  equilibrium.
\newblock {\em Nature}, 563(7730):217--220, 2018.

\bibitem{Schmiedmayer2018}
Sebastian Erne, Robert Bücker, Thomas Gasenzer, Jürgen Berges, and Jörg
  Schmiedmayer.
\newblock Universal dynamics in an isolated one-dimensional bose gas far from
  equilibrium.
\newblock {\em Nature}, 563(7730):225--229, 2018.

\bibitem{Picano2021}
Antonio Picano and Martin Eckstein.
\newblock Accelerated gap collapse in a slater antiferromagnet.
\newblock {\em Phys. Rev. B}, 103:165118, Apr 2021.

\bibitem{Stark2013}
M.~{Stark} and M.~{Kollar}.
\newblock {Kinetic description of thermalization dynamics in weakly interacting
  quantum systems}.
\newblock {\em arXiv:1308.1610}, 2013.

\bibitem{DAlessio2016}
Luca D'Alessio, Yariv Kafri, Anatoli Polkovnikov, and Marcos Rigol.
\newblock From quantum chaos and eigenstate thermalization to statistical
  mechanics and thermodynamics.
\newblock {\em Adv. Phys.}, 65(3):239--362, 2016.

\bibitem{Mallayya2019}
Krishnanand Mallayya, Marcos Rigol, and Wojciech De~Roeck.
\newblock Prethermalization and thermalization in isolated quantum systems.
\newblock {\em Phys. Rev. X}, 9:021027, May 2019.

\bibitem{Balzer2015}
Karsten Balzer, F.~Alexander Wolf, Ian~P. McCulloch, Philipp Werner, and Martin
  Eckstein.
\newblock Nonthermal melting of n\'eel order in the hubbard model.
\newblock {\em Phys. Rev. X}, 5:031039, Sep 2015.

\bibitem{Trotzky2012}
S.~Trotzky, Y-A. Chen, A.~Flesch, I.~P. McCulloch, U.~Schollw{\"o}ck,
  J.~Eisert, and I.~Bloch.
\newblock Probing the relaxation towards equilibrium in an isolated strongly
  correlated one-dimensional bose gas.
\newblock {\em Nature Physics}, 8(4):325--330, 2012.

\bibitem{Kollath2007}
Corinna Kollath, Andreas~M. L\"auchli, and Ehud Altman.
\newblock Quench dynamics and nonequilibrium phase diagram of the bose-hubbard
  model.
\newblock {\em Phys. Rev. Lett.}, 98:180601, Apr 2007.

\bibitem{Rosch2008}
Achim Rosch, David Rasch, Benedikt Binz, and Matthias Vojta.
\newblock Metastable superfluidity of repulsive fermionic atoms in optical
  lattices.
\newblock {\em Phys. Rev. Lett.}, 101:265301, Dec 2008.

\bibitem{Eckstein2011}
Martin Eckstein and Philipp Werner.
\newblock Thermalization of a pump-excited mott insulator.
\newblock {\em Phys. Rev. B}, 84:035122, Jul 2011.

\bibitem{Sensarma2010}
Rajdeep Sensarma, David Pekker, Ehud Altman, Eugene Demler, Niels Strohmaier,
  Daniel Greif, Robert J\"ordens, Leticia Tarruell, Henning Moritz, and Tilman
  Esslinger.
\newblock Lifetime of double occupancies in the fermi-hubbard model.
\newblock {\em Phys. Rev. B}, 82:224302, Dec 2010.

\bibitem{Strohmaier2010}
Niels Strohmaier, Daniel Greif, Robert J\"ordens, Leticia Tarruell, Henning
  Moritz, Tilman Esslinger, Rajdeep Sensarma, David Pekker, Ehud Altman, and
  Eugene Demler.
\newblock Observation of elastic doublon decay in the fermi-hubbard model.
\newblock {\em Phys. Rev. Lett.}, 104:080401, Feb 2010.

\bibitem{Morong2021}
W.~Morong, S.~R. Muleady, I.~Kimchi, W.~Xu, R.~M. Nandkishore, A.~M. Rey, and
  B.~DeMarco.
\newblock Disorder-controlled relaxation in a three-dimensional hubbard model
  quantum simulator.
\newblock {\em Phys. Rev. Research}, 3:L012009, Jan 2021.

\bibitem{Yang1989}
Chen~Ning Yang.
\newblock \ensuremath{\eta} pairing and off-diagonal long-range order in a
  hubbard model.
\newblock {\em Phys. Rev. Lett.}, 63:2144--2147, Nov 1989.

\bibitem{Li2020}
Jiajun Li and Martin Eckstein.
\newblock Nonequilibrium steady-state theory of photodoped mott insulators,
  2020.

\bibitem{Murakami2021}
Yuta Murakami, Shintaro Takayoshi, Tatsuya Kaneko, Zhiyuan Sun, Denis
  Gole\'{z}, Andrew~J. Millis, and Philipp Werner.
\newblock Emergent nonequilibrium phases in the photo-doped one-dimensional
  mott insulator, 2021.

\bibitem{Kaneko2019}
Tatsuya Kaneko, Tomonori Shirakawa, Sandro Sorella, and Seiji Yunoki.
\newblock Photoinduced $\ensuremath{\eta}$ pairing in the hubbard model.
\newblock {\em Phys. Rev. Lett.}, 122:077002, Feb 2019.

\bibitem{Peronaci2020}
Francesco Peronaci, Olivier Parcollet, and Marco Schir\'o.
\newblock Enhancement of local pairing correlations in periodically driven mott
  insulators.
\newblock {\em Phys. Rev. B}, 101:161101, Apr 2020.

\bibitem{Tindall2020}
J.~Tindall, F.~Schlawin, M.~Buzzi, D.~Nicoletti, J.~R. Coulthard, H.~Gao,
  A.~Cavalleri, M.~A. Sentef, and D.~Jaksch.
\newblock Dynamical order and superconductivity in a frustrated many-body
  system.
\newblock {\em Phys. Rev. Lett.}, 125:137001, Sep 2020.

\bibitem{Werner2012}
Philipp Werner, Naoto Tsuji, and Martin Eckstein.
\newblock Nonthermal symmetry-broken states in the strongly interacting hubbard
  model.
\newblock {\em Phys. Rev. B}, 86:205101, Nov 2012.

\bibitem{Eckstein2009}
Martin Eckstein, Marcus Kollar, and Philipp Werner.
\newblock Thermalization after an interaction quench in the hubbard model.
\newblock {\em Phys. Rev. Lett.}, 103:056403, Jul 2009.

\bibitem{Uhrig2009}
G\"otz~S. Uhrig.
\newblock Interaction quenches of fermi gases.
\newblock {\em Phys. Rev. A}, 80:061602, Dec 2009.

\bibitem{Schiro2010}
Marco Schir\'o and Michele Fabrizio.
\newblock Time-dependent mean field theory for quench dynamics in correlated
  electron systems.
\newblock {\em Phys. Rev. Lett.}, 105:076401, Aug 2010.

\bibitem{Eckstein2010b}
Martin Eckstein, Marcus Kollar, and Philipp Werner.
\newblock Interaction quench in the hubbard model: Relaxation of the spectral
  function and the optical conductivity.
\newblock {\em Phys. Rev. B}, 81:115131, Mar 2010.

\bibitem{Gunnarsson2017}
O.~Gunnarsson, G.~Rohringer, T.~Sch\"afer, G.~Sangiovanni, and A.~Toschi.
\newblock Breakdown of traditional many-body theories for correlated electrons.
\newblock {\em Phys. Rev. Lett.}, 119:056402, Aug 2017.

\bibitem{Schaefer2013}
T.~Sch\"afer, G.~Rohringer, O.~Gunnarsson, S.~Ciuchi, G.~Sangiovanni, and
  A.~Toschi.
\newblock Divergent precursors of the mott-hubbard transition at the
  two-particle level.
\newblock {\em Phys. Rev. Lett.}, 110:246405, Jun 2013.

\bibitem{Greiner2002}
Markus Greiner, Olaf Mandel, Theodor~W. H{\"a}nsch, and Immanuel Bloch.
\newblock Collapse and revival of the matter wave field of a bose--einstein
  condensate.
\newblock {\em Nature}, 419(6902):51--54, 2002.

\bibitem{Will2010}
Sebastian Will, Thorsten Best, Ulrich Schneider, Lucia Hackermüller,
  Dirk-Sören Lühmann, and Immanuel Bloch.
\newblock Time-resolved observation of coherent multi-body interactions in
  quantum phase revivals.
\newblock {\em Nature}, 465(7295):197--201, 2010.

\bibitem{Schiro2011}
Marco Schir\'o and Michele Fabrizio.
\newblock Quantum quenches in the hubbard model: Time-dependent mean-field
  theory and the role of quantum fluctuations.
\newblock {\em Phys. Rev. B}, 83:165105, Apr 2011.

\bibitem{Sandri2012}
Matteo Sandri, Marco Schir\'o, and Michele Fabrizio.
\newblock Linear ramps of interaction in the fermionic hubbard model.
\newblock {\em Phys. Rev. B}, 86:075122, Aug 2012.

\bibitem{Hofmann2016}
Felix Hofmann, Martin Eckstein, and Michael Potthoff.
\newblock Nonequilibrium self-energy functional approach to the dynamical mott
  transition.
\newblock {\em Phys. Rev. B}, 93:235104, Jun 2016.

\bibitem{Behrmann2013}
Malte Behrmann, Michele Fabrizio, and Frank Lechermann.
\newblock Extended dynamic mott transition in the two-band hubbard model out of
  equilibrium.
\newblock {\em Phys. Rev. B}, 88:035116, Jul 2013.

\bibitem{Hamerla2013}
Simone~A. Hamerla and G\"otz~S. Uhrig.
\newblock Dynamical transition in interaction quenches of the one-dimensional
  hubbard model.
\newblock {\em Phys. Rev. B}, 87:064304, Feb 2013.

\bibitem{Tsuji2014}
Naoto Tsuji, Peter Barmettler, Hideo Aoki, and Philipp Werner.
\newblock Nonequilibrium dynamical cluster theory.
\newblock {\em Phys. Rev. B}, 90:075117, Aug 2014.

\bibitem{Sciolla20102}
Bruno Sciolla and Giulio Biroli.
\newblock Quantum quenches and off-equilibrium dynamical transition in the
  infinite-dimensional bose-hubbard model.
\newblock {\em Phys. Rev. Lett.}, 105:220401, Nov 2010.

\bibitem{Strand2015}
Hugo U.~R. Strand, Martin Eckstein, and Philipp Werner.
\newblock Nonequilibrium dynamical mean-field theory for bosonic lattice
  models.
\newblock {\em Phys. Rev. X}, 5:011038, Mar 2015.

\bibitem{Brandt1989}
U.~Brandt and C.~Mielsch.
\newblock Thermodynamics and correlation functions of the falicov-kimball model
  in large dimensions.
\newblock {\em Zeitschrift f{\"u}r Physik B Condensed Matter}, 75(3):365--370,
  1989.

\bibitem{Freericks2003}
J.~K. Freericks and V.~Zlati\ifmmode~\acute{c}\else \'{c}\fi{}.
\newblock Exact dynamical mean-field theory of the falicov-kimball model.
\newblock {\em Rev. Mod. Phys.}, 75:1333--1382, Oct 2003.

\bibitem{Eckstein2008}
Martin Eckstein and Marcus Kollar.
\newblock Nonthermal steady states after an interaction quench in the
  falicov-kimball model.
\newblock {\em Phys. Rev. Lett.}, 100:120404, Mar 2008.

\bibitem{Eckstein2011b}
Martin Eckstein and Philipp Werner.
\newblock Damping of bloch oscillations in the hubbard model.
\newblock {\em Phys. Rev. Lett.}, 107:186406, Oct 2011.

\bibitem{Fotso2014}
H.~Fotso, K.~Mikelsons, and J.~K. Freericks.
\newblock Thermalization of field driven quantum systems.
\newblock {\em Scientific Reports}, 4(1):4699, 2014.

\bibitem{Galaiko1972}
V.P. Galaiko.
\newblock Kinetic equations for relaxation processes in superconductors.
\newblock {\em JETP}, 34:203, 1972.
\newblock [Russian original - Zh. Eksp. Teor. Fiz. {\bf 61}, 382 (1971)].

\bibitem{Wang2018}
X.~Yang, C.~Vaswani, C.~Sundahl, M.~Mootz, P.~Gagel, L.~Luo, J.~H. Kang, P.~P.
  Orth, I.~E. Perakis, C.~B. Eom, and J.~Wang.
\newblock Terahertz-light quantum tuning of a metastable emergent phase hidden
  by superconductivity.
\newblock {\em Nature Materials}, 17:586--591, 2018.

\bibitem{Wang2019-A}
X.~Yang, C.~Vaswani, C.~Sundahl, M.~Mootz, L.~Luo, J.~H. Kang, I.~E. Perakis,
  C.~B. Eom, and J.~Wang.
\newblock Lightwave-driven gapless superconductivity and forbidden quantum
  beats by terahertz symmetry breaking.
\newblock {\em Nature Photonics}, 13:707--713, 2019.

\bibitem{Orth2019}
Tianbai Cui, Xu~Yang, Chirag Vaswani, Jigang Wang, Rafael~M. Fernandes, and
  Peter~P. Orth.
\newblock Impact of damping on the superconducting gap dynamics induced by
  intense terahertz pulses.
\newblock {\em Phys. Rev. B}, 100:054504, Aug 2019.

\bibitem{Cavelleri2011}
D.~Fausti, R.~I. Tobey, N.~Dean, S.~Kaiser, A.~Dienst, M.~C. Hoffmann, S.~Pyon,
  T.~Takayama, H.~Takagi, and A.~Cavalleri.
\newblock Light-induced superconductivity in a stripe-ordered cuprate.
\newblock {\em Science}, 331(6014):189--191, 2011.

\bibitem{Cavelleri2014}
W.~Hu, S.~Kaiser, D.~Nicoletti, C.~R. Hunt, I.~Gierz, M.~C. Hoffmann,
  M.~Le~Tacon, T.~Loew, B.~Keimer, and A.~Cavalleri.
\newblock Optically enhanced coherent transport in yba2cu3o6.5 by ultrafast
  redistribution of interlayer coupling.
\newblock {\em Nature Materials}, 13:705--711, 2014.

\bibitem{Cavelleri2016}
M.~Mitrano, A.~Cantaluppi, D.~Nicoletti, S.~Kaiser, A.~Perucchi, S.~Lupi,
  P.~Di~Pietro, D.~Pontiroli, M.~Ricc`o, Clark~S. R., D.~Jaksch, and
  A.~Cavalleri.
\newblock Possible light-induced superconductivity in k3c60 at high
  temperature.
\newblock {\em Nature}, 530:461--464, 2016.

\bibitem{Knap2016}
Michael Knap, Mehrtash Babadi, Gil Refael, Ivar Martin, and Eugene Demler.
\newblock Dynamical cooper pairing in nonequilibrium electron-phonon systems.
\newblock {\em Phys. Rev. B}, 94:214504, Dec 2016.

\bibitem{Millis2016}
D.~M. Kennes, E.~Y. Wilner, D.~R. Reichman, and A.~J. Millis.
\newblock Transient superconductivity from electronic squeezing of optically
  pumped phonons.
\newblock {\em Nature Physics}, 13:479--483, 2017.

\bibitem{Aleiner2018}
Giuliano Chiriac\`o, Andrew~J. Millis, and Igor~L. Aleiner.
\newblock Transient superconductivity without superconductivity.
\newblock {\em Phys. Rev. B}, 98:220510, Dec 2018.

\bibitem{Matsunaga1145}
Ryusuke Matsunaga, Naoto Tsuji, Hiroyuki Fujita, Arata Sugioka, Kazumasa
  Makise, Yoshinori Uzawa, Hirotaka Terai, Zhen Wang, Hideo Aoki, and Ryo
  Shimano.
\newblock Light-induced collective pseudospin precession resonating with
  {H}iggs mode in a superconductor.
\newblock {\em Science}, 345(6201):1145--1149, 2014.

\bibitem{Aoki2015}
Naoto Tsuji and Hideo Aoki.
\newblock Theory of anderson pseudospin resonance with higgs mode in
  superconductors.
\newblock {\em Phys. Rev. B}, 92:064508, Aug 2015.

\bibitem{Cea2016}
T.~Cea, C.~Castellani, and L.~Benfatto.
\newblock Nonlinear optical effects and third-harmonic generation in
  superconductors: Cooper pairs versus higgs mode contribution.
\newblock {\em Phys. Rev. B}, 93:180507, May 2016.

\bibitem{Flaschner2018}
N.~Fläschner, D.~Vogel, M.~Tarnowski, B.~S. Rem, D.~S. Lühmann, M.~Heyl,
  J.~C. Budich, L.~Mathey, K.~Sengstock, and C.~Weitenberg.
\newblock Observation of dynamical vortices after quenches in a system with
  topology.
\newblock {\em Nature Physics}, 14(3):265--268, 2018.

\bibitem{Landig2016}
Renate Landig, Lorenz Hruby, Nishant Dogra, Manuele Landini, Rafael Mottl,
  Tobias Donner, and Tilman Esslinger.
\newblock Quantum phases from competing short- and long-range interactions in
  an optical lattice.
\newblock {\em Nature}, 532(7600):476--479, 2016.

\bibitem{Landin2018}
M.~Landini, N.~Dogra, K.~Kroeger, L.~Hruby, T.~Donner, and T.~Esslinger.
\newblock Formation of a spin texture in a quantum gas coupled to a cavity.
\newblock {\em Phys. Rev. Lett.}, 120:223602, May 2018.

\bibitem{Kroeze2018}
Ronen~M. Kroeze, Yudan Guo, Varun~D. Vaidya, Jonathan Keeling, and Benjamin~L.
  Lev.
\newblock Spinor self-ordering of a quantum gas in a cavity.
\newblock {\em Phys. Rev. Lett.}, 121:163601, Oct 2018.

\bibitem{Kroeze2019}
Ronen~M. Kroeze, Yudan Guo, and Benjamin~L. Lev.
\newblock Dynamical spin-orbit coupling of a quantum gas.
\newblock {\em Phys. Rev. Lett.}, 123:160404, Oct 2019.

\bibitem{Baden2014}
Markus~P. Baden, Kyle~J. Arnold, Arne~L. Grimsmo, Scott Parkins, and Murray~D.
  Barrett.
\newblock Realization of the dicke model using cavity-assisted raman
  transitions.
\newblock {\em Phys. Rev. Lett.}, 113:020408, Jul 2014.

\bibitem{Baumann_2010}
Kristian Baumann, Christine Guerlin, Ferdinand Brennecke, and Tilman Esslinger.
\newblock Dicke quantum phase transition with a superfluid gas in an optical
  cavity.
\newblock {\em Nature}, 464(7293):1301, Apr 2010.

\bibitem{Ritsch_2013}
Helmut Ritsch, Peter Domokos, Ferdinand Brennecke, and Tilman Esslinger.
\newblock Cold atoms in cavity-generated dynamical optical potentials.
\newblock {\em Rev. Mod. Phys.}, 85:553--601, Apr 2013.

\bibitem{Hemmerich_2015}
Jens Klinder, Hans Ke{\ss}ler, Matthias Wolke, Ludwig Mathey, and Andreas
  Hemmerich.
\newblock Dynamical phase transition in the open dicke model.
\newblock {\em Proceedings of the National Academy of Sciences},
  112(11):3290--3295, 2015.

\bibitem{yang2019}
H-X Yang, T~Tian, Y-B Yang, L-Y Qiu, H-Y Liang, A-J Chu, C~B Da${\u{g}}$, Y~Xu,
  Y~Liu, and L-M Duan.
\newblock Observation of dynamical quantum phase transitions in a spinor
  condensate.
\newblock {\em Phys. Rev. A}, 100(1):013622, 2019.

\bibitem{Tian2020}
T.~Tian, H.-X. Yang, L.-Y. Qiu, H.-Y. Liang, Y.-B. Yang, Y.~Xu, and L.-M. Duan.
\newblock Observation of dynamical quantum phase transitions with
  correspondence in an excited state phase diagram.
\newblock {\em Phys. Rev. Lett.}, 124:043001, Jan 2020.

\bibitem{rey2014}
A~M Rey, A~V Gorshkov, C~V Kraus, M~J Martin, M~Bishof, M~D Swallows, X~Zhang,
  C~Benko, J~Ye, N~D Lemke, and A~D Ludlow.
\newblock Probing many-body interactions in an optical lattice clock.
\newblock {\em Ann. Phys.}, 340(1):311, 2014.

\bibitem{Fuchs2002}
J.~N. Fuchs, D.~M. Gangardt, and F.~Lalo\"e.
\newblock Internal state conversion in ultracold gases.
\newblock {\em Phys. Rev. Lett.}, 88:230404, May 2002.

\bibitem{Albiez2005}
Michael Albiez, Rudolf Gati, Jonas F\"olling, Stefan Hunsmann, Matteo
  Cristiani, and Markus~K. Oberthaler.
\newblock Direct observation of tunneling and nonlinear self-trapping in a
  single bosonic {J}osephson junction.
\newblock {\em Phys. Rev. Lett.}, 95:010402, 2005.

\bibitem{Anker2005}
Th. Anker, M.~Albiez, R.~Gati, S.~Hunsmann, B.~Eiermann, A.~Trombettoni, and
  M.~K. Oberthaler.
\newblock Nonlinear self-trapping of matter waves in periodic potentials.
\newblock {\em Phys. Rev. Lett.}, 94:020403, 2005.

\bibitem{Reinhard2013}
Aaron Reinhard, Jean-F\'elix Riou, Laura~A. Zundel, David~S. Weiss, Shuming Li,
  Ana~Maria Rey, and Rafael Hipolito.
\newblock Self-trapping in an array of coupled {1D} {B}ose gases.
\newblock {\em Phys. Rev. Lett.}, 110:033001, 2013.

\bibitem{Levy2007}
S.~Levy, E.~Lahoud, I.~Shomroni, and J.~Steinhauer.
\newblock The a.c. and d.c. {J}osephson effects in a bose-einstein condensate.
\newblock {\em Nature}, 449:579, 2007.

\bibitem{Abbarchi2013}
M.~Abbarchi, A.~Amo, V.~G. Sala, D.~D. Solnyshkov, H.~Flayac, L.~Ferrier,
  I.~Sagnes, E.~Galopin, A.~Lemaître, G.~Malpuech, and J.~Bloch.
\newblock {Macroscopic quantum self-trapping and Josephson oscillations of
  exciton polaritons}.
\newblock {\em Nat. Phys.}, 9:275, 2013.

\bibitem{RevModPhys2010}
Cheng Chin, Rudolf Grimm, Paul Julienne, and Eite Tiesinga.
\newblock Feshbach resonances in ultracold gases.
\newblock {\em Rev. Mod. Phys.}, 82:1225--1286, 2010.

\bibitem{Deutsch:2010ky}
C.~Deutsch, F.~Ramirez-Martinez, C.~Lacro\^ute, F.~Reinhard, T.~Schneider,
  J.~N. Fuchs, F.~Pi\'echon, F.~Lalo\"e, J.~Reichel, and P.~Rosenbusch.
\newblock Spin self-rephasing and very long coherence times in a trapped atomic
  ensemble.
\newblock {\em Phys. Rev. Lett.}, 105:020401, 2010.

\bibitem{Solaro:2016iv}
C~Solaro, A~Bonnin, F~Combes, M~Lopez, X~Alauze, J~N Fuchs, F~Pi{\'e}chon, and
  F~Pereira Dos~Santos.
\newblock Competition between spin echo and spin self-rephasing in a trapped
  atom interferometer.
\newblock {\em Phys. Rev. Lett.}, 117:163003, 2016.

\bibitem{Piechon:2009cr}
F.~Pi\'echon, J.~N. Fuchs, and F.~Lalo\"e.
\newblock Cumulative identical spin rotation effects in collisionless trapped
  atomic gases.
\newblock {\em Phys. Rev. Lett.}, 102:215301, 2009.

\bibitem{bernien2017probing}
Hannes Bernien, Sylvain Schwartz, Alexander Keesling, Harry Levine, Ahmed
  Omran, Hannes Pichler, Soonwon Choi, Alexander~S Zibrov, Manuel Endres,
  Markus Greiner, et~al.
\newblock Probing many-body dynamics on a 51-atom quantum simulator.
\newblock {\em Nature}, 551(7682):579--584, 2017.

\bibitem{ebadi2021quantum}
Sepehr Ebadi, Tout~T Wang, Harry Levine, Alexander Keesling, Giulia Semeghini,
  Ahmed Omran, Dolev Bluvstein, Rhine Samajdar, Hannes Pichler, Wen~Wei Ho,
  et~al.
\newblock Quantum phases of matter on a 256-atom programmable quantum
  simulator.
\newblock {\em Nature}, 595(7866):227--232, 2021.

\bibitem{serbyn2021quantum}
Maksym Serbyn, Dmitry~A Abanin, and Zlatko Papi{\'c}.
\newblock Quantum many-body scars and weak breaking of ergodicity.
\newblock {\em Nature Physics}, 17(6):675--685, 2021.

\bibitem{kohlert2021experimental}
Thomas Kohlert, Sebastian Scherg, Pablo Sala, Frank Pollmann, Bharath~Hebbe
  Madhusudhana, Immanuel Bloch, and Monika Aidelsburger.
\newblock Experimental realization of fragmented models in tilted fermi-hubbard
  chains.
\newblock {\em arXiv preprint arXiv:2106.15586}, 2021.

\bibitem{He2019}
P.~He, M.~A. Perlin, S.~R. Muleady, R.~J. Lewis-Swan, R.~B. Hutson, J.~Ye, and
  A.~M. Rey.
\newblock Engineering spin squeezing in a 3d optical lattice with interacting
  spin-orbit-coupled fermions.
\newblock {\em Phys. Rev. Research}, 1:033075, Nov 2019.

\bibitem{Foss-Feig2017}
M.~Foss-Feig, P.~Niroula, J.~T. Young, M.~Hafezi, A.~V. Gorshkov, R.~M. Wilson,
  and M.~F. Maghrebi.
\newblock Emergent equilibrium in many-body optical bistability.
\newblock {\em Phys. Rev. A}, 95:043826, Apr 2017.

\bibitem{Lewisswan20212}
R.~J. Lewis-Swan, S.~R. Muleady, D.~Barberena, J.~J. Bollinger, and A.~M. Rey.
\newblock Characterizing the dynamical phase diagram of the dicke model via
  classical and quantum probes.
\newblock {\em Phys. Rev. Research}, 3:L022020, Jun 2021.

\bibitem{Gross2017}
Christian Gross and Immanuel Bloch.
\newblock Quantum simulations with ultracold atoms in optical lattices.
\newblock {\em Science}, 357(6355):995--1001, 2017.

\end{thebibliography}

\end{document}